\documentclass[12pt]{article}
\pdfoutput=1

\usepackage{putex}
\usepackage{graphicx}
\usepackage{caption}
\usepackage{amsmath}
\usepackage{array}
\usepackage{subcaption}
\usepackage{epstopdf}
\usepackage{enumerate}
\usepackage{cite}
\usepackage{youngtab}
\usepackage{tensor}
\usepackage{slashed}
\usepackage[aligntableaux=center]{ytableau}
\usepackage[utf8]{inputenc}
\usepackage[
      colorlinks=true,
      linkcolor=blue,
      urlcolor=blue,
      filecolor=black,
      citecolor=red,
      ]{hyperref}
\usepackage{braket}
\usepackage{mathtools}
\usepackage{rotating}
\usepackage{tabularx}
\newcolumntype{Y}{>{\centering\arraybackslash}X}
\newcolumntype{C}[1]{>{\centering\arraybackslash}p{#1}}
\usepackage{todonotes}
\usepackage{color, colortbl}
\definecolor{LightCyan}{rgb}{0.7,1,1}
\definecolor{Gray}{gray}{0.9}
\usepackage{xspace}

\newcommand{\HH}{\mathbb{H}}

\newcommand{\abs}[1]{\left\lvert #1 \right\rvert}

\newcommand {\be} {\begin {equation}}
\newcommand {\ee} {\end {equation}}

\newcommand {\bes} {\begin {equation*}}
\newcommand {\ees} {\end {equation*}}

\newcommand{\es}[2] {\begin{equation} \label{#1} \begin{split} #2 \end{split} \end{equation}}

\newcommand{\R}{\mathbb{R}}

\newcommand{\cL}{{\mathcal L}}
\newcommand{\cN}{{\mathcal N}}
\newcommand{\cO}{{\mathcal O}}

\newcommand{\cM}{{\mathcal M}}

\newcommand{\Q}{\mathbb{Q}}

\newcommand{\beq}{\begin{equation}}
\newcommand{\eeq}{\end{equation}}

\newcommand\De{{\ensuremath{{\Delta}}}}

\def\ie{\begin{equation}\begin{aligned}}
\def\fe{\end{aligned}\end{equation}}

\numberwithin{equation}{section}

\def\<{\langle}
\def\>{\rangle}

\newcommand{\fcy}[1]{\mathcal{#1}}

\newcommand{\dk}{\delta}        \newcommand{\Dk}{\Delta}

       \newcommand{\Lk}{\Lambda}

\newcommand{\nb}{\partial}

\newcommand{\Kahler}{K\"{a}hler\xspace}

\def\beg{\begin{equation}\begin{gathered}}
\def\eeg{\end{gathered}\end{equation}}

\begin{document}

\preprint{PUPT-2626 \\ MIT-CTP/5317}

\institution{PU}{Joseph Henry Laboratories, Princeton University, Princeton, NJ 08544, USA}
\institution{FHI}{Future of Humanity Institute, University of Oxford, Trajan House, Mill Street, \cr  Oxford, OX2 0DJ, UK}
\institution{Stanford}{Stanford Institute for Theoretical Physics, Department of Physics, \cr Stanford University,  Stanford, CA 94305, USA}
\institution{MIT}{Center for Theoretical Physics and Department of Mathematics,  \cr Massachusetts Institute of Technology, Cambridge, MA 02139, USA}

\title{
The Holographic Contributions to the Sphere Free Energy
}

\authors{Damon J.~Binder,\worksat{\PU,\FHI} Daniel Z.~Freedman,\worksat{\Stanford, \MIT} Silviu S.~Pufu,\worksat{\PU} and\\[10pt] Bernardo Zan\worksat{\PU}}

\abstract{
We study which bulk couplings contribute to the $S^3$ free energy $F(\mathfrak{m})$ of three-dimensional ${\cal N} = 2$ superconformal field theories with holographic duals, potentially deformed by boundary real-mass parameters $\mathfrak{m}$.  In particular, we show that $F(\mathfrak{m})$ is independent of a large class of bulk couplings that include  non-chiral F-terms and all D-terms.  On the other hand, in general, $F(\mathfrak{m})$ does depend non-trivially on bulk chiral F-terms, such as prepotential interactions, and on bulk real-mass terms.  These conclusions can be reached solely from properties of the AdS super-algebra, $\mathfrak{osp}(2|4)$.  We also consider massive vector multiplets in AdS, which in the dual field theory correspond to long single-trace  superconformal multiplets of spin zero.  We provide evidence that $F(\mathfrak{m})$ is insensitive to the vector multiplet mass and to the interaction couplings between the massive vector multiplet and massless ones.  In particular, this implies that $F(\mathfrak{m})$ does not contain information about scaling dimensions or OPE coefficients of single-trace long  scalar ${\cal N} = 2$ superconformal multiplets. 
}
\date{July 2021}

\maketitle

\tableofcontents

\section{Introduction}

The round sphere free energy has played several important roles in the study of Quantum Field Theory (QFT) in three space-time dimensions.  First, it was conjectured in \cite{Myers:2010xs,Myers:2010tj,Jafferis:2011zi,Klebanov:2011gs} and later proven in \cite{Casini:2011kv,Casini:2012ei} that the sphere free energy $F = - \log \abs{Z}$ provides a measure of the number of degrees of freedom in 3d conformal field theories (CFTs), in the sense that it obeys a monotonicity property under renormalization group (RG) flow known as the $F$-theorem.  In a preceding related development, it was shown that the ${\cal N} = 2$ superconformal R-symmetry at the infrared fixed point of RG flows that preserve Abelian flavor symmetry can be determined through a procedure called $F$-maximization \cite{Jafferis:2010un, Jafferis:2011zi, Closset:2012vg} (see~\cite{Pufu:2016zxm} for a review), which is the three-dimensional analog of a similar procedure called $a$-maximization in 4d ${\cal N} = 1$ SCFTs \cite{Intriligator:2003jj}.  $F$-maximization means maximizing the sphere free energy $F$ over a set of ``trial'' $U(1)$ R-charges.  For 3d SCFTs with Lagrangian descriptions and at least ${\cal N} = 2$ supersymmetry, it had been shown that $F$ can be calculated exactly using the technique of supersymmetric localization \cite{Kapustin:2009kz,Jafferis:2010un} (see \cite{Willett:2016adv} for a review), as a function of various deformation parameters such as field theory mass or Fayet-Iliopolous parameters.   This observation has led to remarkable tests (see for example \cite{Willett:2011gp,Kapustin:2010mh,Kapustin:2010xq,Amariti:2011uw,Safdi:2012re,Kapustin:2011gh,Jafferis:2011ns,Kapustin:2011vz,Morita:2011cs,Benini:2011mf}) and discoveries \cite{Agmon:2017lga} of various field theory dualities, as well as precision tests of AdS/CFT \cite{Drukker:2010nc,Herzog:2010hf,Freedman:2013oja}.\footnote{For the sphere free energy in 3d holographic theories beyond the tree-level supergravity approximation, see for instance \cite{Marino:2011eh, Bhattacharyya:2012ye, Mezei:2013gqa, Assel:2015hsa, Chester:2020jay, Chester:2021gdw,Minahan:2021pfv} for calculations on the field theory side and \cite{Bobev:2020egg, Bobev:2020zov, Bobev:2021oku} for studies of the bulk dual.}   Lastly, derivatives of the sphere free energy with respect to some of the parameters it depends on, evaluated at the SCFTs values of these parameters, have been related to integrated correlation functions \cite{Closset:2012vg,Binder:2018yvd,Binder:2020ckj} that have served as crucial inputs in  analytic bootstrap studies of ${\cal N} = 6$ and ${\cal N} = 8$ superconformal field theories at strong coupling \cite{Agmon:2017xes,Chester:2018aca,Agmon:2019imm,Binder:2019mpb,Binder:2020ckj,Binder:2021cif,Alday:2021ymb}.\footnote{See \cite{Binder:2019jwn,Chester:2019jas,Chester:2020dja,Chester:2020vyz}  for analogous studies in four-dimensional ${\cal N}= 4$ supersymmetric Yang-Mills theory.} The exact computations of the derivatives of the sphere free energy allowed one to fully evaluate these correlation functions up to the first few non-trivial orders in the derivative expansion on the string-theory / M-theory side.

The last line of work mentioned above raises the following question:  {\em  What information about string theory and, in particular, its weakly-coupled supergravity limit, is contained in the sphere free energy of a CFT with a holographic dual and in its deformations?}  In this paper we begin to answer this question for 3d ${\cal N} = 2$ SCFTs\@.   We choose to focus on 3d ${\cal N} =2$ theories partly because, as mentioned above, the sphere free energy is exactly calculable using supersymmetric localization in this case, so the answer to the question posed above has concrete quantitative consequences.  Intuitively, we should expect that since the sphere free energy is calculable using localization, it is a supersymmetric quantity, and it should therefore contain information only about (certain) supersymmetry-protected quantities in the bulk also.  Our goal in this paper is to make this statement precise.

We consider 3d SCFTs on $S^3$ deformed by field theory ``real mass'' parameters denoted collectively by $\mathfrak{m}$.  On $S^3$, one can introduce a real mass parameter for every Abelian flavor symmetry\footnote{The Abelian flavor symmetry group could be a subgroup of a non-Abelian flavor symmetry group.} of an SCFT by 1) coupling the abelian flavor current multiplet to a background vector multiplet; and 2) giving supersymmetry-preserving expectation values proportional to $\mathfrak{m}$ to the scalars in this background vector multiplet \cite{Closset:2012vg}.  As explained in \cite{Closset:2012vg}, if the real mass parameters $\mathfrak{m}$ are analytically continued to pure imaginary values, they correspond precisely to the trial R-charges used in the $F$-maximization procedure of \cite{Jafferis:2010un}.  

To make precise the statement that $F(\mathfrak{m})$ captures only certain supersymmetry-protected information in the bulk, we use an effective field theory approach\footnote{In Sections~\ref{SUSY} and \ref{CONFORMAL}, our construction can accommodate a large class of interaction terms, but in Section~\ref{ONSHELL} we will restrict to two-derivative interactions.} to construct the 4d bulk theory dual to our mass-deformed deformed 3d SCFT on $S^3$\@.  Thus, we explore the properties of $F(\mathfrak{m})$ in ${\cal N} = 2$ QFT in AdS, which we construct in several ways:  using properties of the $\mathfrak{osp}(2|4)$ supersymmetry algebra, as we do in Section~\ref{SUSY}, or starting from ${\cal N} = 2$ superconformal gravity and using superconformal tensor calculus, as we do in Section~\ref{CONFORMAL}, or starting from on-shell  matter-coupled ${\cal N} = 2$ supergravity, as we do in Section~\ref{ONSHELL}.   

In particular, in Sections~\ref{SUSY} and \ref{CONFORMAL}, we analyze various types of supersymmetric bulk interactions referred to as ``D-terms,'' ``chiral F-terms,'' ``non-chiral F-terms'', ``flavor current terms,'' as well as $1/4$-BPS interactions.  We define all these supersymmetric bulk interactions precisely using the $\mathfrak{osp}(2|4)$ algebra.    As we will show, the $\mathfrak{osp}(2|4)$ algebra implies that the sphere free energy (or derivatives thereof with respect to $\mathfrak{m}$ evaluated at the conformal point, $\mathfrak{m}=0$) is independent of all D-terms, $1/4$-BPS interactions, and non-chiral F-terms, but can in principle depend non-trivially on the chiral F-term couplings or on the coupling constant of the flavor current term, also known as a bulk real mass term.  This stands in contrast with ${\cal N} = 1$ theories in AdS, for which, in general, the sphere free energy has non-trivial dependence even on the ${\cal N} = 1$ D-term couplings.\footnote{As an example, we will consider a free massive ${\cal N} = 1$ vector multiplet in the one-loop approximation in Section~\ref{ONELOOP}.  In this case, the mass term is a D-term, but the one-loop contribution to the sphere free energy does depend non-trivially on the vector multiplet mass.} 

In Section~\ref{SUSY}, we show that the construction of the various supersymmetric interactions and the derivation of their properties can be done abstractly, without explicitly writing the supersymmetry transformation rules in detail.  This analysis relies only on the AdS supersymmetry algebra $\mathfrak{osp}(2|4)$.  As far as we know, this is a novel way of studying the supersymmetry properties of interactions in AdS space.  

It is also instructive to explore a systematic construction of supersymmetric interactions in AdS following the approach pioneered by  Festuccia-Seiberg \cite{Festuccia:2011ws}.  In this method one can obtain theories with global $\text{AdS}_4$ supersymmetry by starting from an off-shell supergravity theory and fixing an appropriate background for the fields of its Weyl multiplet.   For 4d ${\cal N} = 2$ theories, such a construction was considered in \cite{Klare:2013dka} starting from gauge-fixed ${\cal N} = 2$ conformal supergravity \cite{deWit:1980lyi}.  The construction of the AdS background of conformal supergravity is similar to the $S^4$ construction in \cite{Hama:2012bg,Klare:2013dka,Gomis:2014woa} and it requires compensating vector and tensor multiplets.  We review the $\text{AdS}_4$ construction in Section~\ref{CONFORMAL}.  We apply it to several examples and relate it to the discussion in Section~\ref{SUSY}.

While in the approaches developed in Sections~\ref{SUSY} and \ref{CONFORMAL}, it is clear how to construct the various supersymmetric interactions, this construction is often quite cumbersome, especially for D-terms.  One important shortcoming is that, to our knowledge, the simplest long multiplet, namely an ${\cal N} =2$  massive vector multiplet in AdS (consisting of one massive vector field, five real scalar fields, and four Dirac fermions), has not been constructed using either of these approaches.  The study of this multiplet is important because this multiplet arises generically in the effective $\text{AdS}_4$ theories corresponding to ${\cal N} = 2$ $\text{AdS}_4$ backgrounds of M-theory.  Thus, in Section~\ref{ONSHELL}, we develop another approach to obtain ${\cal N} =2$ effective field theories in AdS\@.      This approach is based on taking a certain rigid limit of ${\cal N} = 2$ on-shell supergravity coupled to matter.  In the ${\cal N} = 1$ case, a similar decoupling limit starting with on-shell ${\cal N} = 1$ supergravity was used in \cite{Adams:2011vw,Freedman:2016yue} to obtain ${\cal N} = 1$ theories in AdS\@.  We extend the method to several examples of ${\cal N} = 2$ theories.  In particular, we apply the Stueckelberg mechanism to construct the Lagrangian of the long vector multiplet and some of its interactions.

A shortcoming of the method of Section~\ref{ONSHELL} is that it may not be clear whether a particular construction corresponds to D-terms, F-terms, or $1/4$-BPS interactions.  It is reasonable to conjecture that the bulk theory of the long vector multiplet arises only from interactions that do not affect the $S^3$ free energy.   To support this conjecture, we show in Section~\ref{ONELOOP} that the one-loop partition function of a free bulk ${\cal N} = 2$ massive vector multiplet with mass $m_V$ is indeed independent of $m_V$.   As a more intricate piece of evidence for the conjecture, we show in Section~\ref{INTEGRATED} that the full supersymmetric exchange contribution of a massive vector multiplet to the fourth derivative $\frac{\partial^4 F}{\partial \mathfrak{m}^4} \big|_{\mathfrak{m}=0}$ vanishes.\footnote{We focus on the fourth derivative of $F$ because this is the lowest derivative to which the long vector multiplet contributes.}  These calculations indicate quite explicitly that the sphere free energy $F(\mathfrak{m})$ cannot be used to determine the vector multiplet mass $m_V$, which in the boundary theory corresponds to the scaling dimension of a long single-trace superconformal multiplet with a scalar superconformal primary.  By contrast, the hypermultiplet mass $m_H$ arises from a conserved current term in the bulk, and one might expect that $F(\mathfrak{m})$ depends non-trivially on $m_H$.  Indeed, we calculate the one-loop partition function of a hypermultiplet in Section~\ref{ONELOOP} and exhibit this fact.

We close this introduction with a more compact summary of the rest of the paper.   In Section~\ref{SUSY}, we construct the various supersymmetric interactions using the $\mathfrak{osp}(2|4)$ algebra and show that in theories with holographic duals, $F(\mathfrak{m})$ is independent of the D-terms, the non-chiral F-terms, and the $1/4$-BPS interactions.  In Section~\ref{CONFORMAL} we relate our construction from Section~\ref{SUSY} to the more standard way of obtaining effective field theories on curved manifolds using background off-shell ${\cal N} = 2$ supergravity.  In Section~\ref{ONSHELL} we present  yet another approach for constructing effective field theories in AdS that is based on a rigid limit of on-shell ${\cal N} = 2$ supergravity.  We use this approach to construct supersymmetric actions for a massive vector multiplet.  In Section~\ref{ONELOOP}, we evaluate the bulk one-loop contribution to the $S^3$ partition function coming from free massive vector multiplets and hypermultiplets, and show that, while the hypermultiplet contribution depends non-trivially on the hypermultiplet mass, the massive vector multiplet contribution is independent of the vector multiplet mass.  In Section~\ref{INTEGRATED}, we consider the contributions to the fourth derivative $\frac{\partial^4 F}{\partial \mathfrak{m}^4} \big|_{\mathfrak{m}=0}$ coming from the tree-level exchange of a massive vector multiplet and show that it vanishes.  The observations of Sections~\ref{ONELOOP} and \ref{INTEGRATED} support the conjecture mentioned above that the mass term in the massive vector multiplet action as well as the interactions between massive vector multiplets and massless vector multiplets are either D-terms, non-chiral F-terms, or $1/4$-BPS interactions.  Lastly, we end in Section~\ref{DISCUSSION} with a discussion of our results and future directions.

\section{Supersymmetric interactions in AdS}
\label{SUSY}

The goal of this section is to see what kinds of supersymmetric bulk interactions the sphere free energy $F(\mathfrak{m})$ depends on.  We study this question perturbatively in $\mathfrak{m}$, meaning that we study the dependence of derivatives of $F(\mathfrak{m})$ with respect to $\mathfrak{m}$, evaluated at $\mathfrak{m}=0$, on the bulk interactions.  The advantage of working perturbatively in $\mathfrak{m}$ is that we can expand the action around the AdS background that corresponds to the CFT at $\mathfrak{m}=0$, and thus we can make use of the AdS supersymmetry algebra $\mathfrak{osp}(2|4)$.  

As we explain in more detail below, the supersymmetry algebra in either AdS or flat space has eight supercharges, four of which are left-handed and four right-handed.  In addition half of the right-handed (left-handed) supercharges have R-charge $+1$ while the other half have R-charge $-1$.  Supersymmetric interactions can preserve various amounts of SUSY\@.  In particular, in flat space, we may consider the following interactions:\footnote{In superconformal field theories, these interactions were discussed in \cite{Cordova:2016xhm}.} 
 \begin{itemize}
  \item D-terms, obtained by acting with all $8$ supercharges on some scalar field $X$.
  \item  $1/4$-BPS interactions, obtained by acting with all four supercharges of a given R-charge as well as two supercharges of the opposite R-charge and the same chirality on some scalar field $X$.
  \item Non-chiral F-terms, obtained by acting with all $4$ supercharges of a given R-charge on some scalar field $X$.
  \item Chiral F-terms, obtained by acting with all $4$ supercharges of a given chirality on some scalar field $X$.
  \item Flavor current terms, obtained by acting with $2$ supercharges on some scalar field $X$.
 \end{itemize}
Apart from the D-terms, in all the other cases, it is assumed that $X$ cannot be obtained by acting with supercharges on some other field and that it obeys appropriate shortening conditions.  In AdS, we keep the same terminology above, but we allow, in principle, each supersymmetric term described above to receive corrections in $1/L$, where $L$ is the curvature radius of AdS\@.  Such corrections are in general needed in order to preserve supersymmetry.

As we will argue, of the types of supersymmetric interactions mentioned above, the D-terms, $1/4$-BPS interactions, and the non-chiral F-terms are exact with respect to one of the supercharges preserved by the boundary mass deformation, while the chiral F-terms and the flavor current terms in general are not.  It follows that $F(\mathfrak{m})$ is then independent of the former type of interactions, but it could depend on the latter.

Let us now provide more details, starting with the SUSY algebra and its properties in Sections~\ref{SUSYAlgebraSection} and \ref{IDENTITIES}, following with the construction of the various supersymmetric interactions in Section~\ref{SUSYINTERACTIONS}, and ending with the proof of Euclidean SUSY and SUSY exactness in Sections~\ref{EUCLIDEAN} and \ref{EXACTNESS}.

\subsection{${\cal N} = 2$ supersymmetry algebra in $\text{AdS}_4$}
\label{SUSYAlgebraSection}

${\cal N} = 2$ supersymmetry in $\text{AdS}_4$ is based on the $\mathfrak{osp}(2|4)$ superalgebra that acts locally on the fields.  Since this algebra and its properties are the main ingredients in the construction of supersymmetric interactions, let us take a moment to review it.   Since we will be working with spinors, we should be clear on our conventions:  We will follow the conventions established in \cite{Freedman:2012zz} for spinors, gamma matrices, index placement, etc.

The bosonic subalgebra of $\mathfrak{osp}(2|4)$ contains the AdS spacetime algebra $\mathfrak{so}(3,2) \cong \mathfrak{sp}(4)$.  In our treatment,\footnote{Our treatment of the algebras $\mathfrak{sp}(4)$ and $\mathfrak{osp}(2|4)$ is related to a more standard presentation in Appendix~\ref{OSP}.} the ten generators of $\mathfrak{sp}(4)$  split into four ``momenta'' $P_a$ and six local Lorentz generators $M_{ab}$.  The $P_a$ act on fields as covariant derivatives $P_a=D_a$,\footnote{Here $P_a=D_a= e_a^\mu D_\mu$ where
	$e_a^\mu$ and $D_\mu$ are, respectively, inverse frame fields and standard covariant derivative for $\text{AdS}_4$.} while the $M_{ab}$ transform only the frame indices of vectors and spinors, e.g.~for a spinor field $\Psi(x)$, the $M_{ab}$ act as $M_{ab}\Psi(x) = -\frac12 \gamma_{ab} \Psi(x)$.  In flat space, the $P_a$ commute, but in AdS, they satisfy
 \es{PCommut}{
  [P_a, P_b] = \frac{1}{L^2} M_{ab} \,,
 }
where $L$ is the radius of $\text{AdS}_4$.  (See, for example, Section~12.6.1 of~\cite{Freedman:2012zz}.)  The other commutators between $M_{ab}$ and $P_a$ simply follow from the Lorentz transformation properties of $P_a$ and $M_{ab}$:
 \es{PMCommut}{
  [P_a, M_{bc}] = 2 \eta_{a[b} P_{c]} \,, \qquad
   [M_{ab}, M_{cd}] = 4 \eta_{[a[c} M_{d]b]} \,.
 }

The superalgebra $\mathfrak{osp}(2|4)$ extends the $\mathfrak{sp}(4)$ AdS algebra of $P_a$ and $M_{ab}$ by two left-handed fermionic generators $\Q_i$, $i=1, 2$ (obeying $P_L \Q_i = \Q_i$) and two right-handed fermionic generators $\Q^i$ (obeying $P_R \Q^i = \Q^i$).\footnote{The four-component spinor indices are usually suppressed; when needed they are denoted by $\alpha,\beta,\ldots$.  The $\Q^i,~\Q_i$ may be viewed as the chiral projections of Majorana spinors.}  There is also a $U(1)_R$ symmetry generator $R$ with the property that $Q^1$ and $Q^2$ have R-charges $+1$ and $-1$, respectively, and
$Q_1$ and $Q_2$ have R-charges $-1$ and $+1$.  This information is
incorporated in the commutation relations 
 \es{RCommut}{
  [R, \Q_i] = -\tau_{3i}{}^j \Q_j \,, \qquad [R, \Q^i] = \tau_{3j}{}^i \Q^j \,,
 }
where $\tau_{3i}{}^j = i (\sigma_3)_i{}^j$, where $\sigma_3$ is the third Pauli matrix.\footnote{In Section~\ref{CONFORMAL} we derive the actions of $\mathfrak{osp}(2|4)$-invariant theories from a parent superconformal theory invariant under the algebra $\mathfrak{su}(2,2|2)$ with R-symmetry $\mathfrak{su}(2)\times \mathfrak{u}(1)$.
	The Pauli matrix $\tau_3$ is a convenient choice of a direction within $\mathfrak{su}(2)$ for the residual $\mathfrak{u}(1) \cong \mathfrak{so}(2)$  of the $\mathfrak{osp}(2|4)$ theories.}
The commutators of the $\mathfrak{so}(3, 2)$ generators with the supercharges are
 \es{OtherCommutators}{
  [P_a, \mathbb{Q}_i] &= -\frac 1{2L} \tau_{3 ij} \gamma_a \mathbb{Q}^j \,, \qquad 
   [P_a, \mathbb{Q}^i] = -\frac 1{2L} \tau_3^{ij} \gamma_a \mathbb{Q}_j \,, \\
   [M_{ab}, \mathbb{Q}_i] &= - \frac 12 \gamma_{ab} \mathbb{Q}_i \,,\qquad \ \ \ \  [M_{ab}, \mathbb{Q}^i] = - \frac 12 \gamma_{ab} \mathbb{Q}^i \,. 
 }
Last and most important, the anti-commutator of the supercharges are
 \es{SUSYAlgebra}{
  \{\mathbb{Q}_i, \bar{\mathbb{Q}}_j\} &= \frac{1}{2L} \left( \frac 12  \tau_{3ij} \gamma^{ab} M_{ab} + \varepsilon_{ij} R\right) P_L  \,,\\
  \{\mathbb{Q}^i, \bar{\mathbb{Q}}^j \} &= \frac{1}{2L} \left( \frac 12 \tau_{3}^{ij} \gamma^{ab} M_{ab} + \varepsilon^{ij} R \right) P_R  \,, \\
   \{\mathbb{Q}_i, \bar{\mathbb{Q}}^j\} &= \{\mathbb{Q}^j, \bar{\mathbb{Q}}_i\} =  -\frac 12 \delta_i^j \slashed{P} \,.
 }
Here, $\varepsilon^{12} = -\varepsilon^{21}=  \varepsilon_{12}  = -\varepsilon_{21} = 1$.\footnote{The $i, j, \ldots$ indices are raised and lowered following the NW-SE convention.  In particular, note that $(\tau_3)_{ij} = \tau_{3i}{}^k \varepsilon_{kj} = (i \sigma_1)_{ij}$ and $(\tau_3)^{ij} = \varepsilon^{ik} \tau_{3k}{}^j = (-i \sigma_1)^{ij}$.  For useful information on the $\vec{\tau}_i{}^j$ matrices, see Appendix~20.A of~\cite{Freedman:2012zz}. It is important to note that the upper/lower placement of indices $i,j,\dots$ on all spinors cannot be changed because it indicates their chirality.}    All commutators not explicitly written in \eqref{PCommut}--\eqref{SUSYAlgebra} vanish.

It is useful to note that complex conjugation acts by raising all the lower $i, j$ indices and lowering all the upper ones, without any additional minus signs.  It is a general rule that, in addition to complex conjugating all the numerical factors, complex conjugation interchanges $P_L$ and $P_R$ and hence it transforms left-handed spinors to right-handed ones and vice versa.

In supersymmetric field theories, invariant interaction terms are commonly defined as integrals of the top component of various multiplets of operators.  The SUSY variations of these top components are total spacetime derivatives.  In the remainder of this section we construct several interaction terms for ${\cal N}=2$ SUSY in $\text{AdS}_4$ without the need for a detailed description of the multiplets. It is sufficient to apply properly chosen  products of the supercharges to the lowest weight operators which define each multiplet.

\subsection{Identities obeyed by the supercharges}
\label{IDENTITIES}

To construct the supersymmetric actions directly from the algebra, all we need are a few identities obeyed by the supercharges.  The first is the (Majorana-like) flip property
 \es{FlipProp}{
  \bar \Q_i \Q_j = \bar \Q_j \Q_i - \frac 1L \varepsilon_{ij} R \,,
 }
which can be derived from the first equation in \eqref{SUSYAlgebra}.\footnote{To derive it, note that with spinor indices written out explicitly, the first equation in \eqref{SUSYAlgebra} reads, $\mathbb{Q}_{i\alpha} \mathbb{Q}_j{}^\beta + \mathbb{Q}_j{}^\beta\mathbb{Q}_{i\alpha}  = \frac{1}{4L}  \tau_{3ij} (\gamma^{ab} P_L)_\alpha{}^\beta M_{ab} +  \frac{1}{2L} \varepsilon_{ij} R (P_L)_\alpha{}^\beta$.  Contracting this expression with ${\cal C}^\alpha{}_\beta = - \delta^\alpha_\beta$ gives the desired result. See Section~3.2.2 of \cite{Freedman:2012zz} for more information on the ${\cal C}$ matrix and on using spinor indices.}  For strictly anti-commuting spinors, as in flat space, $\bar \Q_i \Q_j =\bar \Q_j \Q_i$ and the second term in (2.6) would vanish.  This term is a consequence of the curvature of AdS\@.  Similarly, for strictly anti-commuting $\Q_i$, the cubic term $\bar\epsilon \Q_i \bar \Q_j \Q_k  + \text{cyclic permutations in $i, j, k$}$ would vanish (for any spinor parameter $\bar\epsilon$) by the Schouten identity.  But because $\Q_i$ obey the first equation in \eqref{FlipProp}, we have instead the modified Schouten identity
  \es{Schouten}{
  \bar \epsilon \Q_i \bar \Q_j \Q_k 
   + \text{cyclic}
   =  \frac{1}{4L} \tau_{3 ij} \bar \epsilon \gamma^{ab} \Q_k M_{ab} 
    - \frac 1{2L} \varepsilon_{ij} \bar \epsilon \Q_k R 
    + \frac{1}{L} \tau_{3 ij} \bar \epsilon \Q_k + \text{cyclic} \,,
 }
 where ``cyclic'' refers to the two cyclic permutations in $i, j, k$.   Lastly, there are two commutators that will be needed:
 \es{CommutRel}{
  [\bar \epsilon \mathbb{Q}^i, \bar \Q_j \Q_k] &= 
   -\frac 12 \delta^i_j \bar \epsilon \slashed{P} \Q_k 
   - \frac 12 \delta^i_k \bar \epsilon \slashed{P} \Q_j
    - \frac 1L \delta^i_k \tau_{3 jl} \bar \epsilon \Q^l \,, \\
  [\bar \epsilon \Q_i, \bar \Q_j \Q_k]
   &= \frac{ \bar \epsilon \gamma^{ab} \left( \tau_{3 ij} \Q_k + \tau_{3 ik}  \Q_j \right)  M_{ab}}{4L}  
    + \frac{\bar \epsilon ( \varepsilon_{ij} \Q_k + \varepsilon_{ik} \Q_j) R}{2L} \\
     &{}+ \frac{3 \tau_{3 ij} \bar \epsilon \Q_k - \varepsilon_{ij} \tau_{3k}{}^l \bar \epsilon \Q_l }{2L}  \,.
 }
They can be derived straightforwardly using the algebra given in the previous subsection.   In addition to \eqref{FlipProp}--\eqref{CommutRel}, we also have the complex conjugate identities obtained by changing the positions of all the $i$, $j$, $k$ indices.  The relations \eqref{Schouten}--\eqref{CommutRel} are valid for an arbitrary spinor $\epsilon$, but when we apply them to AdS supersymmetry below, $\epsilon$ will be a Killing spinor to be defined shortly.

\subsection{Supersymmetric interactions}
\label{SUSYINTERACTIONS}

Given the ${\cal N} =2$ AdS supersymmetry algebra reviewed above, one can straightforwardly construct supersymmetry-preserving interactions that are based on multiplets of the AdS supersymmetry.  Note that for any (fundamental or composite) field $\Phi$, the supersymmetry variation $\delta \Phi$ is
 \es{deltaPhi}{
  \delta \Phi = \bar \epsilon^i \Q_i \Phi + \bar \epsilon_i \Q^i \Phi \,,
 }
with SUSY parameters $\epsilon^i$  and $\epsilon_i$, which are Majorana spinors, projected by $P_L$ and $P_R$, respectively.  In flat space, the SUSY parameters are constant spinors, but this is not so in AdS, where they must obey Killing spinor equations.  Indeed, the variation $\delta$ should obey $\delta D_a \Phi = D_a \delta \Phi$ for any $\Phi$.  This condition requires\footnote{$\delta D_a \Phi = D_a \delta \Phi$ implies $(D_a \bar \epsilon^i) \Q_i + \bar \epsilon^i [P_a, \Q_i] + (D_a \bar \epsilon_i) \Q^i + \bar \epsilon_i [P_a, \Q^i] = 0$, which, together with the commutation relations in \eqref{OtherCommutators}, implies \eqref{Killing}.  We will derive these Killing spinor equations in a different way from conformal supergravity in Section~\ref{CONFORMAL}.}
  \es{Killing}{
   D_a \epsilon^i &=- \frac 1{2L} \tau_{3}^{ij}  \gamma_a \epsilon_j \,, \qquad
    D_a \epsilon_i = - \frac 1{2L} \tau_{3 ij}\gamma_a   \epsilon^j   \,.
 }
These are the Killing spinor equations for ${\cal N} = 2$ SUSY in AdS\@. The appearance of $\tau_3$ may be unfamiliar, but one can put these equations into a more familiar form after defining $\epsilon = \epsilon^1 - \tau_3^{1i} \epsilon_i$ and $\epsilon' = \epsilon^2 - \tau_3^{2i} \epsilon_i$.  The equations \eqref{Killing} then imply $D_a \epsilon = \frac{1}{2L} \gamma_a \epsilon$ and $D_a \epsilon' = \frac{1}{2L} \gamma_a \epsilon'$, which are the more standard AdS Killing spinor equations.  From $\epsilon$ and $\epsilon'$ we can extract $\epsilon^i$ and $\epsilon_i$ by performing the appropriate chiral projections. 

A supersymmetric interaction is a term $S_\text{int}$ in the action that obeys $\delta S_\text{int} = 0$.  For the F-terms, D-terms, and flavor current terms we will construct, $S_\text{int}$ is an integral of a local operator ${\cal O}_\text{int}$, namely
 \es{SIntegral}{
  S_\text{int} = \int d^4 x\, \sqrt{-g} \, {\cal O}_\text{int} \,,
 }
such that $\delta {\cal O}_\text{int}$ is a total spacetime derivative.

\subsubsection{Chiral F-terms}
\label{FTERMS}

The chiral F-terms are $\mathfrak{osp}(2|4)$-invariant interactions constructed from (composite) AdS chiral multiplets.  A chiral multiplet is a multiplet that contains an R-symmetry neutral scalar field $A$ that is annihilated by all four right-handed supercharges:  
 \es{ChiralShortening}{
   \Q^i A = 0 \,, 
  } 
for $i=1, 2$, as well as all of the other fields that can be obtained by acting with $\Q_i$ on $A$---See Table~\ref{ChiralTable}.\footnote{Acting with $\Q^i$ does not produce any new fields because $\Q^i$ can be (anti)commuted to all the way to the right using the algebra, where they annihilate $A$.  Since $ \{\mathbb{Q}_i, \bar{\mathbb{Q}}^j\} = \{\mathbb{Q}^j, \bar{\mathbb{Q}}_i\} =  -\frac 12 \delta_i^j \slashed{P}$, and since $P_a$ acts on the fields as a covariant derivative $P_a = D_a$, the commutators generated during this process give  covariant derivatives of the fields already present in Table~\ref{ChiralTable}.}   (An anti-chiral multiplet would obey $\Q_i A = 0$ instead.)  The condition $\Q^i A = 0$ is a shortening condition in AdS\@.

\begin{table}[ht]
\begin{center}
\begin{tabular}{c|c|c}
 field & $\mathfrak{su}(2)_L \oplus \mathfrak{su}(2)_R$ spin & $U(1)_R$ charge \\
 \hline
 $A$ & $\left( 0 , 0 \right)$ & $0$ \\
 $\Q_i A$ & $\left( \frac 12 , 0 \right)$ & $\pm 1$ \\
 $\bar \Q_{(i} \Q_{j)} A$ & $\left( 0 , 0 \right)$ & $0$, $\pm 1$ \\
 $\varepsilon^{ij} \bar \Q_{i} \gamma^{ab} \Q_{j} A$ & $\left( 1 , 0 \right)$ & $0$ \\
 $\Q_i \bar \Q_1 \Q_2 A$ & $\left( \frac 12 , 0 \right)$ & $\pm 1$ \\
 $\bar \Q_1 \Q_1 \bar \Q_2 \Q_2 A $ & $\left( 0 , 0 \right)$ & $0$ 
\end{tabular}
\caption{The field content of the chiral multiplet, with all operators being complex.  It contains $16 + 16$ independent real operators, where operators of spin $(j_L, j_R)$ are counted with multiplicity $(2j_L +1)(2j_R + 1)$.\label{ChiralTable}}
\end{center}
\end{table}%

The F-term preserves $\mathfrak{osp}(2|4)$, and therefore it must be R-charge neutral and a Lorentz scalar.  From Table~\ref{ChiralTable}, we see that there are three such fields:
 \es{RNeutral}{
  \text{$R$-charge neutral scalars:} \qquad
   A \,, \quad  \bar{\mathbb{Q}}_1 \mathbb{Q}_2 A \,, \quad \bar{\mathbb{Q}}_1 \mathbb{Q}_1 \bar{\mathbb{Q}}_2 \mathbb{Q}_2 A
 }
Since we would like the F-term to preserve $\mathfrak{osp}(2|4)$, it should be constructed as an integral of some linear combination of the R-charge neutral scalars.   We claim that the F-term is given by the expression 
 \es{FTermDamon}{
  S_F &= \int d^4 x\, \sqrt{-g} \, {\cal O}_F(x) \,, \\
   \text{with}\qquad {\cal O}_F(x) &= \frac{6}{L^2} A(x) + \frac{4i}{L} \bar{\mathbb{Q}}_1 \mathbb{Q}_2 A(x)   + \bar{\mathbb{Q}}_1 \mathbb{Q}_1 \bar{\mathbb{Q}}_2 \mathbb{Q}_2  A(x) \,,
 }
as we will now show.

To prove \eqref{FTermDamon}, it is useful to consider the more general ansatz
 \es{FTermAnsatz}{
   {\cal O}_F(x) =c_1 A(x) + c_2 \bar{\mathbb{Q}}_1 \mathbb{Q}_2 A(x)   + \bar{\mathbb{Q}}_1 \mathbb{Q}_1 \bar{\mathbb{Q}}_2 \mathbb{Q}_2  A(x)\,.
 }
Using $M_{ab}A(x) =RA(x)=0$ and the rearrangements \eqref{Schouten} and \eqref{FlipProp}, one can reverse the indices 1 and 2 in \eqref{FTermAnsatz} and write the equivalent form
 \es{OFAlternative}{
  {\cal O}_F(x) = c_1 A(x) + c_2 \bar{\mathbb{Q}}_2 \mathbb{Q}_1 A(x)   + \bar{\mathbb{Q}}_2 \mathbb{Q}_2 \bar{\mathbb{Q}}_1 \mathbb{Q}_1  A(x) \,.
 }
Since the Killing spinor equations \eqref{Killing} relate $\epsilon^1$ to $\epsilon_2$ and $\epsilon^2$ to $\epsilon_1$, the total variation \eqref{deltaPhi}, with $\Phi = {\cal O}_F$, splits into the two independent terms
 \es{Separate}{
   (\bar \epsilon^1 \Q_1 + \bar \epsilon_2 \Q^2) {\cal O}_F \,, \qquad
    (\bar \epsilon_1 \Q^1 + \bar \epsilon^2 \Q_2) {\cal O}_F \,,
 }
each of which we must separately require to be a total derivative.

First, let us compute $ (\bar \epsilon^1 \Q_1 + \bar \epsilon_2 \Q^2) {\cal O}_F $.  The first part, $\bar \epsilon^1 \Q_1 {\cal O}_F$, is easier because the Schouten identity implies $\bar \epsilon^1 \Q_1 \bar \Q_1 \Q_1 = 0$, and so $\bar \epsilon^1 \Q_1 $ acts only on the first two terms of \eqref{FTermAnsatz}:
 \es{FirstComputation}{
   \bar \epsilon^1 \Q_1  {\cal O}_F 
    = \left( c_1 + \frac{i c_2}{L} \right) \bar \epsilon^1 \Q_1 A - \frac{c_2}{2}  \bar \epsilon^1 \Q_2 \bar \Q_1 \Q_1A
     \,.
 }
On the last term, we used the Schouten identity \eqref{Schouten} and the flip property \eqref{FlipProp} as well as the fact that $A$ is annihilated by both $R$ and $M_{ab}$.  To compute $\bar \epsilon_2 \Q^2 {\cal O}_F$, it is easier to start with \eqref{OFAlternative}.  We use $\bar \epsilon_2 \Q^2 A = 0$ as well as the commutation relation \eqref{CommutRel}:
 \es{Q2OF}{
  \bar \epsilon_2 \Q^2 {\cal O}_F
   = \left( \frac{i}{L} - \frac{c_2}{2} \right) \bar \epsilon_2 \slashed{D}  \Q_1 A 
     -
      \bar \epsilon_2 \slashed{D} \Q_2 \bar{\mathbb{Q}}_1 \mathbb{Q}_1 A \,.
 }
When summing together \eqref{FirstComputation} and \eqref{Q2OF}, we should use the equality $D_a\bar\epsilon_2 \gamma^a  =\frac{2i}{L}\bar\epsilon_1$, which is a contraction of the Majorana conjugate of \eqref{Killing}.  Note that this implies that for any spinor $\Upsilon$, the sum of $c\bar\epsilon_2\slashed{D}\Upsilon + c'\Upsilon$ is the total derivative $c D_a(\bar\epsilon_2\gamma^a\Upsilon)$ provided that the coefficients satisfy $c'=\frac{2i}{L}c$.  Applying this observation to the sum of \eqref{FirstComputation} and \eqref{Q2OF} gives us two equations for the parameters $c_1$ and $c_2$, namely 
 \es{c12Eqs}{
  c_1 + \frac{i c_2}{L} = \frac{2i}{L} \left( \frac{i}L - \frac{c_2}{2} \right) \,, \qquad
   -\frac{c_2}{2} = - \frac{2i}L \,,
 }
which have the solutions 
 \es{Solns}{
  c_1 = \frac{6}{L^2} \,, \qquad c_2 = \frac{4i}{L} \,.
 }
This proves that the F-term must be of the form in Eq.~\eqref{FTermDamon}.  Moreover, we have also shown that
 \es{OFDer}{
   (\bar \epsilon^1 \Q_1 + \bar \epsilon_2 \Q^2) {\cal O}_F
    = D_a \Psi^a \,, \qquad \Psi^a \equiv   -\bar \epsilon_2 \gamma^a \left(  \frac {i}{L}  \Q_1   A
     +
      \Q_2 \bar{\mathbb{Q}}_1 \mathbb{Q}_1  A\right)   \,,
 }
which is a total derivative, as advertised. 
The same calculation, with the indices $1$ and $2$ swapped on the supercharges and on the Killing spinors,  shows that the second quantity in \eqref{Separate} is also a total derivative:
 \es{OFDer2}{
   (\bar \epsilon^2 \Q_2 + \bar \epsilon_1 \Q^1) {\cal O}_F
    = D_a \tilde\Psi_F^a \,, \qquad \tilde\Psi_F^a \equiv  - \bar \epsilon_1 \gamma^a \left( \frac {i}{L}  \Q_2A  
     +
      \Q_1 \bar{\mathbb{Q}}_2 \mathbb{Q}_2 A \right)  \,.
 }
Thus, $\delta {\cal O}_F$ is a total derivative,
 \es{FinalDeltaOF}{
  \delta {\cal O}_F = D_a (\Psi_F^a + \tilde\Psi_F^a) \,,
 }
which shows that the variation $\delta S_F$ of the F-term interaction $S_F = \int d^4 x\, \sqrt{-g} \, {\cal O}_F(x) $ is a boundary term:  $\delta S_F = \int_\partial d^3x \, \sqrt{-\gamma} n_a (\Psi_F^a + \tilde\Psi_F^a)$, where $n_a$ is the outward pointing normal and $\sqrt{-\gamma}$ is the square root of the determinant of the boundary metric. Thus, $S_F$ is supersymmetric provided that $\Psi_F^a + \tilde\Psi_F^a$ vanishes at the boundary sufficiently fast.  Alternatively, it may be possible to add a pure boundary term to $S_F$ whose variation would cancel $\delta S_F$ precisely.  See \cite{Freedman:2016yue} for other cases in which boundary terms are necessary in order to make the AdS action supersymmetric.

\subsubsection{Flavor current term}
\label{FLAVOR}

Another supersymmetric term in the bulk comes from a flavor current multiplet in AdS\@.  We will provide a more systematic description of this multiplet in the next section, but for now we note that that the lowest components of the multiplet are the three $J_{ij}=J_{ji}$, subject to the reality condition $J^{ij}=J_{ij}^*$.  (Indices are raised/lowered with the $\varepsilon$-symbol.) The reality condition implies that the component $iJ_{12}$ is strictly real, while $J_{11}$ and $J_{22}$ are each other's complex conjugates.  The R-charges are determined by the transformation rule $[R, J_{ij}] =-\tau_{3\,i}{}^k J_{kj}-\tau_{3\,j}{}^l J_{il}$.  This tells us that $J_{12}$, $J_{11}$, $J_{22}$ have R-charges $0$, $-2$, and $+2$, respectively.   In addition to these facts, the flavor current multiplet is also defined by shortening conditions that restrict the fermionic components:
 \es{CurrentShortening}{
  \Q_{(i} J_{jk)} = \Q^{(i} J^{jk)} = 0  \,.
 }
 Apart from $J_{ij}$, the other independent fields in the multiplet are constructed by acting with $\Q_i$ and $\Q^i$ on $J_{jk}$ subject to the constraints \eqref{CurrentShortening}---See Table~\ref{ConservedTable} for a list of independent operators.  Importantly, the multiplet contains a conserved current, so it is necessarily associated with a $U(1)$ symmetry of the effective field theory in AdS\@.  (We will see in Section~\ref{FLAVORSUGRA} that a representation of the superconformal algebra on this multiplet's component fields requires that the top component is a conserved current with scale dimension $\Delta =3$.)

\begin{table}[ht]
\begin{center}
\begin{tabular}{c|c|c}
 field & $\mathfrak{su}(2)_L \oplus \mathfrak{su}(2)_R$ spin & $U(1)_R$ charge \\
 \hline
 $J_{12}$ & $\left( 0 , 0 \right)$ & $0$ \\
 $J_{11}, J_{22}$ & $\left( 0 , 0 \right)$ & $-2, 2$ \\
 $\Q_1 J_{12}, \Q_2 J_{12}$ & $\left( \frac 12 , 0 \right)$ & $-1, 1$ \\
 $\Q^1 J_{12}, \Q^2 J_{12}$ & $\left( 0 , \frac 12 \right)$ & $1, -1$ \\
 $\bar \Q_{1} \Q_{2} J_{12}$ & $\left( 0 , 0 \right)$ & $0$ \\
 $\bar \Q^{1} \Q^{2} J_{12}$ & $\left( 0 , 0 \right)$ & $0$ \\
 $(\bar \Q^1 \gamma^{a} \Q_1 - \bar \Q^2 \gamma^{a} \Q_2) J_{12}$ & $\left( \frac 12, \frac 12 \right) \text{ conserved}$ & $0$ \\
\end{tabular}
\caption{The field content of the flavor current multiplet.  It contains $8 + 8$ independent real operators.\label{ConservedTable}}
\end{center}
\end{table}%

A supersymmetry-preserving deformation involving the fields of this multiplet must be R-charge neutral.  In addition to $J_{12}$, there are two other R-charge neutral scalar fields in the multiplet, namely $\bar \Q_1 \Q_2 J_{12}$ and its complex conjugate is $-\bar \Q^1 \Q^2 J_{12}$.  Thus, the flavor current term must be a linear combination of
 \es{currentRNeutral}{
  \text{R-charge neutral scalars:} \qquad
   J_{12}\,, \quad \bar \Q_1 \Q_2 J_{12}\,,  \quad \bar \Q^1 \Q^2 J_{12} \,.
 }
As we will show, the $\mathfrak{osp}(2|4)$-invariant term in the action is
 \es{SJ}{
   S_J &= \int d^4 x\, \sqrt{-g} \, {\cal O}_J(x) \,, \\
   \text{with}\qquad{\cal O}_J(x) &= -3 i  \frac{J_{12}(x)}{L} + \bar \Q_1 \Q_2 J_{12}(x) - \bar \Q^1 \bar \Q^2 J_{12}(x)  \,.
 }

To show that this is the only $\mathfrak{osp}(2|4)$-invariant linear combination of the scalar fields in \eqref{currentRNeutral}, 
we proceed as in the previous section.  We study the action of $\bar \epsilon^i \Q_i$ and $\bar \epsilon_i \Q^i$ on the general linear combination
 \es{OJGen}{
   {\cal O}_J = a J_{12} + b \bar \Q_1 \Q_2 J_{12} + c \bar \Q^1 \Q^2 J_{12} \,,
 }
The quantities $\bar \epsilon^i \Q_i J_{12}$ and $\bar \epsilon_i \Q^i J_{12}$ cannot be simplified further, so we proceed to the second operator, noting that we can write ${\cal O}_b= \bar \Q_1 \Q_2 J_{12}$ in several different ways:
 \es{OEquiv}{
  {\cal O}_b = \bar \Q_1 \Q_2 J_{12}
   = \bar \Q_2 \Q_1 J_{12} = - \frac 12 \bar \Q_1 \Q_1 J_{22} = - \frac 12 \bar \Q_2 \Q_2 J_{11} \,.
 }
The second equality follows from the flip property \eqref{FlipProp} and the others follow from the shortening condition \eqref{CurrentShortening}.  Let's calculate.  First, since $\Q_1 \bar \Q_1 \Q_1 = 0$ due to the Schouten identity \eqref{Schouten}, we have
 \es{e1Q1}{
  \bar \epsilon^1 \Q_1 \cO_b = -\frac 12 \bar \epsilon^1 \Q_1 \bar \Q_1 \Q_1 J_{22} = 0 \,.
 }
Similarly, we also have 
 \es{e2Q2}{
  \bar \epsilon^2 \Q_2 \cO_b =-\frac 12 \bar \epsilon^2 \Q_2 \bar \Q_2 \Q_2 J_{11} = 0 \,.
 }
To calculate $\bar \epsilon_1 \Q^1\cO_b$, we  start with the form ${\cal O}_b = -\frac 12 \Q_1 \Q_1 J_{22} = -\frac 12 \Q_1 \Q_1 J^{11}$. The shortening condition allows us to write
 \es{e1Q1Again}{
 \bar \epsilon_1 \Q^1 {\cal O}_b = - \frac 12 [\bar \epsilon_1 \Q^1, \bar \Q_1 \Q_1] J^{11}
  = - \bar \epsilon_1 \slashed{D} \Q_2 J_{12}
    + \frac i{L}  \bar \epsilon_1 \Q^1 J_{12} \,,
 } 
where we used  \eqref{CommutRel} and shortening again.  A very similar computation gives
 \es{e2Q2Again}{
   \bar \epsilon_2 \Q^2 {\cal O}_b =  - \bar \epsilon_2 \slashed{D} \Q_1 J_{12}
    + \frac i{L} \bar \epsilon_2 \Q^2 J_{12} \,.
 }
The sum of the four previous equations gives the full SUSY variation  of $\cO_b$, namely
\es{deltOb}{
\delta \cO_b = &\frac{i}{L} \left(\bar \epsilon_1 \Q^1 J_{12} + \bar \epsilon_2 \Q^2 J_{12}\right)\\
& {}-\left( \bar \epsilon_1 \slashed{D} \Q_2 J_{12} + \bar \epsilon_2 \slashed{D} \Q_1 J_{12} \right)\,.
}
The variation of the third operator in \eqref{OJGen}, namely $\cO_c =\bar \Q^1 \Q^2 J_{12}$, is the complex conjugate of \eqref{deltOb}:
\es{deltOcz}{
\delta \cO_c= &\frac{-i}{L} \left(\bar \epsilon^1 \Q_1 J_{12} + \bar \epsilon^2 \Q_2 J_{12}\right)\\
& {}- \left( \bar \epsilon^1 \slashed{D} \Q^2 J_{12} + \bar \epsilon^2 \slashed{D} \Q_1 J_{12} \right)\,.
}
The SUSY variation of $\cO_J$ can now be assembled as
\es{varOJ}{
\delta\cO_J & = a\, \delta J_{12} + b\delta \cO_b +c\delta O_c\\
&= \left( a - \frac{ic}{L} \right) \left( \bar \epsilon^1 \Q_1 J_{12} +  \bar \epsilon^2 \Q_2 J_{12}  \right)  
   + \left( a + \frac{ib}{L} \right) \left( \bar \epsilon_1 \Q^1 J_{12} +  \bar \epsilon_2 \Q^2 J_{12}  \right) \\
   &{}- b \left( \bar \epsilon_1 \slashed{D} \Q_2 J_{12} + \bar \epsilon_2 \slashed{D} \Q_1 J_{12} \right) 
   - c \left( \bar \epsilon^1 \slashed{D} \Q^2 J_{12} + \bar \epsilon^2 \slashed{D} \Q^1 J_{12} \right)  \,,
 }
 Using $D_a \bar \epsilon_1 =  \frac{i}{2L} \bar \epsilon^2 \gamma_a$,  $D_a \bar \epsilon^1 = - \frac{i}{2L} \bar \epsilon_2 \gamma_a$, etc.~we see that in order for \eqref{varOJ} to be a total derivative, the following equations must be obeyed
 \es{TotalDerEqns}{
  -\frac{2i}{L} b =  \left( a - \frac{ic}{L} \right) \,, \qquad
   \frac{2i}{L} c =  \left( a + \frac{ i b}{L} \right) \,.
 }
The solution is $c = -b$ and $a = -\frac{ 3 i b}{L}$.  Taking $b=1$ reproduces \eqref{SJ}.  In addition, we have also shown that
 \es{deltaOJDer}{
  \delta {\cal O}_J = D_a \Psi_J^a \,, \qquad
   \Psi_J^{a} \equiv -  \bar \epsilon_1 \gamma^a \Q_2 J_{12} - \bar \epsilon_2 \gamma^a \Q_1 J_{12}  
   +  \bar \epsilon^1 \gamma^a \Q^2 J_{12} + \bar \epsilon^2 \gamma^a \Q^1 J_{12}  \,.
 }

\subsubsection{D-terms and similar interactions}
\label{DTERMS}
In this section we provide a unified treatment of the first three interactions listed at the beginning of this section.
 We study the expression 
 \es{DtermN2}{
  S_\text{int} = \int d^4 x\, \sqrt{-g} \, {\cal O}(x) \,, \qquad
   {\cal O}(x) = \bar \Q_1 \Q_1 \, \bar \Q^2 \Q^2 \, B\,, 
 }
where $B(x)$ is a field with R-charge $+4$ that in \eqref{DtermN2} is acted on by all four supercharges of R-charge $-1$.  All the interactions can be written either in this form or as its complex conjugate.   We focus on \eqref{DtermN2} since  the analysis of the conjugate form is entirely analogous.   

The field $B(x)$ has the following properties in the three cases of interest:
\begin{enumerate}
\item  D-terms: $B = \bar \Q_2 \Q_2\,  \bar \Q^1 \Q^1 X$, for an R-neutral scalar field $X$.
\item  1/4 BPS terms;  $B(x)$ takes either of two possible forms: i) $B = \bar \Q_2 \Q_2 X$ where $X$ satisfies $\Q^1 X=0$, or ii) $B = \bar \Q^1 \Q^1 X$ where $\Q_2 X=0.$
\item Non-chiral F-terms: $\Q_2 B = \Q^1 B = 0$.
\end{enumerate}
In the last two cases we also assume that $B$ cannot be obtained from the SUSY variation of another field.  Note that $\bar \Q^2 \Q^2$ commutes with $\bar \Q_1 \Q_1$, so those operators can be written in any order in \eqref{DtermN2}.  Further, of all fields obtained by acting with the supercharges on $B$, ${\cal O}$ is the only R-charge neutral scalar, so \eqref{DtermN2} is the only supersymmetric invariant candidate. 
 
We now show that \eqref{DtermN2} is supersymmetric.  To be precise, we show that
\es{SeparateD}{
   (\bar \epsilon^1 \Q_1 + \bar \epsilon_2 \Q^2) {\cal O} \,, \qquad
    (\bar \epsilon_1 \Q^1 + \bar \epsilon^2 \Q_2) {\cal O}
 }
are separately total derivatives.  The first condition is easy because the Schouten identity \eqref{Schouten} and its conjugate imply that $\bar \epsilon^1 \Q_1 \, \bar \Q_1 \Q_1 = 0$ and $\bar \epsilon_2 \Q^2 \, \bar \Q^2 \Q^2 = 0$, so
 \es{FirstQuantity}{
   \bar \epsilon^1 \Q_1 {\cal O} =   \bar \epsilon_2 \Q^2 {\cal O} = 0 \qquad
    \Longrightarrow \qquad  (\bar \epsilon^1 \Q_1 + \bar \epsilon_2 \Q^2) {\cal O} = 0 \,.
 }
For the second quantity in \eqref{SeparateD}, we start with $\bar \epsilon_1 \Q^1 {\cal O}$.  Notice first that $\bar \epsilon_1 \Q^1 B = 0$  in the three cases of interest.\footnote{Depending on which case we consider, we may have to use $[\Q^1,\Q_2]=0$ and the conjugate of \eqref{Schouten} or the shortening conditions obeyed by $B$.}  
Thus,
 \es{esp1Q1OD}{
    \bar \epsilon_1 \Q^1 {\cal O} =  [\bar\epsilon_1 \Q^1\,,\,\bar \Q_1 \Q_1\,\bar \Q^2 \Q^2]\,B
    = \left( [  \bar \epsilon_1 \Q^1, \bar \Q_1 \Q_1] \, \bar \Q^2 \Q^2
     +\bar \Q_1 \Q_1 \,  [  \bar \epsilon_1 \Q^1, \bar \Q^2 \Q^2]  \right)  \, B \,.
 }
Using  the conjugate of \eqref{CommutRel} as well as $R B = 4i B$ and the fact that $M_{ab}$ annihilates scalar operators, we obtain
 \es{esp1Q1ODAgain}{
    \bar \epsilon_1 \Q^1 {\cal O} =  \left( - \bar \epsilon_1 \slashed{D} \Q_1 \, \bar \Q^2 \Q^2
     +\bar \Q_1 \Q_1 \,  \frac{2 i \bar \epsilon_1 \Q^2 }{L} \right)  \, B \,. 
 }
We then swap the order of $\bar \epsilon_1 \Q^2$ and $\bar \Q_1 \Q_1$ because they commute:
 \es{esp1Q1ODAgain2}{
    \bar \epsilon_1 \Q^1 {\cal O} =  \left( - \bar \epsilon_1 \slashed{D} \Q_1 \, \bar \Q^2 \Q^2
     +\frac{2 i \bar \epsilon_1 \Q^2 }{L} \bar \Q_1 \Q_1  \right)  \, B  \,.
 }
We can follow the exact same steps to evaluate $\bar \epsilon^2 \Q_2 {\cal O}$, with the result
 \es{esp2Q2ODAgain}{
    \bar \epsilon^2 \Q_2 {\cal O} =  \left( - \bar \epsilon^2 \slashed{D} \Q^2 \, \bar \Q_1 \Q_1
     -  \frac{2 i \bar \epsilon^2 \Q_1 }{L} \bar \Q^2 \Q^2    \right)  \, B  \,.
 }
Adding together \eqref{esp1Q1ODAgain2} and \eqref{esp2Q2ODAgain} and using $D_a \bar \epsilon_1 =  \frac{i}{2L} \bar \epsilon^2 \gamma_a$ and $D_a \bar \epsilon^2 = - \frac{i}{2L} \bar \epsilon_1 \gamma_a$, we exhibit the desired
total derivative:
 \es{Sum}{
 (\bar \epsilon_1 \Q^1 + \bar \epsilon^2 \Q_2) {\cal O}
  = D_a  \Psi_D^{a} \,, \qquad
   \Psi_D^{a} \equiv  \left(  - \bar \epsilon_1 \gamma^a \Q_1 \, \bar \Q^2 \Q^2
   -  \bar \epsilon^2 \gamma^a \Q^2 \, \bar \Q_1 \Q_1 \right)  \, B \,.
 } 
This information, together with \eqref{FirstQuantity}, establishes  that $\delta {\cal O}$ is a total derivative.  Thus
  $S_\text{int} = \int d^4 x\, \sqrt{-g}\,  {\cal O}$ is supersymmetric, provided that $\Psi_D^{a}$ vanishes at the boundary, or that appropriate boundary terms are added to $S_\text{int}$.

\subsection{SUSY in Euclidean AdS and its Killing spinors}
\label{EUCLIDEAN}

Since the physics of interest in this paper involves supersymmetric field theories on $S^3$, we now review the formulation of Euclidean SUSY using the conventions of \cite{Freedman:2013oja}.  In particular, we replace $\gamma^0$ with $\gamma^4$  and keep the other $\gamma^m$, with $m=1, 2, 3$ unchanged: 
 \es{gammai}{
  \gamma^m = \begin{pmatrix} 
 0 & \sigma_m \\
 \sigma_m & 0 
\end{pmatrix} \,, \qquad
 \gamma^4 = \begin{pmatrix} 
 0 & -i I\\
 i I & 0 
\end{pmatrix} \,,
 }
where $\sigma_m$ are the Pauli matrices, and $I$ is the $2\times 2$ identity matrix.   All the equations presented above still hold with the obvious changes, e.g.~$\sqrt{-g} \to \sqrt{g}$.

Since we will consider the effective action expanded around (Euclidean) AdS, also known as the hyperbolic space $\HH^4$, let us present an explicit parameterization of $\HH^4$ and solutions to the Killing spinor equations.  We write the $\HH^4$ metric (set the curvature radius $L=1$ for simplicity) in sphere slicing as
 \es{H4Metric}{
  ds^2 = \frac{4}{(1-r^2)^2} \left( dr^2 + r^2 ds_{S^3}^2 \right) \,,
 }
where $ds_{S^3}^2$ is the line element on the unit-radius $S^3$.  The coordinate $r$ is a radial coordinate ranging from $r=0$ to $r=1$ at the boundary.  The sphere slicing made manifest in \eqref{H4Metric} is convenient for studying CFTs placed on $S^3$.   We can take the frame $e^a = e^a_\mu dx^\mu$ to be
 \es{Frame}{
  e^m = \frac{2 r}{1-r^2}  \hat e^m \,, \qquad e^4 =  \frac{2}{1 - r^2} dr \,,
 }
where $\hat e^m$, $m=1, 2, 3$ are frame vectors on an $S^3$ of unit radius. 

The AdS Killing spinor equation
 \es{KillingSp}{
  D_a T = \frac{1}{2} \gamma_a T 
 }
has 4 linearly independent solutions that can be taken to be
 \es{KillingSpinors}{
  T_1 &= \frac{1}{\sqrt{1 - r^2 }} \begin{pmatrix}
    - ir  \zeta_1 \\
    \zeta_1
  \end{pmatrix} \,, \qquad T_2 = \frac{1}{\sqrt{1 - r^2}} \begin{pmatrix}
    -  i r \zeta_2 \\
    \zeta_2
  \end{pmatrix} \,, \\
   T_3 &= \frac{1}{\sqrt{1 -  r^2}} \begin{pmatrix}
     \xi_1 \\
    ir\xi_1
  \end{pmatrix} \,, \qquad T_4 =\frac{1}{\sqrt{1 - r^2}} \begin{pmatrix}
     \xi_2 \\
    ir \xi_2
  \end{pmatrix} \,,
 } 
where $\zeta_{1, 2}$ are two linearly independent solutions to the $S^3$ Killing spinor equation $\nabla_i \zeta = + \frac{i}{2} \sigma_i \zeta$, and $\xi_{1, 2}$ are two linearly independent solutions to the $S^3$ Killing spinor equation $\nabla_i \xi = - \frac{i}{2} \sigma_i \xi$ on $S^3$.  We normalize these spinors using $\zeta_1^T i \sigma_2 \zeta_2 = 1$ and $\xi_1^T i \sigma_2 \xi_2 = 1$.  Note that, with this normalization, the determinant of the $4 \times 4$ matrix whose columns are the $T_A$ is\footnote{Note that the Killing spinors used in this paper are commuting.}
 \es{DetKS}{
  \det \begin{pmatrix} 
   T_1 & T_2 & T_3 & T_4 \end{pmatrix}
     = 1 \,.
 }
This equation will be very useful shortly.

For our ${\cal N} = 2$ theories, we should solve the Killing spinor equations \eqref{Killing}.  As explained after Eq.~\eqref{Killing}, $\epsilon^1 + i \epsilon_2$ and $\epsilon^2 + i \epsilon_1$ obey the equation \eqref{KillingSp} that we solved above. Thus, we have the following solutions to the ${\cal N} = 2$ Killing spinor equations:  there are 4 solutions of definite chirality
 \es{SolKS1}{
  \epsilon^1 = P_L T_A \,, \qquad \epsilon_2 = -i P_R T_A \,, \qquad \epsilon^2 = 0 \,, \qquad \epsilon_1 = 0 \,,
 }
with $A = 1, \ldots, 4$, and 4 more solutions
 \es{SolKS2}{
   \epsilon^1 = 0 \,, \qquad \epsilon_2 = 0 \,, \qquad \epsilon^2 = P_L T_A \,, \qquad \epsilon_1 = -i P_R T_A \,.
 }
Thus, using $\delta = \bar \epsilon^i \Q_i+ \bar \epsilon_i \Q^i$ we define and use below the eight SUSY variations
 \es{LinCombSuper}{
  \delta_{A+} = \bar T_A \left( \Q_1 - i \Q^2\right) \,, \qquad  \delta_{A-} = \bar T_A \left( \Q_2 - i \Q^1 \right) \,,
 }
with $A = 1, \ldots, 4$.  The $\pm$ index denotes the fact that if we act with $\delta_{A\pm}$ changes the $\mathfrak{u}(1)_R$ R-symmetry by charge by $\pm 1$.  

If we restrict to $A = 1, 2$, then the four such supercharges in \eqref{LinCombSuper} generate an $\mathfrak{su}(2|1)$ subalgebra.  (The same is true for $A = 3, 4$.) We see from \eqref{KillingSpinors} that $T_A$, with $A = 1, 2$ only involve the boundary spinor $\zeta$ and $T_A$ with $A = 3, 4$ only involve the boundary spinor $\xi$.  On $S^3$, it can be checked that bilinears in $\zeta$ and bilinears in $\xi$ give, respectively, $S^3$ Killing vectors corresponding to the $\mathfrak{su}(2)_\ell$ and $\mathfrak{su}(2)_r$ factors of the $\mathfrak{su}(2)_\ell \times \mathfrak{su}(2)_r$ isometry of $S^3$.  Thus, the spinors $\zeta$ extend $\mathfrak{su}(2)_\ell \times \mathfrak{su}(2)_r$ to $\mathfrak{su}(2|1)_\ell \times \mathfrak{su}(2)_r$, while the spinors $\xi$ extend it to $\mathfrak{su}(2)_\ell \times \mathfrak{su}(2|1)_r$.

\subsection{SUSY exactness and independence of interaction couplings}
\label{EXACTNESS}

Our primary goal is to understand what bulk interactions affect the mass-deformed sphere free energy $F(\mathfrak{m})$ of the boundary theory.  However, this observable is a particular case of a more general class of observables with the same property, so we can prove the following more general result:

{\em  In an ${\cal N} = 2$ SCFT on $S^3$ with a holographic dual, the correlation functions of any boundary operators invariant under $\mathfrak{su}(2|1)_\ell \times \mathfrak{su}(2)_r$ are independent of the interactions considered in Section~\ref{DTERMS}, but could depend non-trivially on the other couplings.  (A similar statement holds for operators invariant under $\mathfrak{su}(2)_\ell \times \mathfrak{su}(2|1)_r$.)}

The abstract proof of this statement is very simple given our construction from Section~\ref{DTERMS}, and it is done in two steps.  The first step is to note that since \eqref{DetKS} implies that $\bar \Q_1 \Q_1 \, \bar \Q^2 \Q^2
    =  4 \delta_{1+} \delta_{2+} \delta_{3+} \delta_{4+}$, we can write \eqref{DtermN2} as
 \es{DtermQExact}{
  {\cal O}(x) = \delta_{1+} {\cal P}(x) \,, \qquad
   {\cal P}(x) \equiv  4\, \delta_{2+} \delta_{3+} \delta_{4+}  B(x) \,.
 }
Thus, ${\cal O}$ is $\delta_{1+}$-exact.  A similar argument shows that ${\cal O}$ is also $\delta_{A+}$-exact with any $A = 1, \ldots, 4$.

The second step is to use the Ward identity to show that any correlation functions of ${\cal O}(x)$ and other operators (in the bulk or in the boundary theory) that are $\delta_{1+}$-invariant must vanish:
  \es{GenCorr}{
  \langle {\cal O}(x)  \text{  (any $\delta_{1+}$-invariant operators)} \rangle 
   = \delta_{1+} \langle {\cal P}(x)  \text{  (any $\delta_{1+}$-invariant operators)} \rangle = 0 \,.
 }
Integrating over $x$ then implies that the set of Witten diagrams associated with a given $S_\text{int}$  must vanish.  Since any $\mathfrak{su}(2|1)_\ell \times \mathfrak{su}(2)_r$-invariant operator is in particular invariant under $\delta_{1+}$,  the statement that the CFT correlators of $\mathfrak{su}(2|1)_\ell \times \mathfrak{su}(2)_r$-invariant operators are independent of the class of interactions considered in Section~\ref{DTERMS} follows.

This proof does not apply to the chiral F-term of Section~\ref{FTERMS} or to the flavor current term of Section~\ref{FLAVOR}, as they are not $\dk_{A+}$-exact. No analogue of \eqref{GenCorr} holds for these interactions and so the sphere partition function may depend on the corresponding parameters.  We will see in examples that this is the case.  

As we have seen, because D-terms (and also 1/4-BPS terms and non-chiral F-terms) are both $\dk_{1+}$-invariant and $\dk_{1+}$-exact, they cannot effect $\dk_{1+}$-invariant correlation functions. In Appendix~\ref{CLOSEDEXACT} we consider the more general question of when D-terms for a given superalgebra are exact under some supercharge in the superalgebra. We show that the existence of a nilpotent supercharge (that is a supercharge which, like $\dk_{1+}$, squares to zero), suffices to prove the exactness of D-terms.

The $\cN=1$ supersymmetric algebra $\mathfrak{osp}(1|4)$ does not contain nilpotent supercharges, and so we can ask whether in this case D-terms are exact under any of the supercharges. In Appendix~\ref{CLOSEDEXACT} we show that they are not exact. As a result, no analogue of \eqref{GenCorr} holds in this case, and so the $\cN=1$ $S^3$ partition function will generically depend on bulk D-terms.

\subsection{The real mass deformation on the boundary}
\label{REALMASSSECTION}

Let us now apply the general statement shown in the previous subsection to conclude that $F(\mathfrak{m})$ is independent of the SUSY exact couplings.  To make things concrete, let us consider the case where the 3d ${\cal N} = 2$ SCFT has a $U(1)$ flavor symmetry with flavor current $j_\mu$.  The ${\cal N} = 2$ superconformal multiplet that contains the conserved current also contains a dimension $1$ real scalar $J$ and a dimension $2$ real scalar $K$, and a dimension-$3/2$ complex fermion $\Xi$.\footnote{Note that the conserved current multiplet in the boundary SCFT is of course different from the bulk conserved current multiplet studied in Section~\ref{FLAVOR}.}

Suppose we normalize the bosonic operators so that, in flat space, 
 \es{Normalizations}{
  \langle j_\mu(\vec{x}) j_\nu(0)  \rangle &= \frac{\tau}{16 \pi^2} (\delta_{\mu\nu} \partial^2 - \partial_\mu \partial_\nu) \frac{1}{\abs{\vec{x}}^2} \,, \\
   \langle J(\vec{x}) J(0)  \rangle &= \frac{\tau}{16 \pi^2 \abs{\vec{x}}^2} \,, \\
    \langle K(\vec{x}) K(0)  \rangle &= \frac{\tau}{8 \pi^2 \abs{\vec{x}}^4} \,,
 } 
for some normalization constant $\tau$. The correlation functions on $S^3$ can be obtained from \eqref{Normalizations} by performing a Weyl rescaling.   With these normalizations, the real mass deformation corresponds to adding the following term to the SCFT action on $S^3$: 
 \es{RealMass}{
  \mathfrak{m} \int d^3\vec{x}\, \sqrt{g(\vec{x})}  \left[ \frac{i}{a} J(\vec{x}) + K(\vec{x}) \right]  \,, 
 }
where $a$ is the radius of $S^3$.\footnote{The absence of terms of higher order in $\mathfrak{m}$ appearing explicitly in \eqref{RealMass}  is a regularization choice.  Indeed, it is possible that the supersymmetry Ward identities require the $K(\vec{x}) \times K(\vec{y})$ OPE to contain an operator-valued contact term of the form $\delta^{(3)}(\vec{x}- \vec{y}) L(\vec{x})$, where $L$ is an operator of dimension exactly $1$.  In conformal perturbation theory, the effect of such a contact term would be as if $\mathfrak{m}^2 L$ were included explicitly in the action \eqref{RealMass}, and this is the reason why oftentimes such a term is included explicitly in the action.  (For example, an $\mathfrak{m}^2 \abs{\phi}^2$ term would be present in the theory of a massive chiral multiplet.)  Regardless of whether or not we include $\mathfrak{m}^2 L$ explicitly, we should make sure that our regularization scheme preserves supersymmetry.  See also the discussion in Section~4 of \cite{Cordova:2016xhm}.}   With appropriate conventions, \eqref{RealMass} can be shown to be invariant under $\mathfrak{su}(2|1)_\ell \times \mathfrak{su}(2)_r$ (as opposed to\footnote{One can also consider the deformation $\mathfrak{m} \int d^3\vec{x}\, \sqrt{g(\vec{x})}  \left[- \frac{i}{a} J(\vec{x}) + K(\vec{x}) \right]$, which would then be invariant under $\mathfrak{su}(2)_\ell \times \mathfrak{su}(2|1)_r$.} $\mathfrak{su}(2)_\ell \times \mathfrak{su}(2|1)_r$).  Thus, derivatives of $F(\mathfrak{m})$ with respect to $\mathfrak{m}$ can be interpreted as correlation functions of the $\mathfrak{su}(2|1)_\ell \times \mathfrak{su}(2)_r$-invariant integrated operator $ \int d^3\vec{x}\, \sqrt{g(\vec{x})}  \left[ \frac{i}{a} J(\vec{x}) + K(\vec{x}) \right] $, and so based on the result of the previous section it is independent of the bulk SUSY-exact couplings when computed using Witten diagrams.

\section{$\text{AdS}_4$ as a background of ${\cal N} = 2$ off-shell supergravity}
\label{CONFORMAL}

In this section, we provide a different perspective on the construction of the supersymmetric interactions presented in the previous section.  Here, we obtain the supersymmetric theories in AdS from an appropriate supergravity background solution of conformal supergravity.\footnote{See also Appendix~\ref{REALMASSANOTHER} where the boundary real mass deformation is also phrased in the language of this section.}   Before we begin the technical treatment, let us describe the relation between the various theories which play a role in this section. The parent theory is ${\cal N}=2$ conformal supergravity (see \cite{Lauria:2020rhc,Freedman:2012zz} and references therein) containing two types of supercharges, the ${\bf Q}_i, {\bf Q}^i$, which anti-commute to momenta, and the ${\bf S}_i, {\bf S}^i$, which anti-commute to special conformal transformations. The four-component spinor indices are suppressed in this notation. The indices $i=1,2$ transform under an $SU(2)\times U(1)$ R-symmetry. All symmetries are gauged. The $SU(2)$ generators are proportional to Pauli matrices, i.e.~$\vec{\tau}_{i}{}^j = i \vec{\sigma}_{i}{}^j$.   After partial gauge fixing, the remaining supergravity fields, including compensating multiplets, are fixed to background values which include fixing the metric to that of $\text{AdS}_4$. This process is described in more detail in Section~\ref{ADSBACKGROUND}. The result is a theory invariant under the global superalgebra $\mathfrak{osp}(2|4)$ with supercharges $\Q_i, \,\Q^i$ that are linear combinations of the previous ${\bf Q},\, {\bf S}$. The R-symmetry is reduced to a $U(1)_R$ generated by $\tau_{3\,i}{}^j$.

\subsection{Supergravity background for $\text{AdS}_4$}
\label{ADSBACKGROUND}

As explained in the work of Festuccia and Seiberg \cite{Festuccia:2011ws} in the context of 4d  ${\cal N} = 1$ theories on curved manifolds, a systematic way to write down curved space QFT Lagrangians is to couple a QFT to off-shell background supergravity.  The same procedure was used for 4d ${\cal N} = 2$ theories in \cite{Klare:2013dka,Gomis:2014woa}.  The ${\cal N} = 2$ effective theories in $\text{AdS}_4$ fall under this class.  $\text{AdS}_4$ is realized as a supersymmetric background of off-shell ${\cal N} = 2$ supergravity in a way that is very similar to the $S^4$ construction also explored in \cite{Gomis:2014woa}.

The construction proceeds as follows.  As in \cite{Gomis:2014woa}, one arrives at the appropriate off-shell ${\cal N} = 2$ supergravity by starting with ${\cal N} = 2$ conformal supergravity as well as compensating vector and tensor multiplets.  Thus, we start with a Weyl multiplet with field content
 \es{WeylContent}{
  \text{\underline{Weyl multiplet}:  \hspace{0.5in} bosonic: }& e_\mu^a, b_\mu, \omega_\mu{}^{ab}, f_\mu{}^a, V_{\mu i}{}^j, A_\mu^R, T_{ab}^-, D \\
  \text{fermionic: }& \psi_\mu{}^i , \phi_\mu{}^i, \chi^i \,,
 }
where $ e_\mu^a, b_\mu, \omega_\mu{}^{ab}, f_\mu{}^a, V_{\mu i}{}^j, A_\mu^R$ are respectively, the gauge fields associated with translations, dilatations, Lorentz transformations, special conformal transformations, and the $SU(2)_R$ and $U(1)_R$ symmetries, and $\psi_\mu{}^i , \phi_\mu{}^i$ are the gauge fields associated with Poincar\'e and superconformal symmetries.  The compensator vector multiplet has field content
 \es{VectorContent}{
  \text{\underline{Vector multiplet compensator}:  \hspace{0.5in}bosonic: }& X, \bar X, A_\mu , Y_{ij} \\
  \text{fermionic: }& \Omega_i, \Omega^i \,,
 }
obeying $Y_{ij}^* = Y^{ij}$, and the compensator tensor multiplet has field content
  \es{TensorContent}{
  \text{\underline{Tensor multiplet compensator}:  \hspace{0.5in}bosonic: }& L_{ij}, G, \bar G, E_{ab} \\
  \text{fermionic: }& \phi^i, \phi_i \,,
 }
obeying $L_{ij}^* = L^{ij}$.  Here, $\mu$ is a spacetime coordinate index, $a$, $b$ are frame indices, and $i, j$ are $SU(2)_R$ indices that are raised and lowered with the epsilon symbol as in \cite{Lauria:2020rhc,Freedman:2012zz}.   The off-shell supergravity theory is obtained by imposing curvature constraints that determine the gauge fields $\omega_\mu{}^{ab}$, $f_\mu{}^a$, and $\phi_\mu{}^i$ in terms of the other fields (explicit expressions are given below), and making the gauge choices \cite{deWit:1980lyi}:
 \es{GaugeFixing}{
  b_\mu = 0 \,, \qquad X = M\,,  \qquad \Omega_i = 0\,, \qquad L_{ij} = \tau_{3 ij} \varphi \,,
 }
where $M$ and $\varphi$ are arbitrary dimensionful constants whose values will not be important, and where $\tau_{3}^{ij} = -\tau_{3ij} = (-i \sigma_1)_{ij}$ as in \cite{Lauria:2020rhc,Freedman:2012zz}.  The gauge choice $b_\mu = 0$ fixes the special conformal transformations.  The gauge choice $X = M$ fixes the dilatations as well as the $U(1)_R$ symmetry of conformal supergravity.  The gauge choice $\Omega_i = 0$ fixes the superconformal transformations.  Lastly, the gauge choice $L_{ij} = \tau_{3ij} \varphi$ breaks the $SU(2)_R$ transformations to $SO(2)_R$.  The remaining fields determine an off-shell supergravity theory.

Since we treat the supergravity fields as background fields, we do not need to write down an action for them, but instead set them to background values constrained to preserve global supersymmetry. This means that all supersymmetry variations must vanish. 
This will be important when coupling this theory to matter since it will ensure that the resulting theory is supersymmetric.

The supersymmetric $\text{AdS}_4$ background is achieved as follows.  First and foremost, we are looking for an $\text{AdS}_4$ background metric, so we should set the frame appropriately:
 \es{FrameCond}{
  e_\mu^a &= e_\mu^a \big|_{AdS_4} \,,
 }
where $\text{AdS}_4$ has radius $L$.  Then, we set the following fields to zero:
 \es{VanishingFields}{
    V_{\mu i}{}^j = A_\mu^R = T_{ab}^- = D = A_\mu = E_{ab}  = \psi_\mu{}^i = \chi^i  = \phi^i = 0 \,.
 }
The remaining scalar fields $G$ and $Y_{ij}$ will be non-zero.  Since all the fermions vanish either by \eqref{GaugeFixing} or \eqref{VanishingFields}, the SUSY variations of all the bosons vanish automatically.  The SUSY variations of the fermions then are (see Eqs.~(2.90), (3.15), (3.103) of \cite{Lauria:2020rhc}): 
 \es{FermVariations}{
 \delta \psi_\mu^i &= e_\mu^a \left( D_a   \epsilon^i - \gamma_a \eta^i \right) \,, \\
  \delta \Omega_i &= \slashed{\partial} X \epsilon_i + Y_{ij} \epsilon^j + 2 X \eta_i \,, \\  
   \delta \chi^i &= 0 \,, \\
  \delta \phi^i &= \frac 12 \slashed{D} L^{ij} \epsilon_j - \frac 12 G \epsilon^i + 2 L^{ij} \eta_j  \,.
 }
 To ensure the vanishing of these variations with the gauge choice \eqref{GaugeFixing}, we take
  \es{BackgroundValues}{
   G = - \frac{2 \varphi}{L} \,, \qquad 
   Y_{ij} = \frac{M}{L} \tau_{3 ij} \,.
 }

With these choices, the SUSY variations in \eqref{FermVariations} vanish provided that\footnote{Note that $\epsilon^i$ and $\eta^i$ have opposite chirality.}
 \es{ConditionsVanish}{
   D_a \epsilon^i &= \gamma_a \eta^i \,, \qquad  \eta^i  = - \frac 1{2L} \tau_{3}^{ij} \epsilon_j \,, \\
    D_a \epsilon_i &= \gamma_a \eta_i \,, \qquad  \eta_i  = - \frac 1{2L} \tau_{3 ij} \epsilon^j   \,,
 }
where the second line is the complex conjugate of the first.  These are precisely the Killing spinor equations we also found in the previous section in Eq.~\eqref{Killing}.  Here, we obtained them from the condition that the supersymmetry variations of the fermions in the background supergravity multiplet vanish.   Note that, as mentioned above, in addition to the gauge-fixing constraints, one also has to impose curvature constraints that determine the composite gauge fields $\omega_\mu{}^{ab}$, $f_\mu{}^a$, and $\phi_\mu{}^i$.  Evaluated on our AdS background, one finds
 \es{CompositeFields}{
  \omega_\mu{}^{ab} &= \omega_\mu{}^{ab}  \big|_{AdS_4} \,, \qquad \phi_\mu{}^i = 0 \,, \qquad
  f_\mu{}^a = - \frac 14 R_\mu{}^a + \frac{1}{24} e_\mu{}^a R = \frac{1}{4L^2} e_\mu{}^a \,.
 }

The AdS supersymmetry algebra presented in Section~\ref{SUSYAlgebraSection} is obtained via this construction because it is a subalgebra of the superconformal algebra.  The 4d ${\cal N} = 2$ superconformal algebra has the following generators:\footnote{We use bold letters for these generators.  In \cite{Freedman:2012zz}, these generators are denoted by the same symbols, but not in bold.} momentum generators ${\bf P}_a$, Lorentz generators ${\bf M}_{ab}$, special conformal generators ${\bf K}_a$, dilatation generator ${\bf D}$,  $U(1)_R$ R-symmetry generator ${\bf T}$, $SU(2)_R$ R-symmetry generators ${\bf U}_i{}^j$, Poincar\'e supercharges ${\bf Q}_i$ and ${\bf Q}^i$ and superconformal charges ${\bf S}_i$ and ${\bf S}^i$.  They obey the $\mathfrak{su}(2, 2|2)$ algebra given, for instance, in Chapter~20.2.1 of \cite{Freedman:2012zz} or Chapter 1 of \cite{Lauria:2020rhc}.
 
 In the conformal supergravity construction, the momentum generators ${\bf P}_a$ act on the fields as derivatives ${\bf P}_a = {\cal D}_a$ that are covariant with respect to all ``standard'' gauge fields, meaning all gauge fields except for the ones for translation, $e_\mu{}^a$.  For our AdS background, the only standard gauge fields that are non-vanishing are the spin connection $\omega_\mu{}^{ab}$ and the special conformal gauge field $f_\mu{}^a$ given in \eqref{CompositeFields}.  The spin connection is part of the usual covariant derivative $D_a$, but $f_\mu{}^a$ is not.  Using \eqref{CompositeFields}, we have that
  \es{calDToD}{
   {\cal D}_a = D_a - \frac{1}{4L^2} {\bf K}_a \,.
  }
The action by $D_a$ on the fields is just that of the AdS momentum generators $P_a$ introduced in Section~\ref{SUSYAlgebraSection}.  Thus, 
 \es{PToPK}{
  P_a = {\bf P}_a + \frac{1}{4L^2} {\bf K}_a \,, \qquad M_{ab} = {\bf M}_{ab} \,,
 }
where we also identified the local Lorentz generators in AdS with the superconformal ones.  It is easy to check using the superconformal algebra that the commutation relations \eqref{PCommut} and \eqref{PMCommut} follow from the commutation relations of ${\bf P}_a$, ${\bf K}_a$, and ${\bf M}_{ab}$.

Regarding the rest of the algebra, note that the Poincar\'e supersymmetry transformations have parameters $\epsilon_i$ and $\epsilon^i$ while the superconformal transformations have parameters $\eta_i$ and $\eta^i$.  The SUSY variation corresponds to the action of $\bar \epsilon^i {\bf Q}_i + \bar \eta^i {\bf S}_i + \bar \epsilon_i {\bf Q}^i + \bar \eta_i {\bf S}^i$ on the fields.  In our AdS background, however, the $\eta$ and $\epsilon$ parameters are related by the right-hand equation of \eqref{ConditionsVanish}. This gives
 \es{SUSYVar}{
  \bar \epsilon^i {\bf Q}_i + \bar \eta^i {\bf S}_i + \bar \epsilon_i {\bf Q}^i + \bar \eta_i {\bf S}^i
   = \bar \epsilon^i \left({\bf Q}_i   - \frac 1{2L} \tau_{3 ij}  {\bf S}^j \right) 
    +  \bar \epsilon_i \left({\bf Q}^i   - \frac 1{2L} \tau_{3}{}^{ij}  {\bf S}_j \right)  \,.
 }
Identifying this expression with  $ \bar \epsilon^i \Q_i + \bar \epsilon_i \Q^i$, where $\Q_i$ and $\Q^i$ are the AdS supercharges, we find
 \es{QAdS}{
  \Q_i = {\bf Q}_i   - \frac 1{2L} \tau_{3 ij}  {\bf S}^j \,, \qquad
   \Q^i = {\bf Q}^i   - \frac 1{2L} \tau_{3}{}^{ij}  {\bf S}_j \,.
 }
In addition, the R-symmetry in AdS is the $U(1)$ subalgebra of $SU(2)_R$ generated by $\tau_3$, so we write 
 \es{RSymmAdS}{
  R = {\bf U}_i{}^j \tau_{3j}{}^i \,.
 }
Using the relations \eqref{PToPK}, \eqref{QAdS}, and \eqref{RSymmAdS} between the AdS superalgebra generators and the superconformal ones, one can show that the the remaining (anti)commutation relations \eqref{RCommut}--\eqref{SUSYAlgebra}  in AdS follow from those found in Chapter~20.2.1 of \cite{Freedman:2012zz} or Chapter~1 of \cite{Lauria:2020rhc}.

\subsection{The ${\cal N} = 2$ F-term from conformal supergravity}

We now study the chiral F-term interaction from the perspective of conformal supergravity.  In the ${\cal N} = 2$ AdS background of the previous subsection, we consider the superconformal chiral multiplet.  In applications, this multiplet will most often be composite.  It contains the components
 \es{ChiralField}{
    \text{\underline{Chiral multiplet}:  \qquad bosonic: }& A, B_{ij}, G_{ab}^-, C \\
  \text{fermionic: }& \Psi_i, \Lambda_i \,,
 }
where $A$, $C$, and $B_{ij}$ are complex scalar fields, $G_{ab}^-$ is an anti-self-dual two-form tensor, and $\Psi_i$ and $\Lambda_i$ are left-handed fermions.  We can also consider the complex conjugate anti-chiral multiplet.  The field $A$ is the superconformal primary (it is annihilated by ${\bf S}_i$, ${\bf S}^i$, ${\bf K}_a$), while the other components are each ordinary conformal primaries (they are each annihilated by ${\bf K}_a$).  The $A$ component has scaling dimension (Weyl weight) $w$ as well as an equal $U(1)_R$ charge $w$ as a consequence of the superconformal shortening conditions.  As one goes up in the multiplet, the Weyl weight goes up in $1/2$ units while the $U(1)_R$ charge goes down in $1/2$ units:  the Weyl weights of $\Psi_i$, $B_{ij}$, $G_{ab}^-$, $\Lambda_i$, and $C$ are $w+1/2$, $w+1$, $w+1$, $w+3/2$, and $w+2$, respectively, while their chiral weights are $w-1/2$, $w-1$, $w-1$, $w-3/2$, and $w-2$.   The SUSY variations of a chiral multiplet are given in (3.27) of \cite{Lauria:2020rhc} (see also (B.4) of \cite{Gerchkovitz:2016gxx} but note that the normalization of $B_{ij}$ is different there and here from that in \cite{Lauria:2020rhc} by a factor of~$2$):
 \es{chiral}{
   \delta A &= \frac 12 \bar \epsilon^i \Psi_i \,, \\
   \delta \Psi_i &= \slashed{D} (A \epsilon_i) + \frac 12 B_{ij} \epsilon^j 
      + \frac 14 \gamma^{ab} F_{ab}^- \varepsilon_{ij} \epsilon^j + (2w-4) A \eta_i \\
   \delta B_{ij} &= \bar \epsilon_{(i} \slashed{D} \Psi_{j)} - \bar \epsilon^k \Lambda_{(i} \varepsilon_{j)k}
    + 2 (1 - w) \bar \eta_{(i} \Psi_{j)}  \\  
   \delta F_{ab}^- &= \frac 14 \varepsilon^{ij} \bar \epsilon_i \slashed{D} \gamma_{ab} \Psi_j
     + \frac 14 \bar \epsilon^i \gamma_{ab} \Lambda_i - \frac 12 (1 + w) \varepsilon^{ij} \bar \eta_i \gamma_{ab} \Psi_j  \\ 
   \delta \Lambda_i &= - \frac 14 \gamma^{ab} \slashed{D} (F_{ab}^- \epsilon_i) 
     - \frac 12 \slashed{D} B_{ij} \varepsilon^{jk} \epsilon_k 
      + \frac 12 C \varepsilon_{ij} \epsilon^j - (1 + w) B_{ij} \varepsilon^{jk} \eta_k 
      + \frac {3 - w}2 \gamma^{ab} F_{ab}^- \eta_i   \\   
   \delta C &= -D_a (\varepsilon^{ij} \bar \epsilon_i \gamma^a \Lambda_j) + (2w - 4) \varepsilon^{ij} \bar \eta_i \Lambda_j \,.
 }

The reason why the chiral multiplet is useful for understanding the superconformal origin of the chiral F-term interaction is that, although the chiral multiplet \eqref{ChiralField} forms a representation of the superconformal algebra $\mathfrak{su}(2, 2|2)$, it reduces (for any weight $w$) to the representation of $\mathfrak{osp}(2|4)$ described in Section~\ref{FTERMS}.  Moreover, it is clear from \eqref{chiral} that when the Weyl weight $w=2$, the $C$ component transforms as a total derivative.  Therefore the integral
 \es{FTerm}{
  w=2: \qquad S_F = \int d^4 x\, \sqrt{-g} \, C(x) \,.
 }  
defines an F-term invariant (related to Eq. (3.30) of \cite{Lauria:2020rhc}).

When $w\neq 2$, the chiral superconformal multiplet also reduces to a chiral multiplet of $\mathfrak{osp}(2|4)$.   To show this, first note that when $w =2$ and the relations of \eqref{ConditionsVanish} are used, the transformation rules \eqref{chiral} become
 \es{chiralw2}{
     \delta A &= \frac 12 \bar \epsilon^i \Psi_i \,, \\
   \delta \Psi_i &= \slashed{D} (A \epsilon_i) + \frac 12 B_{ij} \epsilon^j 
      + \frac 14 \gamma^{ab} F_{ab}^- \varepsilon_{ij} \epsilon^j  \\
   \delta B_{ij} &= \bar \epsilon_{(i} \slashed{D} \Psi_{j)} - \bar \epsilon^k \Lambda_{(i} \varepsilon_{j)k}
    + \frac{1}{L} \tau_{3 (i|k|}\bar \epsilon^k \Psi_{j)}  \\  
   \delta F_{ab}^- &= \frac 14 \varepsilon^{ij} \bar \epsilon_i \slashed{D} \gamma_{ab} \Psi_j
     + \frac 14 \bar \epsilon^i \gamma_{ab} \Lambda_i +\frac{3}{4L} \varepsilon^{ij} \tau_{3 ik} \bar \epsilon^k \gamma_{ab} \Psi_j  \\ 
   \delta \Lambda_i &= - \frac 14 \gamma^{ab} \slashed{D} (F_{ab}^- \epsilon_i) 
     - \frac 12 \slashed{D} B_{ij} \varepsilon^{jk} \epsilon_k 
      + \frac 12 C \varepsilon_{ij} \epsilon^j +  \frac{3}{2L} \tau_{3 k l} B_{ij} \varepsilon^{jk} \epsilon^l 
      -   \frac{1}{4L} \tau_{3 ik} \gamma^{ab} F_{ab}^- \epsilon^k   \\   
   \delta C &= -D_a (\varepsilon^{ij} \bar \epsilon_i \gamma^a \Lambda_j) \,,
 } 
Generalizing the F-term \eqref{FTerm} to $w \neq 2$ can be done by noticing that for a chiral multiplet $(A, \Psi_i, B_{ij}, F_{ab}^-, \Lambda_i, C)$ with Weyl weight $w$, the chiral multiplet $(A, \Psi_i, B_{ij}', F_{ab}^-, \Lambda_i', C')$ with 
 \es{Primes}{
  B_{ij}' &= B_{ij} - \frac{2(w-2)}{L} A \tau_{3ij} \,, \\
  \Lambda_i' &= \Lambda_i + \frac{w-2}L \tau_{3ik} \varepsilon^{kj} \Psi_j \,,\\
  C' &= C + \frac{w-2}{L} \tau_3^{ij} B_{ij} - \frac{2 (w-2)(w-3)}{L^2} A
 }
transforms, in the AdS background, as a Weyl multiplet of weight $w=2$.  Thus, the general F-term is
 \es{GeneralF}{
  S_\text{F-term} = \int d^4 x\, \sqrt{-g} \, C'(x)
   = \int d^4 x\, \sqrt{-g} \, \left[ C + \frac{w-2}{L} \tau_3^{ij} B_{ij} - \frac{2 (w-2)(w-3)}{L^2} A \right] \,.
 }
 
We can connect this discussion to the more abstract derivation of the chiral F-term in  Section~\ref{FTERMS}.  In particular, from \eqref{chiral}, we identify
 \es{QonA}{
  \bar{\mathbb{Q}}_1 \mathbb{Q}_2 A &= - \frac 12 \left( B_{12} - \frac{2 i w}{L} A \right) \,, \\
  \bar{\mathbb{Q}}_1 \mathbb{Q}_1 \bar{\mathbb{Q}}_2 \mathbb{Q}_2 A
   &= - \frac{C}{2} + \frac{ i w}{L} B_{12} + \frac{ w(w-1)}{L^2} A \,.
 }
A little algebra shows that the chiral F-term in \eqref{GeneralF} can be written as
 \es{GeneralFAgain}{
  S_\text{F-term} = -2 \int d^4 x\, \sqrt{-g} \,
   \left[\bar{\mathbb{Q}}_1 \mathbb{Q}_1 \bar{\mathbb{Q}}_2 \mathbb{Q}_2 A + \frac{4 i}{L} \bar{\mathbb{Q}}_1 \mathbb{Q}_2 A
    + \frac{6}{L^2} A \right] \,,
 }
which matches the form \eqref{FTermDamon} that we derived in Section~\ref{FTERMS} up to a normalization factor.  We thus see the power of the abstract approach of Section~\ref{FTERMS} at work---we were able to derive the form \eqref{GeneralFAgain} without a detailed understanding of the supersymmetry transformation rules of all the fields in the multiplet.
 
\subsection{An F-term example: prepotential interactions in AdS}
\label{PREPOTENTIAL}
 
We can consider a concrete example of a chiral F-term coming from a prepotential interaction.  Suppose we have $n$ Abelian vector multiplets $(X^I, \bar X^I, A_\mu^I, Y_{ij}^I, \Omega_i^I, \Omega^{iI})$, $I = 1, \ldots, n$, in AdS in addition to the compensating multiplets used in the conformal supergravity construction.  The SUSY transformation rules for the vector multiplets are given in Eq.~(3.15) of \cite{Lauria:2020rhc}.  Specialized to our AdS background, we have
 \es{SUSYVector}{
  \delta X^I &= \frac{1}{2} \bar \epsilon^i \Omega_i^I \,, \\
  \delta \Omega_i^I &= \slashed{\partial} X^I \epsilon_i + \frac 14 \gamma^{ab} F_{ab}^{I -} \varepsilon_{ij} \epsilon^j + Y_{ij}^I \epsilon^j -    \frac 1{L} \tau_{3 ij} X^I \epsilon^j\,, \\
  \delta A_\mu^I &= \frac{1}{2} \varepsilon^{ij} \bar \epsilon_i \gamma_\mu \Omega^I_j + \frac{1}{2} \varepsilon_{ij} \bar \epsilon^i \gamma_\mu \Omega^{Ij} \,, \\
  \delta Y_{ij}^I &= \frac 12 \bar \epsilon_{(i} \slashed{D} \Omega^I_{j)} + \frac 12 \varepsilon_{ik} \varepsilon_{jl}
   \bar \epsilon^{(k} \slashed{D} \Omega^{l)I} \,,
 }
where $F_{ab}^I$ is the field strength of $A_\mu^I$.  The transformation rules for $\bar X^I$ and $\Omega^{iI}$ follow by complex conjugation.   One can then consider a prepotential interaction $F(X)$ that is homogeneous of degree $w$ in $X$.\footnote{Note that in the usual prepotential construction in ${\cal N} = 2$ supergravity, one also includes the compensating multiplet among the $X^I$ and one requires $w=2$.  No such requirement will be needed here.}  The prepotential is in fact the lowest component of a chiral multiplet whose components are given in Eq.~(3.107) of \cite{Lauria:2020rhc}.  On our AdS background, the bosonic fields become 
 \es{ChiralFromF}{
  A &= \frac{i}{2} F\,, \\
  B_{ij} &=i F_I Y_{ij}^I - \frac{i}{4} F_{IJ} \bar \Omega_i^I \Omega_j^J \,, \\
  C &= - i F_I D_a D^a \bar X^I - \frac{i}{2} F_{IJ} Y^{ij I} Y_{ij}^J 
   + \frac{i}{4} F_{IJ} F_{ab}^{-I} F^{-abI}
    + \frac{i}{2} F_{IJ} \bar \Omega_i^I \slashed{D} \Omega^{iJ} \\
     &{}+ \frac{i}{4} F_{IJK} Y^{ijI} \bar \Omega_i^J \Omega_j^K
      - \frac{i}{16} F_{IJK} \varepsilon^{ij} \bar \Omega_i^I \gamma^{ab} F_{ab}^{-J} \Omega_j^K
       + \frac{i}{48} F_{IJKL} \bar \Omega_i^I \Omega_l^J  \bar \Omega_j^K \Omega_k^L 
       \varepsilon^{ij} \varepsilon^{kl} \,,
 }
where $F_I = \partial_I F$, $F_{IJ} = \partial_I \partial_J F$, etc.  The chiral F-term is then given by \eqref{GeneralF}, with $w$ being the degree of homogeneity of $F$ with respect to $X$.   Note that  $C(x)$ agrees (after partial integration of the first term) with the Lagrangian for interacting  gauge multiplets with global supersymmetry in flat spacetime (see (20.15) of~\cite{Freedman:2012zz}.)

\subsection{The flavor current term from conformal supergravity}
\label{FLAVORSUGRA}

We now construct the flavor current interaction of Section~\ref{FLAVOR} by superconformal methods.  We start with a tensor multiplet with components given in \eqref{TensorContent}, and define the conserved current by Hodge duality.
To make this subsection self-contained we repeat \eqref{TensorContent}:  \es{TensorContentAgain}{
  \text{\underline{Tensor multiplet}:  \hspace{0.5in}bosonic: }& L_{ij}, G, \bar G, E_{\mu\nu} \\
  \text{fermionic: }& \phi^i, \phi_i \,,
 }
where the scalars $L_{ij}$ obey the reality condition $L_{ij}^* = L^{ij}$, $E_{\mu\nu}$ is a two-form gauge field, and $G$ is a complex scalar whose conjugate is $\bar G$.  The SUSY transformation rules are (see Section~3.2.5 of \cite{Lauria:2020rhc})
 \es{SUSYTensor}{
  \delta L_{ij} &= \bar \epsilon_{(i} \phi_{j)} + \varepsilon_{ik} \varepsilon_{jl} \bar \epsilon^{(k} \phi^{l)} \,, \\
  \delta \phi^i &= \frac 12 \slashed{\partial} L^{ij} \epsilon_j + \frac 12 \varepsilon^{ij} \slashed{E} \epsilon_j - \frac 12 G \epsilon^i -\frac{1}{L} L^{ij} \tau_{3jk} \epsilon^k \,, \\
  \delta G &= - \bar \epsilon_i \slashed{D} \phi^i - \frac{1}{L} \tau_{3ij} \bar \epsilon^j \phi^i \,, \\ 
  \delta E_{\mu\nu} &= \frac 14 i \bar \epsilon^i \gamma_{\mu\nu} \phi^j \varepsilon_{ij} 
   - \frac 14 i \bar \epsilon_i \gamma_{\mu\nu} \phi_j \varepsilon^{ij} \,,
 }
where we restricted to the AdS background and used the definition $E^\mu \equiv e^{-1} \varepsilon^{\mu\nu\rho \sigma} \partial_\nu E_{\rho\sigma}$, and where the expressions for $\delta \bar G$ and $\delta \phi^i$ are obtained by complex conjugation.

The key observation which gives us the conserved current multiplet is that the divergence of $E^\mu$ above vanishes by Hodge duality, so that the current defined by $j^a= \frac{1}{3!} \varepsilon^{abcd}\partial_b E_{cd}$ is conserved, i.e.~$D_a j^a =0$. Thus $j^a$ has scaling dimension $3$. The other components are simply renamed from those of \eqref{TensorContentAgain}, so the components of the flavor current multiplet are 

  \es{FlavorContent}{
  \text{\underline{Flavor current multiplet}:  \hspace{0.5in}bosonic: }& J_{ij}, K, \bar K, j_a \\
  \text{fermionic: }& \xi^i, \xi_i \,,
 }
where $J_{ij}^* = J^{ij}$.  The transformation rules in AdS are 
 \es{deltaCurrent}{
    \delta J_{ij} &= \bar \epsilon_{(i} \xi_{j)} + \varepsilon_{ik} \varepsilon_{jl} \bar \epsilon^{(k} \xi^{l)} \,, \\
  \delta \xi^i &= \frac 12 \slashed{\partial} J^{ij} \epsilon_j + \frac 12 \varepsilon^{ij} \slashed{j} \epsilon_j - \frac 12 K \epsilon^i -\frac{1}{L} J^{ij} \tau_{3jk} \epsilon^k \,, \\
  \delta K &= - \bar \epsilon_i \slashed{D} \xi^i - \frac{1}{L} \tau_{3ij} \bar \epsilon^j \xi^i \,, \\ 
  \delta j^a &=\frac 12 \bar \epsilon^i \gamma^{ab} D_b \xi^j \varepsilon_{ij} 
  + \frac 12 \bar \epsilon_i \gamma^{ab} D_b \xi_j \varepsilon^{ij} 
  -\frac{3}{4L}  \tau_{3j}{}^k \bar \epsilon_k \gamma^a \xi^j    
  -\frac{3}{4L}  \tau_{3k}{}^j \bar \epsilon^k \gamma^a \xi_j  \,.
 }
Note that $D_a\delta j^a=0$.    Also note the chirality convention $\epsilon^i,\,\xi^i$ are left-handed and $\epsilon_i, \, \xi_i$ are right-handed.  From these transformation rules, it is not hard to check that the flavor current deformation is
 \es{FlavorDef}{
  S_\text{flavor term} = \int d^4x\, \sqrt{-g} \left(K + \bar K -\frac 1 L  \tau_3^{ij} J_{ij}  \right) \,,
 }
because  the quantity in the brackets is a total derivative
 \es{OJ}{
  \delta \left(K + \bar K -\frac 1 L  \tau_3^{ij} J_{ij} \right)   = -D_a (\bar \epsilon_i \gamma^a \xi^i + \bar \epsilon^i \gamma^a \xi_i) \,.
 }

In order to connect this discussion with the discussion in Section~\ref{FLAVOR}, note that the transformation rules \eqref{deltaCurrent} imply
 \es{QQJ}{
  \bar \Q_1 \Q_2 J_{12} = -\frac{K}{2} + \frac{ i J_{12}}{L} \,, \qquad
   \bar \Q^1 \Q^2 J_{12} = \frac{\bar K}{2} - \frac{ i J_{12}}{L}  \,.
 }
Eliminating $K$ and $\bar K$ using these equations, one finds that the flavor current term can be written as
 \es{FlavTermAgain}{
  S_\text{flavor term} = - 2  \int d^4x\, \sqrt{-g}  \left(-3 i  \frac{J_{12}}{L} + \bar \Q_1 \Q_2 J_{12} - \bar \Q^1 \bar \Q^2 J_{12}  \right) \,,
 }
which agrees with \eqref{SJ} up to an unimportant normalization factor.

There is another way to derive the flavor term \eqref{FlavorDef}.  A conserved current multiplet couples to a vector multiplet ${\cal V} = (X, \bar X, A_\mu, Y_{ij}, \Omega_i, \Omega^{i})$ in a supersymmetric way.  This coupling is
 \es{Coupling}{
  S_{A-J} = \int d^4x\, \sqrt{-g} \biggl(A_\mu j^\mu - X K - \bar X \bar K + J_{ij} Y^{ij} 
   - \bar \xi^i \Omega_i  - \bar \xi_i \Omega^i \biggr) \,.
 }
Using the transformation rules \eqref{SUSYVector} for the vector multiplet and \eqref{deltaCurrent} for the current multiplet, one can check that the SUSY variation of the integrand in \eqref{Coupling} is a total derivative.  Now let us think of the vector multiplet ${\cal V}$ as a background multiplet, so that we can set the vector multiplet fields to any values we wish.  In order to preserve supersymmetry, however, we should ensure that the SUSY variations of the vector multiplet fields vanish. From the variations \eqref{SUSYVector}, it is easy to see that, with $m$ a constant with dimensions of mass,
 \es{BackgroundVector}{
  X = \bar X = \frac m2 \,, \qquad Y_{ij} =\frac m2 \frac{ \tau_{3 ij}}{L} \,, \qquad A_\mu = \Omega^i = \Omega_i = 0 \,,
 }
is indeed a supersymmetric background.  Plugging these values into \eqref{Coupling} one obtains
 \es{CouplingExplicit}{
  S_{A-J} =- \frac m2 \int d^4x\, \sqrt{-g} \biggl( K + \bar K - \frac{\tau_{3 ij}}{L} J^{ij}  \biggr) \,,
 }
which again agrees with the flavor term \eqref{FlavorDef} up to an unimportant overall normalization constant.

\subsection{A flavor current term example:  the hypermultiplet mass term}
\label{MASSIVEHYPEREXAMPLE}

We now will consider a theory of $N_H$ free massless hypermultiplets  with global flavor symmetry $USp(2N_H)$.  We will couple this multiplet to a background  abelian vector multiplet which gauges a $U(1)$ subgroup of  $USp(2N_H)$.  With proper choice of the background the hypermultiplet will become massive.

The $N_H$ hypermultiplets consist of scalars $q_{iA}$, with $i=1, 2$, and $A = 1, \ldots 2 N_H$ and fermions $\zeta_A$ and $\zeta^A$:
  \es{HyperContent}{
  \text{\underline{$2 N_H$ hypermultiplets}:  \hspace{0.5in}bosonic: }& q_{iA}, q^{iA} \\
  \text{fermionic: }& \zeta_A,  \zeta^A \,.
 }
The scalars obey the reality condition $q_{iA}^* = q^{iA} \equiv \varepsilon^{ij} \Omega^{AB} q_{jB}$, where $\Omega^{AB}$ is the anti-symmetric symplectic form of $USp(2 N_H)$.  For concreteness, we can consider a basis where $\Omega^{AB}$ is block diagonal, with each block being equal to $i \sigma_2$:
 \es{OmAB}{
  \Omega^{AB} = \begin{pmatrix}
   i \sigma_2 & & \\
   & i \sigma_2 & \\
   & & \ddots
  \end{pmatrix} \,.
 }
The fermions $\zeta_A$ are left-handed, while their conjugates $\zeta^A$ are right-handed.  The symplectic indices can be raised and lowered with $\Omega^{AB} = \Omega_{AB}$  using the NW-SE convention on all bosonic quantities, but we will not raise and lower the indices on the fermions since they are used to also indicate the chirality. 

Let us consider a $U(1)$ subgroup of $USp(2N_H)$ whose generator is $T_A{}^B$.  This means that under infinitesimal $U(1)$ transformations with parameter $\theta$, we have $\delta q^{iA} = \theta T_B{}^A q^{iB}$ and $\delta q_{iA} = - \theta T_A{}^B q_{iB}$.  Consistent raising and lowering rules require that $T^{AB} = \Omega^{AC} T_C{}^B$ and $T_{AB} = \Omega_A{}^C T_{CB}$ are symmetric matrices. As an example, we can consider the case where these matrices are block diagonal, with each $2 \times 2$ block being the same:
 \es{TExample}{
  T_A{}^B = \begin{pmatrix}
   i \sigma_3 & & \\
   & i \sigma_3 & \\
   & & \ddots
  \end{pmatrix}  \,, \qquad
   T_{AB} = - T^{AB} =  \begin{pmatrix}
   i \sigma_1 & & \\
   & i \sigma_1 & \\
   & & \ddots
  \end{pmatrix} \,.
 }

 The gauge covariant derivatives are 
 \es{GaugeCov}{
  D_\mu q^{i A} = \partial_\mu q^{i A} - A_\mu q^{i B} T_B{}^A \,, \qquad
   D_\mu \zeta^A = \partial_\mu \zeta^A + \frac 14 \omega_\mu{}^{ab} \gamma_{ab} \zeta^A - A_\mu \zeta^B T_B{}^A \,.
 }
The hypermultiplet Lagrangian is then a particular case\footnote{To obtain \eqref{SHyp} from (3.163) of \cite{Lauria:2020rhc}, proceed as follows.  First, note that what we call $q^{iA}$ is the same as the section $A^{iA} = f^{iA}{}_X q^X$ in \cite{Lauria:2020rhc}, and we choose frame vectors such that $q^{iA} = \frac{1}{\sqrt{2}} \begin{pmatrix}
 i q^3 + q^4 & i q^1 - q^2 \\
 i q^1 + q^2 & - i q^3 + q^4 
\end{pmatrix}$.  The dilatation Killing vector is chosen to be $k_D^X = q^X$.  The Killing vector associated with the $U(1)$ isometry is $k^X = T_B{}^A q^{iB} f^X{}_{iA}$ and the corresponding triplet of moment maps is $\vec{P} = \frac 12 q^{iA} \vec{\tau}_i{}^j T_A{}^C q_{jC}$.  We also have $t_{AB} = T_{AB}$.  Lastly, we take $g_{XY} = \delta_{XY}$ and $d^A{}_B = \delta^A_B$.} of Eq.~(3.163) of \cite{Lauria:2020rhc}:
 \es{SHyp}{
  S_\text{gauged hyp} &= \int d^4 x\, \sqrt{-g} \biggl[ 
   - \frac 12 D_\mu q^{i A} D_\mu q_{i A} + \frac{1}{L^2}  q^{iA} q_{iA} + 2 \abs{X}^2 T_B{}^A T_A{}^C q^{i B} q_{i C} 
    \\
   &{}+  q^{i A} T_A{}^B Y_i{}^j  q_{jB} + \left( -\bar \zeta_A \slashed{D} \zeta^A + 2 X \bar \zeta^A \zeta^B T_{AB} + 2i T_B{}^A q^{iB}  \bar \zeta_A \Omega^j \varepsilon_{ij} + \text{c.c.} \right)  
  \biggr] \,.
 }
It is invariant under the SUSY rules (3.101) of \cite{Lauria:2020rhc}, which, when restricted to our AdS background, become:
 \es{SUSYHyper}{
  \delta q^{iA} &= - i \bar \epsilon^i \zeta^A + i \bar \epsilon_j \zeta_B \varepsilon^{ji} \Omega^{BA} \,,\\
  \delta \zeta^A &= \frac 12 i \slashed{D} q^{iA} \epsilon_i + i \bar X T_B{}^A q^{iB} \varepsilon_{ij} \epsilon^i -\frac{1}{2L} i q^{i A} \tau_{3 ij} \epsilon^j \,.
 }

If we treat the vector multiplet as a background vector multiplet and give the fields of this multiplet the supersymmetric values  in \eqref{BackgroundVector}, we obtain the action for $N_H$ massive hypermultiplets in AdS:
 \es{SHypMassive}{
  S_\text{massive hyp} &= \int d^4 x\, \sqrt{-g} \biggl[ 
   - \frac 12 \partial_\mu q^{i A} \partial^\mu q_{i A} + \frac{1}{L^2}  q^{iA} q_{iA} + \frac{m^2}{2} T_B{}^A T_A{}^C q^{i B} q_{i C} 
    \\
   &{}+  \frac{ m}{2L} q^{i A}  \tau_{3 i}{}^j T_A{}^B q_{jB} + \left( -\bar \zeta_A \slashed{D} \zeta^A + m \bar \zeta^A \zeta^B T_{AB} + \text{c.c.} \right)  
  \biggr] \,.
 }
The case where all the hypermultiplets have equal masses corresponds to the choice \eqref{TExample}.  To make this clearer, note that we can write the terms involving the scalar fields only in terms of $q_{1A}$, because the $q_{2A}$ are related by complex conjugation to the $q_{1A}$.  The massive hypermultiplet action becomes 
 \es{SHypMassiveAgain}{
  S_\text{massive hyp} &= \int d^4 x\, \sqrt{-g} \biggl[ 
   \sum_A \biggl( - \abs{\partial_\mu q_{1 A}}^2  - \left( m^2 - \frac{2}{L^2} \right)  \abs{q_{1 A} }^2 \\
      &{}- \bar \zeta_A \slashed{D} \zeta^A -  \bar \zeta^A \slashed{D} \zeta_A + i m \bar \zeta^A \zeta^A  - i m \bar \zeta_A \zeta_A    \biggr) 
    \\
   &{}-\frac{m}{L}  \sum_{A \text{ odd}}  \abs{  q_{1A}}^2  + \frac{m}{L} \sum_{A \text{ even}}  \abs{  q_{1A}}^2  
  \biggr]  \,.
 }
Thus, all fermions have the same mass $m$, while the scalars $q_{1A}$ have squared masses equal to $m^2 - \frac{2}{L^2} - \frac{m}{L}$ when $A$ is odd and $m^2 - \frac{2}{L^2} + \frac{m}{L}$ when $A$ is even.  In the more general case \eqref{SHypMassive}, from comparing with \eqref{CouplingExplicit}, we can identify the flavor current deformation as the term proportional to $m$ in \eqref{SHypMassive}:
 \es{SFlavorHyp}{
  S_\text{flavor term} =  m  \int d^4 x\, \sqrt{-g} \biggl[ 
    \frac{ 1}{2L} q^{i A}  \tau_{3 ij} T_{AB} q^{jB} +   \bar \zeta^A \zeta^B T_{AB} +   \bar \zeta_A \zeta_B T^{AB}
  \biggr]  \,.
 }

We can arrive at the same result by first identifying the composite flavor multiplet fields from comparing the linear term in the vector multiplet fields in \eqref{SHyp} with the general coupling \eqref{Coupling} between a vector multiplet and a flavor current multiplet.  This comparison gives
 \es{GotCurrent}{
  j_a &= (\partial_a q_{i A}) T_B{}^A q^{iB}  -   q_{i A} T_B{}^A (\partial_a q^{iB})  
   + \bar \zeta_A \gamma_a T_B{}^A \zeta^B - \bar \zeta^A \gamma_a T_A{}^B \zeta^B \,, \\
   J_{ij} &=  q_{iA} T^{AB} q_{jB} \,, \qquad
  K = - 2 \bar \zeta^A \zeta^B T_{AB} \,, \qquad \bar K =  - 2 \bar \zeta_A \zeta_B T^{AB} \,, \\
  \xi_j &= 2 i T_B{}^A q^{i B} \zeta_A \varepsilon_{ij} \,, \qquad \xi^j = -2 i T^B{}_A q_{i B} \zeta^A \varepsilon^{ij} \,.
 }
Then, Eq.~\eqref{CouplingExplicit} with the values \eqref{GotCurrent} reproduces \eqref{SFlavorHyp}.

\subsection{Other supersymmetric interactions from conformal supergravity}

We will not perform a detailed analysis of the conformal supergravity origin of the SUSY-exact supersymmetric interactions from Section~\ref{DTERMS}.  However, let us point out that the unconstrained multiplet in AdS can be obtained from an unconstrained superconformal multiplet. Such a multiplet was analyzed in \cite{Hayashi:1985gc}.  It starts with a field ${\cal C}$ with Weyl weight $w$ and chiral weight $n$ obeying no constraints.  As in the case of the chiral F-term, we expect that when all the supergravity fields are set to their background AdS values, the parameters $w$ and $n$ can be redefined away.  When $n=0$, the long superconformal multiplet can be taken to be real, and it should reduce precisely to the unconstrained AdS multiplet that was used in Section~\ref{DTERMS} to construct the D-term interactions.\footnote{One should set ${\cal C} = X$, where $X$ is the R-neutral scalar below \eqref{DtermN2}.}  When $n \neq 0$ the long superconformal multiplet is necessarily complex, so it should reduce to two copies of the (real) unconstrained multiplet in AdS\@.

While the D-term can be obtained from a long superconformal multiplet of spin $0$, the non-chiral F-term and the $1/4$-BPS interactions mentioned in Section~\ref{DTERMS} can be obtained from other superconformal multiplets that obey shortening conditions.  In the study of supersymmetric deformations of superconformal field theories in flat space, it was observed in~\cite{Cordova:2016xhm} that analogous $1/2$-BPS and $1/4$-BPS deformations also exist in that case.  For instance, see Table 22 of \cite{Cordova:2016xhm}, where the non-chiral F-terms are given on the second line and the $1/4$-BPS deformations are given on the fifth and sixth lines.  That table also contains the flavor current deformation on the first line, the chiral F-terms on the third and fourth lines, and the D-terms on the last line.

\section{Effective field theory in AdS from on-shell supergravity and the ${\cal N} = 2$ massive vector multiplet}
\label{ONSHELL}

So far, we constructed effective Lagrangians in $\text{AdS}_4$ in two ways:  using the $\mathfrak{osp}(2|4)$ algebra, in Section~\ref{SUSY}, and starting from conformal supergravity in Section~\ref{CONFORMAL}.  In the latter case, $\text{AdS}_4$ was realized as a supersymmetric background of a certain ${\cal N} = 2$ off-shell supergravity that is obtained from conformal supergravity with compensating vector and tensor multiplets.  In Section~\ref{RIGID}, we will explore another approach, namely the decoupling limit of matter-coupled ${\cal N} = 2$ {\em on-shell} supergravity.  Several applications of this formalism then follow:  first to the free massive hypermultiplet in Section~\ref{MASSIVEHYPER}, then to the F-term interactions of a vector multiplet in Section~\ref{MASSLESSVECTOR}, to the quadratic Lagrangian of a massive vector multiplet in Section~\ref{MASSIVEVECTOR}, and finally to the cubic interactions of a massive and massless vector multiplet in Section~\ref{INTERACTIONS}.  These applications prepare the way for Sections~\ref{ONELOOP} and~\ref{INTEGRATED} in which we undertake calculations of the sphere free energy.

\subsection{A rigid limit of matter-coupled ${\cal N} = 2$ supergravity}
\label{RIGID}

${\cal N} = 2$ on-shell supergravity is described in Section~21 of \cite{Freedman:2012zz}.  As emphasized there, while the action is derived using the superconformal gravity formalism, the final result can be derived in other ways and is more general than the derivation would indicate.  Let us describe briefly the data that goes into the ${\cal N} = 2$ supergravity theory, and then how to obtain the rigid $\text{AdS}_4$ theory from it.

The ${\cal N} =2$ supergravity theory with $n_V$ vector multiplet and $n_H$ hypermultiplets is constructed from the following data:
\begin{itemize}
	\item A prepotential $F(X^I)$, which is a holomorphic function of $n_V + 1$ variables $X^I$ (with $I = 0, \ldots, n_V$) that is homogeneous of degree $2$.  This determines the scalar target space, which is an $n_V$-dimensional special K\"ahler manifold parameterized by complex coordinates $z^\alpha$, with $\alpha = 1, \ldots, n_V$.
	\item A hypermultiplet scalar manifold, which is quaternionic-K\"ahler and has $4 n_H$ real dimensions or, equivalently, $n_H$ quaternionic dimensions.  It is parameterized by the real coordinates $q^u$ (with $u = 1, \ldots 4 n_H$), and it is negatively curved, with its Ricci scalar curvature proportional to $-\kappa^2$.
	\item The structure constants $f_{IJ}{}^K$ of the gauge algebra.  These structure constants also determine the couplings of the vector fields to the scalars and fermions in the vector multiplets.  For simplicity, we  focus on the case of an Abelian gauge theory, so the $f_{IJ}{}^K$ vanish.
	
	\item The gauging describing the coupling of the $n_V + 1$ vector fields $A_\mu^I$ to the hypermultiplet matter fields. This coupling is encoded in Killing vectors $k_I^u$, or equivalently the triplet of moment maps denoted by $\vec{P}_I$.  The Killing vectors and moment maps must obey compatibility conditions with the quaternionic structure.
	
\end{itemize} 

The action for rigid   ${\cal N} = 1$ SUSY in $\text{AdS}_4$ was obtained by decoupling supergravity fields in \cite{Adams:2011vw}.  The ${\cal N} = 2$ procedure is much trickier;  it proceeds via an expansion at small $\kappa$:
\begin{enumerate}
	\item First, we assume scalings in $\kappa$ of the various fields so that at leading order, namely order $1/\kappa^2$, one obtains pure ${\cal N} = 2$ supergravity with negative cosmological constant adjusted to produce a classical solution of the equations of motion describing $\text{AdS}_4$ with radius $L$.
	\item We then observe that this solution with metric tensor of $\text{AdS}_4$ and vanishing gravitino and graviphoton fields is supersymmetric.
	\item Next, we consider fluctuations around this background at order $\kappa^0$.  At this order, the matter fields decouple from the supergravity fluctuations, so we can focus on the matter fields alone.  In addition, several simplifications occur, e.g.~the hypermultiplet manifold becomes flat since its curvature was proportional to $\kappa^2 \to 0$.
\end{enumerate}

We now go through the steps in detail.  Although we write the steps as an expansion in $\kappa$, the dimensionless expansion parameter is $\kappa / L$.

\subsubsection{Step 1:  pure ${\cal N} = 2$ supergravity with negative cosmological constant}

At leading order, take the prepotential
\es{FLeading}{
	F(X^I) = -\frac{i}{4 \kappa^2} (X^0)^2 + O(\kappa^0) \,,
}
and take $X^0 = 1$.  This implies that the K\"ahler potential for the vector multiplet scalars is $K = O(\kappa^0)$.  We choose a direction in $SU(2)$ space, say the third direction, and set the $\vec{P}_0$ moment map to be a constant of order $O(1/\kappa^2)$,
\es{P0Leading}{
	\vec{P}_0 = \frac{1}{\kappa^2 L} \left(0 , 0, 1 \right) + O(\kappa^0) \,,
}
and all other moment maps to be $O(\kappa^0)$.  We scale the scalar and fermion fields with $\kappa$ in such a way that their kinetic terms are $O(\kappa^0)$.   Lastly, we take all gauge fields $A_\mu^I$, the gravitini, and the metric to be of order $O(\kappa^0)$.

With these simple choices, the Lagrangian (21.34) of \cite{Freedman:2012zz} becomes
\es{ActionLeading}{
	e^{-1} {\cal L} 
	= \frac{1}{\kappa^2} \left[ \frac{R}{2} - \bar \psi_{i \mu} \gamma^{\mu\nu\rho} D_\nu \psi_\rho^i
	- \frac{1}{8}  F_{\mu\nu}^{0} F^{\mu\nu0}
	-  \frac{\tau_{3\, ij} \bar \psi_\mu^i \gamma^{\mu\nu} \psi_\nu^j + \text{h.c.} }{2 L }       + \frac{3}{L^2} \right] 
	+ O(\kappa^0) \,.
} 
Note that the quantity $S^{ij}$ (see  (21.39) of \cite{Freedman:2012zz}) that gives the gravitino mass term above evaluates to 
\es{SijLeading}{
	S^{ij} = P_{I}^{ij} \bar X^I =  \frac{1}{\kappa^2 L } \tau_3^{ij} + O(\kappa^0) \,.
}
Eq.~\eqref{ActionLeading} is the pure ${\cal N} = 2$ supergravity Lagrangian containing as dynamical fields only the metric, the gravitini, and the gauge field which represents the graviphoton.   

\subsubsection{Step 2:  supersymmetric $\text{AdS}_4$ solution}

The equations of motion following from \eqref{ActionLeading} are solved by the $\text{AdS}_4$ solution of radius $L$, with vanishing gravitini and graviphoton: 
\es{SolLeading}{
	g_{\mu\nu} = g_{\mu\nu} \Big|_{AdS_4} \,, \qquad
	\psi_i = 0 \,, \qquad A^0_\mu = 0 \,.
}
This solutions is supersymmetric as can be checked from the vanishing supersymmetry variation of the fields.  The SUSY variations of the bosonic fields automatically vanish because the fermions vanish.  For the SUSY variation, note that ${\cal A}_\mu$, ${\cal V}_\mu{}^i{}_j$, and $T_{ab}^-$ are $O(\kappa^2)$,  while $S^{ij}$ is $O(\kappa^{-2})$ as in \eqref{SijLeading}.  Thus, the gravitino variation is  
\es{GravitinoVarLeading}{
	\delta \psi_\mu^i 
	= D_\mu \epsilon^i + \frac{1}{2 L } \tau_3^{ij} \gamma_\mu \epsilon_j  + O(\kappa^2)\,,
}
with a similar expression for $\delta \psi_{\mu i}$.  These expressions vanish provided that the SUSY parameters $\epsilon^i$ and $\epsilon_i$ obey the Killing spinor equations
\es{KSEqsLeading}{
	D_\mu \epsilon^i = - \frac{1}{2 L } \tau_3^{ij}  \gamma_\mu\epsilon_j + O(\kappa^2) \,, \qquad
	D_\mu \epsilon_i = - \frac{1}{2 L } \tau_{3ij}  \gamma_\mu \epsilon^j + O(\kappa^2)  \,,
}
which are the same equations as the ones encountered in the previous sections.  As indicated, these equations hold only at leading order in $\kappa$, but this is sufficient for our purposes. 

\subsubsection{Step 3:  effective theory in $\text{AdS}_4$}

We now expand to the next order in $\kappa$.  We proceed in pragmatic fashion, emphasizing the facts needed for the expansion of the action (21.34)\footnote{All equations numbered (21.xy) refer to Ch. 21 of \cite{Freedman:2012zz}.}.  In this discussion  the index $I$ will take values from $0$ to $n_V$, while ${\cal I}$ will take values from $1$ to $n_V$, and so will $\alpha$.  As a temporary notation, quantities before expansion will carry the asterisk as a subscript, e.g.  $F_*(X^I_*)$.

\paragraph{Vector multiplets:}  We consider a theory with $n_V$ Abelian vector multiplets with the prepotential
\es{FSG}{
	F_*(X_*^I) =  -\frac{i}{4} (X_*^0)^2 \left( \frac{1}{\kappa^2}  - F(Y^{\cal I}) + O(\kappa^2) \right)\,, \qquad
	Y^{\cal I} \equiv \frac{X_*^{\cal I}}{X_*^0} \,,
}
for some arbitrary function $F(Y^{\cal I})$ that will become the prepotential of the rigid AdS theory.  We then define the homogeneous coordinates  $X_*^I = y_*(z^\alpha) Z_*^I(z^\alpha)$ in terms of $n_V$ variables $z^\alpha$.  We assume that $Z_*^0=1$, as is common, and that the $Z_*^{\cal I}(z^\alpha)$ do not receive corrections in $\kappa$.  The variable $y_*$ will have the expansion $y_*=1+\kappa^2 y + O(\kappa^4)$, where $y$ is determined from the conventional constraint, see (21.1):
\es{constraint}{
	N_{*\,IJ} X_*^I \bar X_*^J = - \frac{1}{\kappa^2} \,, \qquad 
	N_{*\, IJ} = 2 \Im F_{* IJ}
}
One solves the constraint for $y_*$ and then expands to find $y$.  The K\"ahler  potential, see (21.2), is  obtained from
\es{Kpot}{
	{\cal K}_*(z, \bar z) = {\cal K} + O(\kappa^2) = \frac{2}{\kappa^2} \log( y_*) =\frac{2}{\kappa^2}(1 + \kappa^2 y +O(\kappa^2)).
}
The result is
\es{Kahler}{
	{\cal K} \equiv \frac{F + \bar F}{2} + \frac 14 \left(z^{\cal I} - \bar z^{\cal I}\right) \left( {\bar \partial}_{\cal I} \bar F -  \partial_{\cal I} F\,,  \right)  \,.
}
which determines the K\"ahler metric $g_{{\cal I} {\cal J}} = \partial_{\cal I} \bar \partial_{\cal J} {\cal K}$.  The expansion of ${\cal N}_{*IJ}$ in (21.5) determines gauge field kinetic terms. Since we have set $A^0_\mu =0$, we need only  the components ${\cal I}{\cal J}$ components:
\es{GotNMR}{
	{\cal N}_{{\cal I} {\cal J}} = - \frac{i}{4}  \bar \partial_{\cal I} \bar \partial_{\cal J} \bar F \,.
} 
Henceforth, we simplify the notation by identifying the indices ${\cal I}$ and $\alpha$ by taking $Z_*^{\cal I}= z^{\cal I}=z^\alpha$.

\paragraph{Gauged Hypermultiplets:}  The Abelian gauge fields $A^{\cal I}$ couple to the hypermultiplet matter fields.  In supergravity, the hypermultiplet scalar manifold is a quaternionic K\"ahler manifold parameterized by coordinates $q^u$ and with curvature proportional to $-\kappa^2$.\footnote{We lower and raise the $u$ indices with the metric $h_{uv}$ and its inverse.}
 The quaternionic \Kahler manifold is described by a
	frame field $f_*^{iA}{}_u$ and its inverse $f_*^{u}{}_{iA}$, with indices $u=1,\ldots,4n_H$ and indices $i,A=1,\ldots,2n_H$. From these we construct the metric $h_{uv}$ and the hypercomplex structure $\vec{J}$:
 \es{eq:frame_properties}{
	f_*^{iA}{}_v f_*^{u}{}_{iA} &= \delta^u_v\,, \qquad  f_*^{iA}{}_u f_*^{u}{}_{jB} = 
	\delta_j^i\\
	\vec{J}_{*\, u}{}^v &= -f_*^{iA}{}_u f_*^{v}{}_{jA} \vec{\tau}\indices{_i^j}\,, \qquad
	h_{*uv}={f_*^{iA}}_{u} \varepsilon_{ij} C_{AB} {f_*^{jB}}_v\,,
	}
	with $C_{AB}$ an antisymmetric matrix (see Section~20.3.4 of \cite{Freedman:2012zz} for details).
	The hypercomplex structures $\vec{J}_{*\, uv}$ obey $J^1_{* u}{}^v J^2_{*\, v}{}^w = J^3_{*\, u}{}^w$.
	The coupling to vector multiplets is described by the Killing vectors $k_{*\, I}^u$ and moment maps $\vec{P}_{*\, I}$.  These obey the condition 
	\es{ConditionKilling}{
		\partial_u \vec{P}_{*\, I} + 2 \vec{\omega}_{*\, u} \times \vec{P}_{*\, I}
		= \vec{J}_{*\, uv} k_{*\, I}^v \,,
	}
	where $\vec{\omega}_{*\, u}$ are functions defined through $\nabla_w \vec{J}_{*\, u}{}^v + 2 \vec{\omega}_{*\, w} \times\vec{J}_{*\, u}{}^v = 0$.

In the limit $\kappa \to 0$, the hypermultiplet scalar manifold becomes flat, but this limit has to be taken with care because we want  $\vec{P}_{*\, 0 }$ to approach the value given in \eqref{P0Leading} which is of order $O(1/\kappa^2)$, so subleading corrections in $\kappa$ will be important.  At the end of the day, what we will need is that the frame fields, the metric, and the hypercomplex structures approach corresponding quantities on a flat $\R^{4 n_H}$ space
\es{HyperComplexLimit}{
	f_*^{iA}{}_u = f^{iA}{}_u + O(\kappa^2) \,, \qquad
	h_{*\, uv} = \delta_{uv} + O(\kappa^2) \,, \qquad
	\vec{J}_{*\, uv} = \vec{J}_{uv} + O(\kappa^2) \,,
}
while the functions $\vec{\omega}_{*\, u}$ vanish as $\kappa \to 0$ as
 \es{omegaLeading}{
  \vec{\omega}_{*\, u} = \frac{\kappa^2}{4} \vec{J}_{uv} q^v + O(\kappa^4) \,.
 } 
(For more details, see Appendix~\ref{FLATLIMIT}.)  The Killing vectors $k_{*\, I}^u$ approach
\es{KillingLimit}{
	k_{*\, 0}^u =- \frac{1}{2L} J^{3\, u}{}_{v} q^v + k_0^u + O(\kappa^2) \,, \qquad 
	k_{*\, {\cal I}}^u = k_{\cal I}^u + O(\kappa^2)
}
where $k_0^u$ and $k_{\cal I}^u$ are (commuting) flat space Killing vectors, and the moment maps approach
\es{MomentLimit}{
	\vec{P}_{*\, 0 }
	=  \left(0 , 0, \frac{1}{\kappa^2 L} + \frac{q^u q_u}{4L} \right) 
	+ \vec{P}_0 + O(\kappa^2) \,, \qquad
	\vec{P}_{*\, {\cal I} } = \vec{P}_{\cal I} + O(\kappa^2) \,.
}
Here, $\vec{P}_0$ and $\vec{P}_{\cal I}$ are the flat space moment maps related to the flat space Killing vectors via
\es{MomentKillingFlat}{
	\partial_u \vec{P}_{I} 
	= \vec{J}_{uv} k_{ I}^v \,, \qquad \text{for } I = 0, {\cal I} \,.
}
Note that the formulas for $k_{*\, 0}^u$ and $\vec{P}_{*\, 0 }$ differ from those for $k_{*\, {\cal I}}^u$ and $\vec{P}_{*\, {\cal I} }$.  The Killing vectors are triholomorphic. They obey both the flat space Killing equations and must be compatible with the hypercomplex structure, viz.
\es{Triholo}{
	\partial_u k_{I v} + \partial_v k_{I u} = 0  \,, \qquad
	\partial_u k_I^v \vec{J}_v{}^w - \vec{J}_u{}^v \partial_v k_I^w = 0 \,, \qquad
	\text{for } I = 0, {\cal I} \,.
} 
In addition, in supergravity, Abelian Killing vectors must obey the condition $k_{*\, I}^u \vec{J}_{*\, uv} k_{*\, J}^v = -\kappa^2 \vec{P}_{*\, I} \times \vec{P}_{*\, J}$.  For us, this implies
\es{ConditionsKilling}{
	k_{\cal I}^u \vec{J}_{uv} k_{\cal J}^v = 0 \,, \qquad
	k_{\cal I}^u \vec{J}_{uv} k_0^v = -\frac{1}{L} \left(  \frac{1}{2}k_{\cal I}^u J^2_{uv} q^v + P_{\cal I}^2 ,  
	-   \frac{1}{2}k_{\cal I}^u J^1_{uv} q^v - P_{\cal I}^1 , \frac{1}{2} k_{\cal I}^u q_u \right) \,.
}

Finally, in order to write down the supergravity potential as well as the supersymmetry variations later on, it is useful to define the following quantities: 
\es{Defs}{
	S_{ij} &= P_{{\cal I} ij} \bar z^{\cal I} + P_{0 ij}
	+ \frac{q^u q_u}{4L} \tau_{3 ij} +  \frac{1}{2 L } \tau_{3\, ij}  {\cal K} \,, \\
	W_{\cal I}{}^{ij} &= -  P_{\cal I}{}^{ij}  
	-  \frac{1}{ L} \tau_3^{ij}  \partial_{{\cal I}} {\cal K} \,, \\
	N^i{}_A &= i f^{iB}{}_u \left(k_{\cal I}^u z^{\cal I} + k_0^u -  \frac{1}{2L} J^{3\, u}{}_{v} q^v \right) \Omega_{BA} \,,
}
which are the terms proportional to $\kappa^0$ in the expansions of the corresponding expressions in supergravity, see (21.39-.40).  Then, after writing
\es{ScalarPot}{
	V_* = -\frac{3}{ \kappa^2 L^2} + V + O(\kappa^2) \,,
}
(21.46)  gives        
\es{Reln}{
	- \frac{3}{L} \tau_3^{ik} S_{jk} 
	- \frac{3}{L} S^{ik} \tau_{3jk} 
	+ W_{\cal I}{}^{ik} g^{{\cal I}{\cal J}} \bar W_{{\cal J} jk}
	+ 4 N^i{}_A \bar N_j{}^A = \delta^i_j\,  V \,.
}

At order $\kappa^0$, the Lagrangian is then
\es{Action}{
	e^{-1} {\cal L} \big|_{\kappa^0} &= - g_{{\cal I} {\cal J}} \partial_\mu z^{\cal I} \partial^\mu \bar z^{\cal J}
	- \frac 12 D_\mu q^u D^\mu q_u - V \\
	&{}+ \biggl\{ 
	- \frac{i}{4} {\cal N}_{{\cal I} {\cal J}} F_{\mu\nu}^{+{\cal I}} F^{+\mu\nu {\cal J}}  
	-\frac 14 g_{{\cal I} {\cal J}}  \bar \chi_i^{\cal I} \slashed{D} \chi^{i {\cal J}}
	- \bar \zeta_A \slashed{D} \zeta^A \\
	&{}- 2 i \bar k_{\cal I}^u \varepsilon_{ij} f^{jA}{}_u  \bar \chi^{i {\cal I}} \zeta_A
	+ 2  t_{0AB} \bar \zeta^A \zeta^B + 2 z^{\cal I} t_{{\cal I} AB} \bar \zeta^A \zeta^B + \text{h.c.} 
	\biggr\}  \,.
}
where the covariant derivatives are
\es{partialDefs}{
	D_\mu q^u& = \partial_\mu q^u - A_\mu^{\cal I} k_{\cal I}^u \,, \\
	D_\mu \chi_i^{\cal I} &= \left(\partial _\mu 
	+ \frac 14 \omega_\mu{}^{ab} \gamma_{ab} \right) \chi_i^{\cal I}
	+ \Gamma^{{\cal I}}_{{\cal J}{\cal K}}  \chi_i^{\cal K} \partial_\mu z^{\cal J}  \,, \\
	D_\mu \zeta^A &=   \left(\partial _\mu 
	+ \frac 14 \omega_\mu{}^{ab} \gamma_{ab} \right) \zeta^A
	- A_\mu^{\cal I} t_{{\cal I} B}{}^A  \zeta^B  \,, \qquad
	t_{{ I} A}{}^B \equiv \frac 12 f^v{}_{iA} \partial_v k_{ I}{}^u f^{iB}{}_u \,.
}

The supersymmetry variations of the bosonic fields are
\es{SUSYVarsBos}{
	\delta z^{\cal I} &= \frac 12 \bar \epsilon^i \chi_i^{\cal I} \,, \\
	\delta q^u &= - i \bar \epsilon^i \zeta^A f^u{}_{iA} \,, \\
	\delta A_\mu^{\cal I} &= \frac 12 \varepsilon^{ij} \bar \epsilon_i \gamma_\mu 
	\chi_j^{\cal I} + \text{h.c.} \,. 
}
The supersymmetry variations of the fermions are
\es{SUSYVarsFerm}{
	\delta \chi_i^{\cal I} &= \slashed{\partial} z^{\cal I} \epsilon_i 
	- g^{{\cal I}{\cal J}} (\Im {\cal N}_{{\cal J}{\cal K}}) F_{ab}^{-{\cal K}} \gamma^{ab} \varepsilon_{ij} \epsilon^j 
	+ g^{{\cal I} {\cal J}} \bar W_{{\cal J} ji} \epsilon^j \,, \\
	\delta \zeta^A &= \frac 12 i f^{iA}{}_u \slashed{D} q^u \epsilon_i
	- \bar N_i{}^A \epsilon^i \,.   
}

\subsubsection{$U(1)_R$ symmetry}
\label{U1RSECTION}

The on-shell supergravity theory has a local $U(1)_R$ symmetry, gauged by the graviphoton field  $A_\mu^0$.  Since we set $A_\mu^0$ to zero, the rigid theory in AdS we constructed above will have  $U(1)_R$ as a global symmetry.  The $U(1)_R$ transformation properties of the various fields are described by the Killing vectors $k^u_{*\, 0}$ as defined in \eqref{KillingLimit} for the scalar fields and $ t_{0 A}{}^B \equiv \frac 12 f^v{}_{iA} \partial_v k_{*\, 0}^u f^{iB}{}_u$ for the fermions.  In particular, under $U(1)_R$ transformations with parameter $\theta$, we have
\es{U1R}{
	\delta_\theta z^{\cal I} = 0 \,, \qquad
	\delta_\theta q^u = 2L\,  \theta\,  k^u_{*\, 0} \,, \qquad
	\delta_\theta \chi^{i {\cal I}} =\theta\,   \tau_{3\, j}{}^{i} \chi^{j {\cal I}} \,, \qquad
	\delta_\theta \zeta^A = 2L\,  \theta \,  t_{0B}{}^A \zeta^B \,,
} 
as well as the conjugates of these transformations for the hermitian conjugate fields.  In \eqref{U1R}, the normalization is chosen such that the vector multiplet fermions transform as $\delta \chi^{1 {\cal I}}= i \theta \chi^{1 {\cal I}}$ and $\delta \chi^{2 {\cal I}}= -i \theta \chi^{2 {\cal I}}$, which we take to mean that $\chi^{1 {\cal I}}$ have $U(1)_R$ charge $+1$ while $\chi^{2 {\cal I}}$ have $U(1)_R$ charge $-1$.

\subsection{Quadratic action for a massive hypermultiplet}
\label{MASSIVEHYPER}

One of the simplest applications of the formalism above is to the case $n_H = 1$, $n_V = 0$ of a (massive) hypermultiplet.  The only freedom we have is to choose the Killing vector $k_0^u$, which, per the discussion in Section~\ref{U1RSECTION}, determines the $U(1)_R$ charges of the hypermultiplet fields. 

	To be concrete, let us choose the frame fields to be\footnote{These are the same expressions as in (20.47) of \cite{Freedman:2012zz} rescaled by a factor of $1/\sqrt{2}$.}
	\es{FrameFields}{
	f^{12}{}_1 &= f^{21}{}_1 = f^{11}{}_3 = -f^{22}{}_3 = \frac{i}{\sqrt{2}} \,, \\
	f^{21}{}_2 &= f^{11}{}_4 = f^{22}{}_4 = -f^{12}{}_2 = \frac{1}{\sqrt{2}} \,,
	}
	with the other components vanishing. Using \eqref{eq:frame_properties}, with $C_{AB}=\varepsilon_{AB}$, we find $h_{uv}=\delta_{uv}$ and the hypercomplex structures then turn out to be
	\es{HypercomplexExplicit}{
		J^1 = \begin{pmatrix}
			0 & 0 & 0 & 1 \\
			0 & 0 & -1 & 0 \\
			0 & 1 & 0 & 0 \\
			-1 & 0 & 0 & 0 
		\end{pmatrix} \,, \quad
		J^2 = \begin{pmatrix}
			0 & 0 & 1 & 0 \\
			0 & 0 & 0 & 1 \\
			-1 & 0 & 0 & 0 \\
			0 & -1 & 0 & 0 
		\end{pmatrix} \,, \quad
		J^3 = \begin{pmatrix}
			0 & -1 & 0 & 0 \\
			1 & 0 & 0 & 0 \\
			0 & 0 & 0 & 1 \\
			0 & 0 & -1 & 0 
		\end{pmatrix}  \,.
	}

One can check that these anti-symmetric matrices obey the relation $J^1 J^2 = J^3$, and its cyclic permutations.  With a vanishing $k_0^u$, we would have $k_{\text{SG}\, 0}^u = \frac{1}{2L} (q^2, -q^1, -q^4, q^3) + O(\kappa^2)$, which, by \eqref{U1R}, would imply that $q^1 - i q^2$ has R-charge $+1$ and $q^3 - i q^4$ has R-charge $-1$.  In order for the conditions in \eqref{Triholo} to be obeyed, one should choose a $k_0^u$ such that the matrix $Q$ with entries $Q_u{}^v = \partial_u k_0^v$ commutes with the three matrices in \eqref{HypercomplexExplicit}.  Since the three matrices \eqref{HypercomplexExplicit} represent the generators of one of the two $SU(2)$ subgroups of the $SO(4) \cong SU(2) \times SU(2)$ group that rotates the $q^u$'s, it follows that $Q$ must be one of the generators of the $SU(2)$ that commutes with \eqref{HypercomplexExplicit}.  Up to a redefinition of the $q^u$'s, such a generator can always be put into the form
\es{FormQ}{
	Q = \frac{r-1}{2L}  \begin{pmatrix}
		0 & -1 & 0 & 0 \\
		1 & 0 & 0 & 0 \\
		0 & 0 & 0 & -1 \\
		0 & 0 & 1 & 0
	\end{pmatrix} \,,
} 
for some proportionality constant $r$. Assuming that the gauge symmetry acts linearly on the $q^u$, this means that 
\es{Gotk0u}{
	k_0^u = \frac{r-1}{2L} ( q^2 , -q^1, q^4, -q^3) \,,
}
which implies $k_{*\, 0}^u = \frac{1}{2L} ( r q^2, -r q^1, (r-2) q^4, -(r-2) q^3) + O(\kappa^2)$.  From \eqref{U1R}, it follows that the R-charge of $q^1 - i q^2$ is $r$ and that of $q^3 - i q^4$ is $r-2$.  One can also determine $t_{0A}{}^B = \frac{r-1}{2L} \tau_{3A}{}^B$, which implies that the R-charge of $\zeta^1$ is $r-1$ while that of $\zeta^2$ is $-(r-1)$.  Consequently, $\zeta_2$ and $\zeta^1$ also have R-charges $r-1$ and $-(r-1)$, respectively.  One can then compute the moment map associated with \eqref{Gotk0u}:
\es{P0Gen}{
	\vec{P}_0 = -\frac{r-1}{2L} \left(
	q_1 q_3 + q_2 q_4, q_2 q_3 - q_1 q_4, \frac{q_3^2 + q_4^2 - q_1^2 - q_2^2}{2} \right)  \,.
} 

The formulas presented above\footnote{In particular, $S_{ij} = \begin{pmatrix}
		\frac{(r-1)(i q^1 + q^2) (q^3 + i q^4)}{2L} & \frac{i r [  (q^1)^2 + (q^2)^2] + i (2-r)[  (q^3)^2 + (q^4)^2] }{4 L} \\
		\frac{i r [  (q^1)^2 + (q^2)^2] + i (2-r)[  (q^3)^2 + (q^4)^2] }{4 L} &   \frac{(r-1)(-i q^1 + q^2) (q^3 - i q^4)}{2L}
	\end{pmatrix}$ and $N^i{}_A = \begin{pmatrix}
		\frac{r(q^2 - i q^1)}{2 \sqrt{2}L } & \frac{i (2-r) (q^3 - i q^4)}{2 \sqrt{2}L } \\
		\frac{i (r-2) (q^3 + i q^4)}{2 \sqrt{2}L } &  \frac{-ir  (q^1 - i q^2)}{2 \sqrt{2}L }
	\end{pmatrix}$.} determine the Lagrangian
\es{ActionGen}{
	e^{-1} {\cal L}  &=   - \frac 12 D_\mu q^u D^\mu q_u - \frac{r(r-3)}{2L^2} \left[  (q^1)^2 + (q^2)^2 \right] - \frac{(r+1)(r-2)}{2L^2} \left[  (q^3)^2 + (q^4)^2 \right] \\
	&{}+ \biggl\{ 
	- \bar \zeta_A \slashed{D} \zeta^A + 2  \frac{i r}{L} \bar \zeta^1 \zeta^2 + \text{h.c.} 
	\biggr\}  \,.
}
The spectrum of this theory is as follows.  In the bosonic sector, the masses of the two scalar fields $q^1$ and $q^2$ are thus $r(r-3)/L^2$, while the masses of $q^3$ and $q^4$ are $(r+1)(r-2)/L^2$.  In the fermion sector, the Dirac fermion $\zeta^1 + i \zeta_2$ has mass $r/L$, while $\zeta_1 - i \zeta^2$ has mass $-r/L$.

By the AdS/CFT dictionary, a scalar field of mass $m_B$ in AdS is dual to a boundary operator of dimension $\Delta_B$, where $m_B^2 L^2 = \Delta_B(\Delta_B - 3)$.\footnote{In the range $-\frac 94 < m_B^2 L^2 < -2$, two possible values of $\Delta_B$ are possible, namely $\Delta_{B+} = \frac 32 + \sqrt{ \frac{9}{4} + m^2}$ in the usual quantization or $\Delta_{B-} = \frac 32 - \sqrt{ \frac{9}{4} + m^2}$ in the alternate quantization.}  Similarly, a Majorana/Dirac fermion of mass $m_F$ is dual to a Majorana/Dirac fermionic operator in the 3d CFT of dimension $\Delta_F$, with $\Delta_F =  \abs{m_F} L  + \frac 32$.\footnote{If $0< \abs{m_F} L < \frac 12$, it is in principle also possible to also have $\Delta_F = - \abs{m_F} L  + \frac 32$, but this situation does not arise in supersymmetric theories.} In addition, the $U(1)_R$ charge in the bulk should match the $U(1)_R$ charge of the dual operators in the ${\cal N} = 2$ boundary SCFT\@.  

It follows that the bulk hypermultiplet described above is dual to an ${\cal N} = 2$ chiral multiplet on the SCFT side consisting of a scalar operator $\Phi$ of dimension $r$ and R-charge $r$ (dual to $q^1 - i q^2$), a fermionic operator $\Psi$ of dimension $r + 1/2$ and R-charge $r-1$ (dual to $ \zeta^1 + i \zeta_2 $), and another  scalar operator $\Phi'$ of dimension $r+1$ and R-charge $r-2$ (dual to $q^3 - i q^4$).  See Table~\ref{HyperDual} for a summary.
\begin{table}[htp]
	\begin{center}
		\begin{tabular}{c||c|c|c||c}
			SCFT operator & scaling dimension & spin & R-charge & bulk dual field \\
			\hline
			$\Phi$ & $r$ & $0$ & $r$ & $q^1 - i q^2$  \\
			\hline
			$\Psi$ & $r + \frac 12$ & $\frac 12$ & $r - 1$ & $\zeta^1 + i \zeta_2$  \\
			\hline
			$\Phi'$ & $r + 1$ & $0$ & $r - 2$ & $q^3 - i q^4$  
		\end{tabular}
	\end{center}
	\caption{Operator content of a boundary chiral multiplet and its bulk dual. Note that each of the operators/fields is complex.}
	\label{HyperDual}
\end{table}

\subsection{Action for massless vector multiplet}
\label{MASSLESSVECTOR}

Another application of the formalism above is to a single massless vector multiplet.  In top-down AdS/CFT constructions, a massless vector multiplet would appear if the dual ${\cal N} = 2$ SCFT has a flavor $U(1)$ symmetry.   Thus, we take $n_V = 1$ and $n_H = 0$ above.
Since there is only one vector multiplet, we drop the index ${\cal I}$ and denote  the components by ${\cal V} = (z, \bar z, \chi^i , \chi_i, A_\mu)$, with gauge field strength $F_{\mu\nu}$.   The quadratic action is obtained from the prepotential
\es{PrepotChoiceMassless}{
	F(z) = z^2 \,,
}
which gives ${\cal K} = \abs{z}^2$,  $g_{11} = 1$, and ${\cal N}_{11} = -\frac{i}{2}$.   In addition, one finds $S_{ij} = \frac{\abs{z}^2}{2L} \tau_{3\, ij}$, $W_{1}{}^{ij} = -\frac{\bar z}L \tau_{3}^{ij}$, and $N^i{}_A = 0$, from which one can derive the potential $V = -2 \abs{z}^2/L^2$.   The full Lagrangian is then
\es{LagMassless}{
	e^{-1} {\cal L}
	= -  \partial_\mu z \partial^\mu \bar z
	- \frac{1}{8}  F_{\mu\nu} F^{\mu\nu}   
	-\frac 14   \bar \chi_i \slashed{D} \chi^{i}
	-\frac 14   \bar \chi^i \slashed{D} \chi_{i}
	+ \frac{2 \abs{z}^2}{L^2}  \,.
}
It is convenient to define $z = (A + i B)/\sqrt{2}$, so that the Lagrangian is
\es{LagMasslessAgain}{
	e^{-1} {\cal L}
	= -  \frac 12 (\partial_\mu A)^2 - \frac 12 (\partial_\mu B)^2
	- \frac{1}{8}  F_{\mu\nu} F^{\mu\nu}   
	-\frac 14   \bar \chi_i \slashed{D} \chi^{i}
	-\frac 14   \bar \chi^i \slashed{D} \chi_{i}
	+ \frac{ A^2 + B^2}{L^2}  \,.
}
The fields $A$ and $B$ are scalar fields with mass $-2/L^2$, and they correspond to SCFT operators $J$ and $K$ of scaling dimension $1$ and $2$, respectively, that belong to the dual conserved current multiplet---see Table~\ref{ConservedSCFTTable} for a complete list of conformal primaries in the conserved current multiplet and their corresponding bulk dual fields.
\begin{table}[ht]
	\begin{center}
		\begin{tabular}{c||c|c|c||c}
			SCFT operator &  scaling dimension & spin & R-charge & bulk dual field\\
			\hline
			$J$  & $1$ & $0$ & $0$ & $A$\\
			\hline
			$\Xi$ & $\frac 32 $ &  $\frac 12$ & $1$ & $\chi^1 - i \chi_2$ \\
			$\tilde \Xi$ & $\frac 32 $ &  $\frac 12$ & $-1$ & $\chi_1 + i \chi^2$\\
			\hline
			$K$ & $2$ & $0$ & $0$ & $B$ \\
			$j_\mu$ & $2$ & $1$ & $0$ & $A_\mu$
		\end{tabular}
		\caption{The conformal primaries of an ${\cal N} = 2$ flavor current multiplet.\label{ConservedSCFTTable}}
	\end{center}
\end{table}%

One can consider interactions by including non-linear terms in the AdS prepotential $F(z)$.  As an example, if we require a $z\to -z$ symmetry, then the next allowed prepotential interaction is a quartic term:
\es{FInt}{
	F(z) = z^2 + \beta z^4 \,,
}
for some complex coupling constant $\beta$.  It is straightforward to compute the corrections in $\beta$ to all the quantities determining the action. Focusing only on the terms involving the scalar field $z$, the Lagrangian now has the following expansion at small $\abs{\beta}$:
\es{LagzInt}{
	e^{-1} {\cal L} &= -  \partial_\mu z \partial^\mu \bar z +  \frac{2 \abs{z}^2}{L^2} 
	+\left[  \frac{ \beta (z^4 + 4 z^3 \bar z )}{2L^2} - 3 \beta z^2 \partial_\mu z \partial^\mu \bar z + \text{h.c.} \right] 
	+ O(\abs{\beta}^2) \\
	&{}+ \text{other fields} \,.
}
As we will see in Section~\ref{INTEGRATED}, the term in the square bracket will give a non-vanishing contribution to the fourth derivative of the sphere free energy in the boundary SCFT\@.   In preparation for that computation, we note that it is possible to significantly simplify the form of the interaction term by adding a total derivative to the Lagrangian and using the equations of motion.  Indeed, if we add
\es{LTotalDer}{
	e^{-1} {\cal L}_\text{der} = \partial_\mu \left( \beta z^3 \partial^\mu  \bar z + \text{h.c.}  \right) \,,
}
and use the equations of motion to discard terms involving $\Box z$ and $\Box\bar z$, we can write ${\cal L}' = {\cal L} + {\cal L}_\text{der}$ as
\es{LagzIntAgain}{
	e^{-1} {\cal L}' &= -  \partial_\mu z \partial^\mu \bar z +  \frac{2 \abs{z}^2}{L^2}  
	+ \beta \frac{z^4}{2L^2} + \bar \beta \frac{\bar z^4}{2L^2} 
	+ O(\abs{\beta}^2) + \text{other fields} \,.
} 
We can easily write this expression in terms of the fields $A$ and $B$ that are dual to operators of well-defined scaling dimensions.

\subsection{Quadratic action for massive vector multiplet}
\label{MASSIVEVECTOR}

Let us now use the formalism introduced above to construct a massive vector multiplet, first at the quadratic level and then with  interactions.  Without supersymmetry, the quadratic Lagrangian for a massive vector field $B_\mu$ of mass $m_V$ is the Proca Lagrangian
\es{Proca}{
	e^{-1} {\cal L} = \frac 18 G_{\mu\nu} G^{\mu\nu} + \frac{m_V^2}4 B_\mu B^\mu \,,
}
where $G_{\mu\nu} = \partial_\mu B_\nu - \partial_\nu B_\mu$, and the unconventional normalization is chosen so that it matches the normalization in the supersymmetric case below.  The massive vector field $B_\mu$ can be written in terms of a (massless) gauge field $A_\mu$, with the usual gauge symmetry $A_\mu \to A_\mu + \partial_\mu \lambda$. This gauges the shift symmetry of a real scalar field $\phi$, which transforms as $\phi \to \phi + \frac{m_V}{\sqrt{2}} \lambda$.  The quadratic action consistent with this gauge symmetry is the Stueckelberg Lagrangian
\es{LagStuck}{
	e^{-1} {\cal L} = \frac 18 F_{\mu\nu} F^{\mu\nu} + \frac 12 \left( \partial_\mu \phi- \frac{m_V}{\sqrt{2}} A_\mu \right) \left( \partial^\mu \phi - \frac{m_V}{\sqrt{2}} A^\mu \right)  \,,
}
where $F_{\mu\nu} = \partial_\mu A_\nu - \partial_\nu A_\mu$ is the gauge field strength.  Denoting 
\es{BToA}{
	B_\mu = A_\mu - \frac{\sqrt{2}}{m_V} \partial_\mu \phi \,,
} 
and noticing that $G_{\mu\nu} = F_{\mu\nu}$, it is then easy to see that \eqref{LagStuck} is equivalent to \eqref{Proca}.   This analysis extends to interactions of $B_\mu$ with other fields, provided that in the $(A_\mu, \phi)$ description all such interactions are gauge invariant, so they depend on $A_\mu$ and $\phi$ only through the gauge-invariant combination $B_\mu = A_\mu - \frac{\sqrt{2}}{m_V} \partial_\mu \phi$.

A massive ${\cal N} = 2$ vector multiplet can be constructed by supersymmetrizing the Stueckelberg trick presented above.  In this case, one starts with one ${\cal N} = 2$ vector multiplet and one hypermultiplet.  One then  gauges a shift symmetry of one of the hypermultiplet scalars, as we now explain.  We take $n_V = n_H = 1$.  As in the case of the massless vector field of Section~\ref{MASSLESSVECTOR}, we will drop the index ${\cal I}$ and denote the vector multiplet fields simply by $(z, \bar z, \chi^i , \chi_i, A_\mu)$, with gauge field strength $F_{\mu\nu}$.  The hypermultiplet fields will be the real scalar fields $q_u$, with $u= 1, \ldots, 4$, and the fermions $\zeta^A$ and $\zeta_A$.  Note that $\chi_i$ and $\zeta^A$ are left-handed, while $\chi^i$ and $\zeta_A$ are right-handed. In order to obtain a quadratic action, we take again the quadratic prepotential
\es{PrepotChoice}{
	F(z) = z^2 \,,
}
which gives  $g_{11} = 1$ and ${\cal N}_{11} = -\frac{i}{2}$, so that the kinetic terms in the Lagrangian are
\es{LagKin}{
	e^{-1} {\cal L}_\text{kin} 
	= -  \partial_\mu z \partial^\mu \bar z
	- \frac {1}{2} D_\mu q_u D^\mu q_u
	- \frac{1}{8}  F_{\mu\nu} F^{\mu\nu}   
	-\frac 14   \bar \chi_i \slashed{D} \chi^{i}
	-\frac 14   \bar \chi^i \slashed{D} \chi_{i}
	- \bar \zeta_A \slashed{D} \zeta^A 
	- \bar \zeta^A \slashed{D} \zeta_A   \,.
}
The covariant derivative $D_\mu q_u$ must correspond to a gauged shift symmetry of one of the hypermultiplet scalars, say $q_4$. This will give us a  massive vector field via the Stuckelberg mechanism as in \eqref{Proca}--\eqref{BToA}, with $\phi = q_4$.  Thus we choose
\es{CovDer}{
	D_\mu q_u = \partial_\mu q_u - \delta_{u4} \frac{m_V}{\sqrt{2}} A_\mu \,.
}
This implies that the Killing vector corresponding to this gauging is
\es{KillingGauging}{
	k_1^u =  \frac{m_V}{\sqrt{2}} ( 0, 0, 0, 1) \,.
} 
Then, with the choice of hypercomplex structures in \eqref{HypercomplexExplicit}, the moment map that solves \eqref{MomentKillingFlat} is
\es{GotKillingShift}{
	\vec{P}_1 =\frac{m_V}{\sqrt{2}} (q_1, q_2, q_3) \,.
}
The consistency conditions \eqref{Triholo}--\eqref{ConditionsKilling} also require a non-zero $k_0^u$: 
\es{Gotk0}{
	k_0^u = \frac{1}{2L} (q_2, -q_1, q_4, -q_3) \,, \quad
	\vec{P}_0
	= -\frac{1}{2L} \left(
	q_1 q_3 + q_2 q_4, q_2 q_3 - q_1 q_4, \frac{q_3^2 + q_4^2 - q_1^2 - q_2^2}{2} \right)  \,.
}
These choices imply $t_{1A B} = 0$ and $t_{0A}{}^B = \frac{1}{2L} (\tau_3)_A{}^B$.  Then, the mass terms in the Lagrangian are 
\es{LagMass}{
	e^{-1} {\cal L}_\text{mass} 
	&= - \left( m_V^2 - \frac{2}{L^2} \right) \left(  \frac{q_1^2}{2} + \frac{q_2^2}{2} 
	+ \abs{z}^2 \right) - m_V^2 \frac{q_3^2}{2} + \frac{2 m_V }{L} \frac{z + \bar z}{\sqrt{2}} q_3 \\
	&{}+ 
	i m_V \bar \chi^{2} \zeta_1 -  i m_V \bar \chi^{1} \zeta_2 -  \frac{2i}{L} \bar \zeta_1 \zeta_2   
	-  i m_V \bar \chi_{2} \zeta^1 +  i m_V \bar \chi_{1} \zeta^2
	+  \frac{2i}{L} \bar \zeta^1 \zeta^2 \,.
}
The sum of \eqref{LagKin} and \eqref{LagMass} is the quadratic Lagrangian of a massive vector multiplet in AdS\@.  Notice that there is no potential for the field $q_4$, as required for a scalar with a shift symmetry.  

The next task is to rewrite the Lagrangian in terms of  physical fields of definite mass, obtained  by diagonalizing ${\cal L}_\text{mass}$ in \eqref{LagMass}.  For orientation, we first point out that a massive vector multiplet in AdS is dual in the dual SCFT to a generic long scalar superconformal multiplet  of single-trace operators.  Such a multiplet, ${\cal S}_\Delta$, starts with a scalar superconformal primary with scale dimension $\Delta$ and R-charge 0.  In addition it contains the following conformal primaries as superconformal descendents:  fermionic operators $\chi$ and $\tilde \chi$ of dimension $\Delta + 1/2$ and R-charges $\pm 1$;  three scalar operators $\phi$, $\tilde \phi$, and $\cO'$ of dimension $\Delta + 1$ and R-charges $0$ and $\pm 2$;  a vector operator $V_\mu$ of dimension $\Delta_V = \Delta + 1$ and R-charge $0$;  two more fermionic operators $\xi$ and $\tilde \xi$ of dimension $\Delta + 3/2$ and R-charges $\pm 1$;  and lastly, a scalar operators ${\cal O}''$ of scaling dimension $\Delta + 2$ and R-charge $0$.  This information is recorded in  Table~\ref{LongMultiplet}.  In a holographic setup, each of these operators would have its own bulk field. 
\begin{table}[htp]
	\begin{center}
		\begin{tabular}{c||c|c|c||c}
			SCFT operator & scaling dimension & spin & R-charge & bulk dual field  \\
			\hline
			$\cO$ & $\Delta$ & $0$ & $0$ & $C_1$ \\
			\hline
			$\chi_\alpha$ & $\Delta + \frac 12$ & $\frac 12$ & $1$  \\
			$\tilde \chi_\alpha$ & $\Delta + \frac 12$ & $\frac 12$ & $-1$  \\
			\hline
			$\phi$ & $\Delta + 1$ & $0$ & $2$ & $C_3 + i C_4$ \\
			$\tilde \phi$ & $\Delta + 1$ & $0$ & $-2$ & $C_3 - i C_4$ \\   
			$\cO'$ & $\Delta + 1$ & $0$ & $0$ & $C_2$  \\
			$V_\mu$ & $\Delta + 1$ & $1$ & $0$ & $B_\mu$ \\
			\hline
			$\xi_\alpha$ & $\Delta + \frac 32$ & $\frac 12$ & $1$ \\
			$\tilde \xi_\alpha$ & $\Delta + \frac 32$ & $\frac 12$ & $-1$ \\
			\hline
			$\cO''$ & $\Delta + 2$ & $0$ & $0$  & $C_5$
		\end{tabular}
	\end{center}
	\caption{Operator content of a generic long scalar multiplet.  Such a multiplet is dual to a massive vector multiplet.}
	\label{LongMultiplet}
\end{table}%

In particular, the vector operator $V_\mu$ of dimension $\Delta_V = \Delta + 1$ is dual to the massive vector $B_\mu$ of mass $m_V$, and the standard AdS/CFT dictionary identifies $m_V^2 = (\Delta_V - 1) (\Delta_V - 2) / L^2$.  In terms of the scaling dimension $\Delta$ of the superconformal primary, we have
\es{mVDeltaRel}{
	m_V^2 = \frac{\Delta (\Delta-1)}{L^2} \,.
}
In order to identify the scalar fields of definite mass, let us define
\es{DefBosonic}{
	C_1 &= \sqrt{\frac{\Delta}{2 \Delta - 1}} \frac{z + \bar z}{\sqrt{2}} + q_3 \sqrt{\frac{\Delta - 1}{2 \Delta - 1}} \,, \\
	C_2 &= \frac{z - \bar z}{\sqrt{2} i } \,, \\
	C_3 &= q_1 \,, \\
	C_4 &= q_2 \,, \\
	C_5 &= \sqrt{\frac{\Delta-1}{2 \Delta - 1}} \frac{z + \bar z}{\sqrt{2}} - q_3 \sqrt{\frac{\Delta}{2 \Delta - 1}} \,.
}
Writing the sum of \eqref{LagKin} and \eqref{LagMass} in terms of the new fields, we have 
\es{LagMassive}{
	e^{-1} {\cal L} 
	=  -\frac 18 G_{\mu\nu} G^{\mu\nu} - \frac {m_V^2}4 B_\mu B^\mu -   \frac 12\sum_{k=1}^5 \left[ 
	(\partial_\mu C_k)^2 + m_k^2 C_k^2\right] + \text{fermions} \,,
}
where the squared masses of the scalar fields are
\es{ScalarMasses}{
	m_1^2 = \frac{\Delta(\Delta - 3)}{L^2} \,, \quad m_2^2 = m_3^2 = m_4^2 = \frac{(\Delta + 1)(\Delta-2)}{L^2} \,, \quad
	m_5^2 = \frac{(\Delta + 2)(\Delta - 1)}{L^2} \,.
}
Thus, by the AdS CFT dictionary, the field $C_1$ is dual to a scalar operator of dimension $\Delta$ (the superconformal primary ${\cal O}$ of the multiplet), $C_2$, $C_3 + i C_4$, and $C_3 - i C_4$ are each dual to scalar conformal primary operators of dimension $\Delta +1$ (namely ${\cal O}'$, $\phi$, and $\tilde \phi$), and $C_5$ is dual to a scalar conformal primary of dimension $\Delta + 2$ (namely ${\cal O}''$).   Note that the choice of $k_0^u$ in \eqref{Gotk0} corresponds to $r=2$ in \eqref{Gotk0u}, and thus $q^1 + i q^2$ has R-charge $2$ while $q^3 + i q^4$ and the vector multiplet scalars have R-charge $0$.  From \eqref{DefBosonic}, it follows that $C_3 \pm i C_4$ have R-charge $\pm 2$ while all other scalar fields have vanishing R-charge.  See Table~\ref{LongMultiplet}. 

A similar analysis in the fermion sector reveals two fermions dual to fermionic operators of dimension $\Delta + 1/2$, and two fermions dual to fermionic operators of dimension $\Delta + 3/2$, as expected from the operator content of the long superconformal multiplet.  We will not describe the diagonalization in the fermion sector in more detail because it will not be needed in our analysis.

As an aside, let us point out that the Stuckelberg mechanism by which we constructed the long vector multiplet corresponds to multiplet recombination in the boundary SCFT\@.  From the SCFT perspective, we started off with a conserved current multiplet (dual to a massless vector multiplet) and a chiral multiplet of dimension $2$ (equal to the R-charge of the superconformal primary), and the two multiplets combined into a generic long multiplet whose dimension can then vary.  For the recombination to be possible, it is important that the chiral multiplet has dimension $2$ because only then does this multiplet contain a scalar operator of dimension $3$ and vanishing R-charge that can become the conformal descendant $\partial_\mu j^\mu$ after the multiplets recombine.  It is very nice that these requirements appear naturally in the supergravity construction.

\subsection{Interactions with massless vector multiplet}
\label{INTERACTIONS}

Let us now consider a massive vector multiplet interacting with a massless vector multiplet via a cubic prepotential.  We construct the massive vector multiplet as in the previous section from a massless vector multiplet $(z_1, \bar z_1, \chi^1_i, \chi^{1i}, A_\mu^1)$ coupled to a hypermultiplet $(q_u, \zeta^A, \zeta_A)$ by gauging the shift symmetry of the scalar field $q_4$.  The massless vector multiplet will be $(z_2, \bar z_2, \chi^2_i, \chi^{2 i}, A_\mu^2)$.\footnote{We choose to write the indices on the $z$ fields as lower in order to avoid confusion with powers of the fields.} Before considering the interactions, the prepotential is
\es{PrepotBefore}{
	F_2(z_1, z_2) = z_1^2 + z_2^2\,.
}
as well as the hypermultiplet gauging as in \eqref{GotKillingShift}--\eqref{Gotk0}.  The quadratic Lagrangian is simply a sum of \eqref{LagMassless} (with $(z, \bar z, \chi_i, \chi^{i}, A_\mu) \to (z_2, \bar z_2, \chi^2_i, \chi^{2i}, A_\mu^2)$) and \eqref{LagKin} and \eqref{LagMass} (with $(z, \bar z, \chi_i, \chi^{i}, A_\mu) \to (z_1, \bar z_1, \chi^1_i, \chi^{1i}, A_\mu^1)$).  Equivalently, it can be written as the sum of \eqref{LagMasslessAgain} and \eqref{LagMassive}.  Focusing only on the scalar fields, we have
\es{L2}{
	e^{-1} {\cal L}_2 &= 
	- \frac 12 (\partial_\mu A)^2 - \frac 12 (\partial_\mu B)^2  
	+ \frac{ A^2 + B^2}{L^2} 
	-  \sum_{k=1}^5 \left[ 
	\frac 12 (\partial_\mu C_k)^2 + m_k^2 C_k^2\right] + \text{other fields}  \,,
} 
where the masses are as in \eqref{ScalarMasses}, and where the ellipses denote terms involving the fermions and/or the vector fields.

We consider a cubic prepotential interaction such that the total prepotential is  
\es{F3}{
	F(z_1, z_2) = F_2(z_1, z_2) + \alpha z_1 z_2^2 \,,
}
for some complex coupling constant $\alpha$.  This prepotential generates interaction terms in the Lagrangian.  For scalar fields In particular, we have both cubic terms, specifically
\es{GotL3}{
	e^{-1} {\cal L}_3 &= 
	- \frac {\alpha}2 \left[ z_2  \partial_\mu z_2 \partial^\mu \bar z_1 + \partial_\mu (z_1 z_2) \partial^\mu \bar z_2 
	- \frac{\bar z_1 z_2^2 + 2 \abs{z_2}^2 z_1  
		+ \frac{1}{\sqrt{2}} m_V L q_3 z_2^2}{L^2}  \right] + \text{h.c.}\,,
}
and quartic terms 
\es{GotL4}{
	e^{-1} {\cal L}_4 &= - \frac{\alpha^2 z_2^2}{16 L^2} \left[ 2 m_V^2 L^2 (q_1^2 + q_2^2 + q_3^2) + 4 \sqrt{2} m_V L z_1 q_3 + 4 z_1^2 + z_2^2  \right]  + \text{h.c.} \\
	&{}-\frac{\abs{\alpha}^2 \abs{z_2}^2}{8 L^2}   \left[ 2 m_V^2 L^2 (q_1^2 + q_2^2 + q_3^2) 
	+ 2 \sqrt{2} m_V L q_3 (z_1 + \bar z_1) + 4 \abs{z_1}^2 + \abs{z_2}^2 \right] \,,
}
together with higher-order terms we do not need.

It is possible to simplify the expressions for the interaction vertices by adding a total derivative and using the equations of motion to eliminate the terms involving the Laplacians of the various fields.  Indeed, instead of ${\cal L} = {\cal L}_2 + {\cal L}_3 + {\cal L}_4 + \cdots$, we can consider
\es{Lp}{
	{\cal L}' = {\cal L} + {\cal L}_\text{der}
}
where in this case
\es{Lder}{
	{\cal L}_\text{der} = \partial^\mu \left[ 
	\frac{\alpha}{4} z_2^2 \partial_\mu \bar z_1
	+ \frac{\alpha}2 z_1 z_2 \partial_\mu \bar z_2 + \frac{\abs{\alpha}^2}{8} \abs{z_2}^2 \bar z_2 \partial_\mu z_2 + \text{h.c.}
	\right] \,.
}
Then, after using the equations of motion, we can expand ${\cal L}'$ as
\es{LpExpansion}{
	{\cal L}' = {\cal L}_2 + {\cal L}_3' + {\cal L}_4' + \cdots \,,
} 
with the new cubic and quartic vertices being
\es{CubicNew}{
	e^{-1} {\cal L}_3' = \frac{\alpha m_V^2}{4} z_2^2 \bar z_1 +  \frac{\bar \alpha m_V^2}{4} \bar z_2^2 z_1 \,,
}
\es{QuarticNew}{
	e^{-1} {\cal L}_4' = - \frac{\alpha^2 z_2^4 + \bar \alpha^2 \bar z_2^4}{16 L^2} 
	+ \frac{\abs{\alpha}^2}{8} \left[ 
	- \frac{3}{L^2} \abs{z_2}^4 + 4 \abs{z_2}^2 \partial_\mu z_2 \partial^\mu \bar z_2
	\right] + \text{terms involving $z_1$ and $q_u$} \,.
}
The vertices written explicitly in \eqref{CubicNew} and \eqref{QuarticNew} are the only ones that contribute at tree level to the fourth mass derivative of the $S^3$ free energy.

\section{One-loop free energy for free multiplets}
\label{ONELOOP}

In this section we restrict our attention to a free massive hypermultiplet, described in Section~\ref{MASSIVEHYPER}, as well as a free massive vector multiplet, described in Section~\ref{MASSIVEVECTOR}, and investigate whether the sphere free energy of the dual field theory, computed in the one-loop approximation, depends on the mass parameters.

Consider a free field in AdS$_4$ dual to an operator of spin $\ell$ and conformal dimension $\Delta$.  The contribution of such a field to the $S^3$ free energy comes solely from the determinant factor obtained after performing the Gaussian integral in the bulk.   This contribution is divergent and requires regularization.  After subtracting the power divergences (or equivalently, using zeta function regularization \cite{Hawking:1976ja}), one is still left with a logarithmic divergence as well as a finite contribution to the sphere free energy given by~\cite{Camporesi:1993mz,Liu_2016}
 \es{FFree}{
{\cal F}_{\Delta, \ell} = -\frac12H_\ell\left(\Dk-\frac32\right) - \frac12G_\ell\left(\Dk-\frac32\right)\log L^2\Lk^2\,,
 }
where $L$ is the AdS$_4$ radius, $\Lk$ is the UV cutoff. For real bosons, the functions $G_\ell(x)$ and $H_\ell(x)$ take the form
 \es{GotGH}{
G_\ell(x) &= \frac{2\ell+1}{24}\left[x^4 - \left(\ell+\frac12\right)^2\left(2x^2+\frac16\right)-\frac 7{240}\right]\,, \\
H_\ell(x) &= \frac{2\ell+1}{72}\biggl[ 3x^2(2\ell+1)^2 - 8x^4 + 24\zeta'\left(-3,x+\frac12\right) - 72 x \zeta'\left(-2,x+\frac12\right) \\
& \  - 6\left((2\ell+1)^2-12x^2\right)\zeta'\left(-1,x+\frac12\right) + 6x\left((2\ell+1)^2-4x^2\right)\zeta'\left(0,x+\frac12\right)\biggr]\,,
 }
where $\zeta'(a,b) = \frac{\nb \zeta(a,b)}{\nb a}$ is the derivative of the Hurwitz zeta function.  For Majorana fermions, the functions $G_\ell(x)$ and $H_\ell(x)$  take the form: 
\begin{equation}\begin{split}
G_\ell(x) &= -\frac{2\ell+1}{24}\left[x^4 -\left(\ell+\frac12\right)^2\left(2x^2-\frac13\right)+\frac 1{30}\right]\,, \\
H_\ell(x) &= \frac{2\ell+1}{72}\biggl[8x^4 - 3x^2(2\ell+1)^2 - 24\zeta'\left(-3,x\right) + 72 x \zeta'\left(-2,x\right) \\
& \  +6\left((2\ell+1)^2-12x^2\right)\zeta'\left(-1,x\right) - 6x\left((2\ell+1)^2-4x^2\right)\zeta'\left(0,x\right)\biggr]\,.
\end{split}\end{equation}

To compute the free-energy for a SUSY multiplet, we must simply sum over the fields of that multiplet.  For a hypermultiplet with mass\footnote{A massless hypermultiplet is such that the scalar operators are conformally coupled and the fermions are massless, thus being dual to scalar operators of dimensions $1$ and $2$ and to fermionic operators of dimension $3/2$.} $m_H \equiv r -1$, we can use Table~\ref{HyperDual} to write the sphere free energy as
 \es{FFreeHyper}{
  F_{\substack{\text{${\cal N} = 2$ hyper}}}(r)
   = 2 {\cal F}_{r, 0}  + 2 {\cal F}_{r + \frac 12, \frac 12} + 2 {\cal F}_{r + 1, 0} \,.
 }
Plugging in \eqref{FFree}--\eqref{GotGH}, we find 
 \es{FFreeHyperAgain}{
  F_{\substack{\text{${\cal N} = 2$ hyper}}}(r) &= - \left[ \frac{r(r-2)}{2}  + \frac{5}{12} \right] \log (\Lambda L) 
   +   (r-1) \zeta'( 0, r)  - \zeta'(-1, r)  \\
    &{}+ \frac{7r(r-2)}{12} + \frac{83}{144}  \,.
 }
Clearly, this expression is a non-trivial function of $r$, which shows that changing the mass $m_H = r-1$ of the hypermultiplet cannot be SUSY exact.

One can do a similar calculation for a massive vector multiplet.  From Table~\ref{LongMultiplet} we can read off the field content and identify the sphere free energy as
 \es{FFreeVector}{
  F_{\substack{\text{${\cal N} = 2$ vector}}}(\Dk) = {\cal F}_{\Dk, 0} + 2 {\cal F}_{\Dk + \frac 12, \frac 12} 
   + 3 {\cal F}_{\Dk+1, 0} + {\cal F}_{\Dk + 1, 1} + 2 {\cal F}_{\Dk + \frac 32, \frac 12} + {\cal F}_{\Dk+2, 0} \,.
 }
Summing these all together, we find that the free energy for a massive vector is UV-finite and independent of $\Dk$:
 \es{FFreeVectorAgain}{
F_{\text{${\cal N} = 2$ vector}}(\Dk) = \frac 5{24}\,.
 }
This supports our conjecture that changing the mass of the long vector multiplet is a SUSY exact deformation.
 
For comparison, let us discuss briefly two multiplets of ${\cal N} = 1$ supersymmetry.  First, a chiral multiplet of mass $m = \Delta - 1$ is dual to a scalar ${\cal N} = 1$ superconformal multiplet consisting of scalar conformal primaries operators of dimension $\Delta$ and $\Delta + 1$ as well as a fermionic operator of dimension $\Delta +1/2$.  The one-loop free energy is simply half that for a hypermultiplet, namely 
 \es{N1Chiral}{
  F_{\substack{\text{${\cal N} =1$ chiral}}}(\Delta) = \frac 12 F_{\substack{\text{${\cal N} =2$ hyper}}}(\Delta) \,.
 }  
Again, this is a non-trivial function of $\Delta$ and hence a non-trivial function of the hypermultiplet mass.  In this case, it is known that the chiral multiplet mass is a superpotential deformation, hence it is an ${\cal N} = 1$ F-term.  Based on this example, the $S^3$ partition function must depend in general on the ${\cal N} = 1$ F-term couplings.

Let us also discuss the case of an ${\cal N} = 1$ massive vector multiplet.  Such a multiplet is dual to an ${\cal N} = 1$ superconformal multiplet with a fermionic superconformal primary.  The conformal primaries in the multiplet are:  a fermionic operator of dimension $\Delta$, a scalar operator of dimension $\Delta + 1/2$;  a vector operator also of dimension $\Delta + 1/2$;  and another fermionic operator of dimension $\Delta + 1$.  The $S^3$ free energy is then
 \es{FN1Vector}{
  F_{\substack{\text{${\cal N} = 1$ vector}}}(\Delta) =  {\cal F}_{\Dk, \frac 12} +  {\cal F}_{\Dk + \frac 12, 0}
   + {\cal F}_{\Dk + \frac 12, 1} 
   +  {\cal F}_{\Dk+1, \frac 12} \,.
 }
This expression evaluates to  
 \es{FN1VectorAgain}{
  F_{\substack{\text{${\cal N} = 1$ vector}}}(\Delta) &=  \left[ \frac{\Delta(\Delta-2)}{2}  + \frac{13}{24} \right] \log (\Lambda L)
     - \frac{42 \Delta(\Delta - 2) + 37 }{72}\\
   &{}- (\Delta-1) \zeta'\left( 0, \Delta - \frac 32\right)  + \zeta'\left(-1, \Delta - \frac 32\right) 
    - \frac 12 \log \left( \Delta - \frac 32 \right)  \,.
 }
This expression is again a non-trivial function of $\Delta$ and hence of the vector multiplet mass.  In this case, however, it can be shown that the ${\cal N} = 1$ vector multiplet mass term is a D-term \cite{VanProeyen:1979ks}.  This computation therefore shows that the $S^3$ partition function cannot be independent of the ${\cal N}= 1$ D-term couplings.

\section{Integrated four-point correlators}
\label{INTEGRATED}

In this section we will compute directly certain holographic contributions to the fourth derivative $\frac{\partial^4 F}{\partial \mathfrak{m}^4} \big|_{\mathfrak{m}=0}$.\footnote{It follows from the work of \cite{Closset:2012vg} that the fourth derivative $d^4F/d\mathfrak{m}^4$ of the free energy is an unambiguous quantity.}  As explained in Section~\ref{REALMASSSECTION}, the real mass deformation of the boundary theory is $\mathfrak{m} \int d^3 \vec{x}\,\sqrt{g(\vec{x})} [i J(\vec{x}) + K(\vec{x})]$, where we set the radius of the sphere to one and parameterized the sphere with coordinates $\vec{x}$.  Thus, the fourth mass derivative is
 \es{FourthMass}{
  {\cal I} \equiv -\frac{\partial^4 F}{\partial \mathfrak{m}^4} \bigg|_{\mathfrak{m}=0}
   = \left \langle \left( \int d^3 \vec{x}\, \sqrt{g(\vec{x})}\,  [i J(\vec{x}) + K(\vec{x})] \right)^4 \right\rangle \,.
 }
Since the scaling dimension of $J$ is equal to $1$ and that of $K$ is equal to $2$, in order to evaluate \eqref{FourthMass} in a holographic setup, we should first calculate 
 \es{IntArbitrary}{
  \left \langle \prod_{i=1}^4  \int d^3 \vec{x}\, \sqrt{g(\vec{x})}\, \phi_i(\vec{x})  \right\rangle
 }
for scalar operators $\phi_i$ with scaling dimensions $\Delta_i = 1, 2$.  In Section~\ref{EXCHANGE}, we will evaluate the contribution to \eqref{IntArbitrary} coming from a tree-level scalar exchange, and in Section~\ref{CONTACT}, we will explain how to determine contributions coming from certain four-point contact diagrams.  Afterwards, in Section~\ref{FOURTH}, we will apply these results to evaluate the quantity \eqref{FourthMass} at tree-level that we will use in two examples.  In Section~\ref{PREPCONTACT}, we will evaluate the contribution to \eqref{FourthMass} from the prepotential four-point contact interaction described in Section~\ref{MASSLESSVECTOR} and show that it does not vanish.  This result is consistent with the prepotential being an F-term interaction.  Lastly, in Section~\ref{LONGVECTOR}, we determine the contribution to \eqref{FourthMass} from a long vector multiplet exchange and show that it vanishes exactly.  This fact supports our conjecture that the massive vector multiplet interactions are SUSY exact.

 \subsection{Exchange diagram contribution to integrated correlator}
\label{EXCHANGE}

Consider first the contribution to the separated-point correlator $\langle \phi_1(\vec{x}_1) \phi_2(\vec{x}_2) \phi_3(\vec{x}_3) \phi_4(\vec{x}_4) \rangle$ of operators $\phi_i$ of dimension $\Delta_i$ coming from the $s$-channel exchange of an operator of dimension $\Delta$.  Let us consider this correlator first in flat space, and pass to the correlator on $S^3$ later.  Similarly, let us consider (Euclidean) $\text{AdS}_4$ parameterized in Poincar\'e coordinates
 \es{Poincare}{
  ds^2 = \frac{dy_0^2 + d\vec{y}^2}{y_0^2}  \,.
 }
For unit three point couplings, the exchange Witten diagram is
 \es{ExchangeWitten}{
  E_{\De_1 \De_2 \De_3 \De_4}^\De(\vec{x}_i) \equiv \int \frac{d^4y\, d^4 w}{y_0^4 w_0^4} \,  G^{\Delta_1}_{B\partial}(y; \vec{x}_1) G_{B\partial}^{\Delta_2}(y; \vec{x}_2)
   G_{B\partial}^{\Delta_3}(w; \vec{x}_3) G_{B\partial}^{\Delta_4}(w; \vec{x}_4) G_{BB}^\Delta(y; w) \,,
 }
where $y = (y_0, \vec{y})$ and $w = (w_0, \vec{w})$ are the two bulk points and $\vec{x}_i$ are the four boundary points.  Let us choose the normalization of the bulk-to-boundary propagator such that
 \es{GBulkBdry}{
  G_{B\partial}^\Delta(y; \vec{x}) = \Gamma(\Delta) \left( \frac{y_0}{y_0^2 + (\vec{y} - \vec{x})^2} \right)^\Delta \,.
 }
 In this normalization, the split representation of the bulk-bulk propagator is
  \es{GBulkBulkSplit}{
   G_{BB}^\Delta(y; w)
    = \frac{1}{2 \pi^3} \int_{-i \infty}^{i \infty}  \frac{d\lambda}{2 \pi i} \int d^3 \vec{z} \, \frac{1}{\left( \Delta - \frac 32 \right)^2 - \lambda^2} \frac{1}{\Gamma(\lambda) \Gamma(-\lambda)} G_{B \partial}^{\frac 32 + \lambda} (y; \vec{z}) 
     G_{B \partial}^{\frac 32 - \lambda} (w; \vec{z})  \,,
  } 
where the integration contour is the imaginary axis provided that $\Delta > 3/2$.  In the following discussion, we will assume $\Delta > 3/2$ and analytically continue the results to the alternate-quantization range $\Delta < 3/2$ afterwards.  Note that the scalar bulk-bulk propagator obeys the equation
  \es{BulkBulk}{
  (\nabla^2 - m^2 ) G_{BB}(y, w) = -\frac{1}{\sqrt{g(y)}} \delta^{(4)}(y - w)  \,, \qquad
   m^2 = \Delta(\Delta - 3) \,.
 }
Also, in the normalization \eqref{GBulkBdry} of the bulk-boundary propagator, if we compute correlators of the $\phi_i$ operators as in \eqref{ExchangeWitten} without any additional normalization factors, then the $\phi_i$ are normalized as given by
 \es{NormTwoPoint}{
  \langle \phi_i(\vec{x}) \phi_i(0) \rangle = \frac{\cN_{\Delta_i}}{\abs{\vec{x}}^{2 \Delta_i}} \,, \qquad
    \cN_{\Delta_i} \equiv 2^{3 - 2 \Delta_i} \pi^2 \Gamma(2\Delta_i - 1) \,.
 } 
This normalization factor will be important when applying these results to the correlators of $J$ and $K$, which are normalized as in \eqref{Normalizations}. 

After plugging \eqref{GBulkBulkSplit} into \eqref{ExchangeWitten}, we can perform the integrals over $y$ and $w$ using
 \es{ThreePoint}{
  \int \frac{d^4 y}{y_0^4} \, G_{B\partial}^{\Delta_1} (y; \vec{x}_1) 
   G_{B\partial}^{\Delta_2} (y; \vec{x}_2) G_{B\partial}^{\Delta_3} (y; \vec{x}_3) 
    = \frac{c_{\De_1, \De_2, \De_3}}{\abs{\vec{x}_{12} }^{\De_{123}}
     \abs{\vec{x}_{23} }^{\De_{231}} \abs{\vec{x}_{31} }^{\De_{312}} } 
    }
where we defined $\De_{ijk} \equiv \De_i + \De_j - \De_k$, and where
 \es{Gotc}{
  c_{\De_1,  \De_2,  \De_3} \equiv \frac{\pi^{3/2}}{2} \Gamma \left( \frac{\De_1 + \De_2 + \De_3 - 3}{2} \right) 
   \Gamma(\Delta_{123}/2) \Gamma(\Delta_{231}/2) \Gamma(\Delta_{312}/2)  \,.
 }
After renaming $\vec{z} \to \vec{x}_5$, we find
 \es{ExchangeAgain}{
  E_{\De_1 \De_2 \De_3 \De_4}^\De(\vec{x}_i) &= \frac{1}{2 \pi^3} \int_{-i \infty}^{i \infty}  \frac{d\lambda}{2 \pi i} \int d^3 \vec{x}_5 \, \frac{1}{\left( \Delta - \frac 32 \right)^2 - \lambda^2} \frac{c_{\De_1, \De_2, \frac 32 + \lambda} c_{\De_3, \De_4, \frac 32 - \lambda} }{\Gamma(\lambda) \Gamma(-\lambda)} \\
   &\times \frac{1}{\abs{\vec{x}_{12} }^{\De_{125}}
     \abs{\vec{x}_{25} }^{\De_{251}} \abs{\vec{x}_{51} }^{\De_{512}}
      \abs{\vec{x}_{34} }^{\De_{346}}
     \abs{\vec{x}_{45} }^{\De_{463}} \abs{\vec{x}_{53} }^{\De_{634}} } 
 }
where $\De_5 \equiv \frac 32 + \lambda$ and $\De_6 \equiv \frac 32 - \lambda$.

So far, we have worked in flat space in the boundary CFT\@.  To map the correlator to $S^3$, we use the stereographic map.  In stereographic coordinates, the metric on $S^3$ is 
 \es{dsS3}{
  ds^2 = \Omega(\vec{x})^2 d\vec{x}^2 \,, \qquad \Omega(\vec{x}) \equiv \frac{1}{1 + \frac{\vec{x}^2}{4}} \,.
 }
The general rule for going from $\R^3$ to $S^3$ is that the distance between two points $\abs{\vec{x}_{ij}}$ is replaced by the chordal distance $s_{ij} = \Omega(\vec{x}_i)^{1/2} \Omega(\vec{x}_j)^{1/2} \abs{\vec{x}_{ij}}$.  Thus
 \es{ES3}{
  E_{S^3}(\vec{x}_i) = E(\vec{x}_i) \prod_{i=1}^4 \Omega(\vec{x}_i)^{-\Delta_i} 
 }
The integrated correlator we want to compute is
 \es{IntE}{
  I_{\De_1 \De_2 \De_3 \De_4}(\De) = \int \prod_{i=1}^4 \left( d^3\vec{x}_i \,  \sqrt{g(\vec{x}_i)}   \right) E_{S^3}(\vec{x}_i) \,, \qquad
   \sqrt{g(\vec{x})} \equiv \Omega(\vec{x})^3 \,,
 }
where $d^3\vec{x} \, \sqrt{g(\vec{x})}$ is the volume element on $S^3$.  Quite nicely, the integrated correlator becomes
 \es{IntEExplicit}{
  I_{\De_1 \De_2 \De_3 \De_4}(\De) &=  \frac{1}{2 \pi^3} \int_{-i \infty}^{i \infty}  \frac{d\lambda}{2 \pi i} \int \prod_{i=1}^5 \left( d^3 \vec{x}_i \, \sqrt{g(\vec{x}_i)}  \right)  \, \frac{1}{\left( \Delta - \frac 32 \right)^2 - \lambda^2} \frac{c_{\De_1, \De_2, \frac 32 + \lambda} c_{\De_3, \De_4, \frac 32 - \lambda} }{\Gamma(\lambda) \Gamma(-\lambda)} \\
   &\times \frac{1}{s_{12}^{\De_{125}}
     s_{25}^{\De_{251}} s_{51}^{\De_{512}}
      s_{34}^{\De_{346}}
     s_{45}^{\De_{463}} s_{53}^{\De_{634}} } \,.
 }

We can easily do all the $\vec{x}_i$ integrals in \eqref{IntEExplicit}.  Indeed, because of rotational symmetry, we can set $\vec{x}_5$ to any value we want (everywhere except for the measure) and multiply the answer by the volume of $S^3$, $\Vol(S^3) = 2 \pi^2$.   The simplest choice is $\abs{\vec{x}_5} \to \infty$, in which case
 \es{s5Expressions}{
  \frac{1}{ s_{12}^{\De_{125}} s_{25}^{\Delta_{251}} s_{51}^{\Delta_{512}}
    s_{34}^{\De_{346}} s_{45}^{\Delta_{453}} s_{53}^{\Delta_{634}}} 
     \to \frac{\prod_{i=1}^4 \Omega(\vec{x}_i)^{-\Delta_i} }{\abs{\vec{x}_{12}}^{\De_{125}} 
      \abs{\vec{x}_{34}}^{\De_{346}}  4^{\Delta_5 + \Delta_6}} = \frac{1}{64} 
       \frac{\prod_{i=1}^4 \Omega(\vec{x}_i)^{-\Delta_i} }{\abs{\vec{x}_{12}}^{\De_{125}} 
      \abs{\vec{x}_{34}}^{\De_{346}}}\,.
 }
Thus,
 \es{IntEExplicit2}{
   I_{\De_1 \De_2 \De_3 \De_4}(\De) &=  \frac{1}{2 \pi^3} \frac{2\pi^2}{64} \int_{-i \infty}^{i \infty}  \frac{d\lambda}{2 \pi i} \int \frac{\prod_{i=1}^4 d^3 \vec{x}_i \, \Omega(\vec{x}_i)^{3-\Delta_i} }{\abs{\vec{x}_{12}}^{\De_{125}} 
      \abs{\vec{x}_{34}}^{\De_{346}}}
        \, \frac{1}{\left( \Delta - \frac 32 \right)^2 - \lambda^2} \frac{c_{\De_1, \De_2, \frac 32 + \lambda} c_{\De_3, \De_4, \frac 32 - \lambda} }{\Gamma(\lambda) \Gamma(-\lambda)}  \,.
 }
The integrals in $\vec{x}_{1, 2}$ and $\vec{x}_{3, 4}$ factorize.  The first one is
 \es{intx12}{
  \int d^3\vec{x}_1 \, d^3 \vec{x}_2 \frac{\Omega(\vec{x}_1)^{3 - \Delta_1} \Omega(\vec{x}_2)^{3 - \Delta_2} }{\abs{\vec{x}_{12}}^{\De_{125}}}
   = \frac{64}{2^{\De_{125}}} \int 
    \frac{d^3\vec{x}_1 \, d^3 \vec{x}_2}{(1 + \abs{\vec{x}_1}^2)^{3 - \Delta_1} (1 + \abs{\vec{x}_2}^2)^{3 - \Delta_2}  
    \abs{\vec{x}_{12}}^{\De_{125}} }
 }
Integrals of this type were evaluated in \cite{Fei:2015oha}.  In their notation, the integral in \eqref{intx12} (without the prefactor) equals $\Gamma_0(3 - \De_1, 3 - \De_2, \De_{125}/2)$.  In general, the resulting expression for \eqref{IntEExplicit2} is pretty complicated, but it simplifies if $\Delta_i \in \{1, 2\}$, where it takes the form
 \es{IFinal}{
  I_{\De_1 \De_2 \De_3 \De_4}(\De) = \frac{I(\Delta)}{\prod_i \Delta_i} \qquad 
   I(\Delta) = \int \frac{d \lambda}{2 \pi i} \, 
   8 \pi^{13} \frac{\lambda (4 \lambda^2 - 1) \sin(\pi \lambda)}{\left[(\Delta - 3/2)^2 - \lambda^2 \right] \cos^3 (\pi \lambda) } \,. 
 }
This integral can be performed by contour integration assuming $\Delta > 3/2$.  The result is
 \es{IExplicit}{
   I(\De) \equiv 16 \pi^{10} \biggl[1 
   + (\De-1)(\De-2) \psi^{(2)} (\De-1)   \biggr] \,,
 }
where $\psi^{(2)}$ denotes a polygamma function defined as the third derivative of the logarithm of the gamma function.  (In general, $\psi^{(m)}(z) = \frac{d^{m+1}}{dz^{m+1}} \log \Gamma(z)$.) While we derived the expression \eqref{IExplicit} for $\Delta > 3/2$, it also holds in the range $1 < \Delta < 3/2$ by analytic continuation.

 \subsection{Contact diagram contributions to integrated correlator}
 \label{CONTACT}
 
 Let us now move on to the contribution to the integrated  $\langle \phi_1(\vec{x}_1) \phi_2(\vec{x}_2) \phi_3(\vec{x}_3) \phi_4(\vec{x}_4) \rangle_{S^3}$ correlator from contact Witten diagrams.   We do not have to do any new computations, because the contact Witten diagrams relevant for our applications below can be obtained from the $m\to \infty$ limit of the $s$-channel scalar exchange that we evaluated in the previous section.  Indeed, the equation \eqref{BulkBulk} for the bulk-bulk propagator implies that, at large $m$, we have the expansion
 \es{LargemExpansion}{
  G_{BB}(y, w) = \left( \frac{1}{m^2} + \frac{1}{m^4} \nabla^2 + O(m^{-6}) \right) \delta^{(4)}(y - w) \,,
 }
where the Laplacian acts on the point $y$. This equation implies that
 \es{ExchangeLimit}{
  E_{\De_1 \De_2 \De_3 \De_4}^\De
   = \frac{1}{m^2} C^{(0)}_{\De_1 \De_2 \De_3 \De_4}  + \frac{1}{m^4} C^{(2)}_{\De_1 \De_2 \De_3 \De_4} + O(m^{-6}) \,,
 }
where $E_{\De_1 \De_2 \De_3 \De_4}^\De$ was defined in \eqref{ExchangeWitten} and $C^{(2n)}$ are contact Witten diagrams with $2n$ derivatives:
 \es{ContactExplicit}{
  C_{\De_1 \De_2 \De_3 \De_4}^{(0)}(\vec{x}_i) &= \int \frac{d^4y}{y_0^4} \,  G^{\Delta_1}_{B\partial}(y; \vec{x}_1) G_{B\partial}^{\Delta_2}(y; \vec{x}_2)
   G_{B\partial}^{\Delta_3}(y; \vec{x}_3) G_{B\partial}^{\Delta_4}(y; \vec{x}_4)  \,, \\
   C_{\De_1 \De_2 \De_3 \De_4}^{(2)}(\vec{x}_i) &= \int \frac{d^4y}{y_0^4} \,  \nabla^2 \left[ G^{\Delta_1}_{B\partial}(y; \vec{x}_1) G_{B\partial}^{\Delta_2}(y; \vec{x}_2) \right] 
   G_{B\partial}^{\Delta_3}(y; \vec{x}_3) G_{B\partial}^{\Delta_4}(y; \vec{x}_4)  \,, 
 } 
 and so on.  Let us denote the contribution of the diagram $C_{\De_1 \De_2 \De_3 \De_4}^{(2n)}(\vec{x}_i)$ to the integrated sphere correlator by $J^{(2n)}_{\De_1 \De_2 \De_3 \De_4}$.  By considering the large $m$ expansion of \eqref{IFinal} with $\Delta = (3 +\sqrt{9+4m^2})/2$, namely
  \es{IExchangeExpanded}{
  I(\Delta) = 8 \pi^{10} \left[ \frac{1}{m^2} - \frac{8}{3 m^4} + \frac{22}{3 m^6} - \frac{316}{15 m^8} + O(m^{-10})  \right] \,,
 }
 we find that for $\Delta_i = 1,2$, 
  \es{J2n}{
   J^{(2n)}_{\De_1 \De_2 \De_3 \De_4} = \frac{J^{(2n)}}{\De_1 \De_2 \De_3 \De_4}
    \qquad \text{with} \quad J^{(0)} = 8 \pi^{10} \,, \qquad J^{(2)} = -\frac{64 \pi^{10}}{3}  \,.
  }
Note that because the bulk-boundary propagator $G_{B\partial}^{\Delta_i}$ obeys the equation $\nabla^2 G_{B\partial}^{\Delta_i}  = \Delta_i(\Delta_i-3) G_{B\partial}^{\Delta_i}$, and for $\Delta_i = 1,2$, we have $\Delta_i(\Delta_i-3) = -2$, we can expand the Laplacian in the second line of \eqref{ContactExplicit} and find that the contact diagram
 \es{tildeC}{
  \tilde C_{\De_1 \De_2 \De_3 \De_4}^{(2)}(\vec{x}_i) &= \int \frac{d^4y}{y_0^4} \,  \partial_\mu G^{\Delta_1}_{B\partial}(y; \vec{x}_1) \partial^\mu G_{B\partial}^{\Delta_2}(y; \vec{x}_2) 
   G_{B\partial}^{\Delta_3}(y; \vec{x}_3) G_{B\partial}^{\Delta_4}(y; \vec{x}_4)
  }
contributes to the integrated correlator an amount equal to
 \es{JTilde}{
  \tilde J^{(2)}_{\De_1 \De_2 \De_3 \De_4}  = \frac{J^{(2)}}{\De_1 \De_2 \De_3 \De_4} \,, \qquad
   \tilde J^{(2)}  = \frac{J^{(2)} + 4 J^{(0)}}{2} = \frac{16 \pi^{10}}{3} \,.
 }

 \subsection{Fourth derivative of the sphere free energy}
 \label{FOURTH}
 
 To go from the normalization in \eqref{NormTwoPoint} to the normalization in \eqref{Normalizations}, we should proceed as follows.  For every $\phi_i = J$, we take $\Delta_i = 1$ and multiply the correlator by 
  \es{GotcalN}{
   {\cal N} \equiv \frac{\sqrt{\tau}}{4 \sqrt{2} \pi^2} \,.
  }  
Similarly, for every $\phi_i = K$, we take $\Delta_i = 2$ and multiply the correlator by $\frac{\sqrt{\tau}}{2 \sqrt{2} \pi^2} =2 {\cal N}$.  It follows that in order to calculate the integrated four-point correlator of any combination of $J$ and $K$ operators, we should multiply the expressions \eqref{IFinal} and \eqref{J2n} by ${\cal N}^4 \Dk_1 \Dk_2 \Dk_3 \Dk_4$ as well as by the interaction vertices.  Quite nicely, the product $\Dk_1 \Dk_2 \Dk_3 \Dk_4$ cancels the denominators of the expressions \eqref{IFinal} and \eqref{J2n}, so we simply have to multiply $I(\Delta)$ or $J^{(2n)}$ by ${\cal N}^4$ and the interaction vertex coefficients if we want to find the integrated four-point correlator of any combination of four $J$ and $K$ operators.  

In particular, let us consider a real field $C$ dual to an operator of dimension $\Delta$ with leading order kinetic Lagrangian $-\frac 12 (\partial_\mu C)^2 - \frac{\Delta(\Delta-3)}{2L^2} C^2$.  Let us assume this field interacts with the fields $A$ and $B$ that are the bulk fields dual, respectively, to the operators $J$ and $K$ in \eqref{FourthMass}, through the three-point contact vertices\footnote{These interaction vertices, as well as those in \eqref{Vertex4}, below are written in the Lorentzian-signature Lagrangian in mostly plus signature.}
  \es{Vertex3}{
   e^{-1} {\cal L}_\text{3-pt} = \frac 12 \lambda_{AAC} A^2 C + \lambda_{ABC} A B C + \frac 12 \lambda_{BBC} B^2 C \,.
  }
Then, the $s$-channel exchange contribution to the integrated correlator ${\cal I}$ in \eqref{FourthMass} equals
 \es{SChannel}{
  {\cal I} \Big|_{\substack{\text{$s$-channel} \\ \text{$C$ exch}}}
   = {\cal N}^4 \left( \lambda_{AAC}^2 - 4 \lambda_{ABC}^2 - 2 \lambda_{AAC} \lambda_{BBC} + \lambda_{BBC}^2
    - 4 i \lambda_{AAC}\lambda_{ABC} + 4 i \lambda_{BBC}\lambda_{ABC}  \right) I(\Delta) 
 }
where the $ \lambda_{AAC}^2$ term comes from the $\langle JJJJ \rangle$ part of \eqref{FourthMass}, the $\lambda_{AAC}\lambda_{ABC}$ term comes from $\langle JJJK \rangle$, etc.  One can recognize the quantity in the brackets of \eqref{SChannel} as the perfect square $(\lambda_{AAC} - 2 i \lambda_{ABC} - \lambda_{BBC})^2$.  In fact, there is a further simplification that is manifest when we write the vertex \eqref{Vertex3} in terms of the complex field $z = (A + i B)/\sqrt{2}$ and its conjugate, 
 \es{Vertex3Again}{
  e^{-1} {\cal L}_\text{3-pt} = \frac 12 \lambda_{zzC} z^2 C + \lambda_{z \bar z C} \abs{z}^2 C + \frac 12 \lambda_{\bar z \bar z C} \bar z^2 C \,.
 }
Then $\lambda_{zzC} = \frac 12 (\lambda_{AAC} - 2 i \lambda_{ABC} - \lambda_{BBC})$ and so only the square of the holomorphic vertex coefficient $\lambda_{zzC}$ contributes to \eqref{SChannel}.  The total exchange contribution is then
  \es{FullExchange}{
  {\cal I} \Big|_{\substack{\text{$C$ exchange}}}
   = 12\,  {\cal N}^4 \lambda_{zzC}^2  I(\Delta)  \,,
 }
where we multiplied \eqref{SChannel} by a factor of $3$ in order to account for the $t$-channel and $u$-channel exchanges as well.

A similar analysis can be done for contact four-point interactions.  The most general interaction with up to two derivatives takes the form
 \es{Vertex4}{
  e^{-1} {\cal L}_\text{4-pt}
   &= \frac{1}{4!} \lambda_{zzzz} z^4 + \frac{1}{3!} \lambda_{zzz\bar z} z^2 \abs{z}^2 +\frac{1}{4} \lambda_{zz\bar z \bar z} \abs{z}^4 +  \frac{1}{3!} \lambda_{z\bar z\bar z\bar z} \bar z^2 \abs{z}^2 + \frac{1}{4!} \lambda_{\bar z\bar z\bar z\bar z} \bar z^4 \\
    &{}+ \frac 12 \left[ \frac 12 \lambda_{zz,zz}'  z^2 + \lambda_{z\bar z,zz}'  \abs{z}^2
     + \frac 12 \lambda_{\bar z \bar z,zz}'  z^2 \right]  (\partial_\mu z)^2 \\
    &{}+ \left[ \frac 12 \lambda_{zz,z\bar z}'  z^2 +  \lambda_{z\bar z,z\bar z}'  \abs{z}^2
     + \frac 12 \lambda_{\bar z \bar z,z\bar z}'  z^2 \right]  \abs{\partial_\mu z}^2  \\
    &{}+\frac 12 \left[ \frac 12 \lambda_{zz,\bar z\bar z}'  z^2 + \lambda_{z\bar z,\bar z\bar z}'  \abs{z}^2
     + \frac 12 \lambda_{\bar z \bar z,\bar z\bar z}'  z^2 \right]  (\partial_\mu \bar z)^2   \,.
 }
As in the case of the exchange diagram, only the holomorphic vertices contribute to the fourth derivative of the sphere free energy:
 \es{FullContact}{
  {\cal I} \Big|_{\substack{\text{contact}}} 
   = 4 {\cal N}^4 \lambda_{zzzz} J^{(0)} +  24 {\cal N}^4 \lambda_{zz,zz}' \tilde J^{(2)}  \,.
 }

In order to apply these results to a specific situation, we need to compare the interaction vertices in the setup of interest to \eqref{Vertex3Again} and \eqref{Vertex4},  and then sum together \eqref{FullExchange} and \eqref{FullContact}.

 \subsection{Application:  Quartic prepotential interaction}
 \label{PREPCONTACT}
 
 As a first application, let us examine the case of the quartic prepotential interaction in Section~\ref{MASSLESSVECTOR}.  In this case, from comparing \eqref{LagzIntAgain} to \eqref{Vertex4} we see that, after setting $L=1$, we can identify $\lambda_{zzzz} = 12 \beta $ and $\lambda'_{zz, zz} = 0$.  Thus, in this case
  \es{IContribPropot}{
  {\cal I} \Big|_{\substack{\text{contact}}} = 48 {\cal N}^4  J^{(0)} \beta = \frac{3 \pi^{2} }{8} \tau^2 \beta \,.
  } 
 The fact that this expression does not vanish shows that this prepotential interaction is not SUSY exact.  This is consistent with it being an F-term.

 \subsection{Application:  Long vector multiplet exchange}
 \label{LONGVECTOR}
 
 We can now analyze the more complicated case of a long vector multiplet exchange.  In this case, the relevant interaction vertices are in \eqref{CubicNew} for the 3-point ones and in \eqref{QuarticNew} for the 4-point ones.  In those expressions, $z_2$ is the vector multiplet scalar playing the role of $z$ in \eqref{Vertex3Again} and \eqref{Vertex4}, and $z_1$ is a scalar from the massive vector multiplet.  However $z_1$ is not a field of definite mass.  Expressing it in terms of the definite-mass fields $C_i$ using \eqref{DefBosonic} (with $z \to z_1$), we have\footnote{The scalar fields $C_3, C_4$ do not contribute to \eqref{L3Again} because they carry R-charge. (See Table~\ref{LongMultiplet}.)}
  \es{L3Again}{
   e^{-1} {\cal L}_3' = \frac{\alpha \Delta(\Delta-1)}{4 \sqrt{2} L^2} z_2^2 \left(-i C_2 
    + \frac{C_1 \sqrt{\Delta} + C_5 \sqrt{\Delta - 1} }{\sqrt{2 \Delta - 1}} \right)  +  \text{h.c.} \,.
  }
 Comparing with \eqref{Vertex3Again}, we extract the interaction vertices (with $L=1$)
  \es{ExtractedVertices3}{
   \lambda_{zzC_1} &= \frac{\alpha \Delta(\Delta-1)}{2 \sqrt{2}} \sqrt{\frac{\Delta}{2 \Delta - 1}} \,, \\
   \lambda_{zzC_2} &=-i  \frac{\alpha \Delta(\Delta-1)}{2 \sqrt{2}}  \,, \\
   \lambda_{zzC_5} &= \frac{\alpha \Delta(\Delta-1)}{2 \sqrt{2}} \sqrt{\frac{\Delta-1}{2 \Delta - 1}} \,,
  }
and from comparing \eqref{QuarticNew} with \eqref{Vertex4}, we have
 \es{ExtractedVertices4}{
  \lambda_{zzzz} = - \frac{3\alpha^2}{2} \,, \qquad \lambda_{zz, zz}' = 0  \,.
 }
Plugging these expressions into \eqref{FullExchange} and \eqref{FullContact} and using the fact that the scaling dimensions of the operators dual to $C_1$, $C_2$, and $C_5$ are $\Delta$, $\Delta + 1$, and $\Delta + 2$, respectively, we find the final answer
 \es{LongExchange}{
   {\cal I} \Big|_{\substack{\text{long exch}}} 
    &=   \frac{3\alpha^2 \Delta^2(\Delta-1)^2  {\cal N}^4}{2}  
     \left( \frac{\Delta}{2 \Delta-1} I(\Delta) 
      - I(\Delta + 1) + \frac{\Delta - 1}{2 \Delta - 1} I(\Delta + 2) \right) \\
     &{}- 6 \alpha^2   {\cal N}^4 J^{(0)} \,. 
 } 
Using the expression for $I(\Delta)$ in \eqref{IExplicit} as well as the identity $\psi^{(2)}(x+1) - \psi^{(2)}(x) = 2/x^3$ we finally find
 \es{LongExchangeAgain}{
   {\cal I} \Big|_{\substack{\text{long exch}}}  = 0 \,,
 }
in support of our conjecture that the interactions between a massless vector and massive vector multiplet are SUSY exact.

\section{Discussion}
\label{DISCUSSION}

In this paper, we studied ${\cal N} = 2$-preserving supersymmetric interactions in AdS and how they affect the (mass-deformed) $S^3$ free energy in the boundary dual SCFT\@.  In particular, in Section~\ref{SUSY} we showed that certain ${\cal N} = 2$ D-terms, non-chiral F-terms, and $1/4$-BPS interactions are SUSY exact, and they do not affect the $S^3$ free energy.  By contrast, chiral F-terms and flavor current terms in the bulk do affect the $S^3$ free energy.  Later on, in Sections~\ref{ONELOOP} and~\ref{INTEGRATED}, we provided evidence that the mass parameter and the interaction couplings of a massive vector multiplet are SUSY exact, although the precise type of interaction was not determined.  Our calculations also showed that ${\cal N} = 1$ D-terms are not exact and do affect the sphere free energy.  

Our results are fully consistent with the observation made in \cite{Lee:2014rca,Hosseini:2016tor,Hosseini:2017fjo} (see also \cite{Zaffaroni:2019dhb, prepotential}) that, at the two-derivative level in the bulk, the sphere free energy is directly related to the prepotential of the bulk supergravity theory.  Indeed, as shown in Section~\ref{PREPOTENTIAL}, prepotential interactions are particular cases of bulk chiral F-terms.   As far as higher-derivative interactions are concerned, the higher-derivative interactions constructed in \cite{Bobev:2021oku} using superconformal tensor calculus are other examples of chiral F-terms.  The fact that they affect the $S^3$ free energy \cite{Bobev:2021oku} is again consistent with our results.

Several different methods were used to construct ${\cal N} = 2$ supersymmetric interactions in AdS\@.  The first approach, in Section~\ref{SUSY}, used only manipulations of generators of the $\mathfrak{osp}(2|4)$ superalgebra.  This was sufficient to prove the statements above about SUSY exact interactions.  Then, in Sections~\ref{CONFORMAL} and \ref{ONSHELL}, we presented two other methods to construct supersymmetric Lagrangians in AdS\@.  The first one used off-shell superconformal theories with the ${\cal N} = 2$ Weyl multiplet and its compensator multiplets as backgrounds for matter multiplets, and the second one involved the decoupling limit of on-shell ${\cal N} = 2$ supergravity coupled to matter.

There are various questions that we leave for future work.  The first is whether the massive vector multiplet mass term and interactions are indeed SUSY exact. To settle this question, one would perhaps need to derive the kinetic and interaction Lagrangians from an appropriate off-shell formulation in which the supersymmetry variations are independent of the vector multiplet mass.  In such a formulation, it would then become clear whether the vector multiplet mass term and its interactions with other multiplets are particular cases of the supersymmetry-preserving deformations we have studied that do not affect the $S^3$ free energy.

Another question left for future work is whether the $S^3$ free energy of ${\cal N} = 2$ SCFTs captures a protected sector of these theories.  For ${\cal N} = 4$ SCFTs, it is known that the $S^3$ free energy equals the free energy of a 1d topological sector that is built from the $1/2$-BPS operators of the 3d theory \cite{Dedushenko:2016jxl,Dedushenko:2017avn,Dedushenko:2018icp}.  Thus, the $S^3$ free energy captures information only about these $1/2$-BPS operators.   It is not known whether analogous statements continue to hold true for ${\cal N} = 2$ SCFTs.  In the present work, we have shown that, in the dual $\text{AdS}_4$ theory, the $S^3$ free energy of the boundary theory does capture a protected sector consisting of special types of bulk interactions.  It would be very interesting to determine whether such a protected sector also exists more generally for all ${\cal N} =2$ SCFTs, not necessarily just for those with holographic duals.

Lastly, it would be interesting to generalize our analysis to a different numbers of spacetime dimensions and other amounts of supersymmetry.  Of particular interest is the case of ${\cal N} = 2$ superconformal field theories in 4d and their 5d holographic duals, where again the $S^4$ free energy of the mass-deformed theory can be computed using supersymmetric localization.  Another equally interesting generalization would be to supersymmetric theories on squashed spheres and other curved manifolds.  In particular, one can ask whether the partition functions of such theories also receive contributions only from a very special class of bulk interactions.  Generically, the analog of the analysis performed in Section~\ref{INTEGRATED} would involve Witten diagrams for spinning correlators.  Such calculations would be greatly simplified with the help of the bispinor embedding space formalism introduced in \cite{Binder:2020raz}.

\section*{Acknowledgments}

This research is supported in part by the Simons Foundation Grant No.~488653 (DJB, SSP, BZ) and by the US NSF under Grant Nos.~PHY-1620045 (DZF), PHY-1820651 (SSP) and~PHY-1914860 (BZ)\@.

\appendix

\section{$\mathfrak{osp}(2|4)$ algebra}
\label{OSP}

The $\mathfrak{osp}(2|4)$ algebra has as bosonic generators the $\mathfrak{sp}(4)$ generators ${\cal M}_{\alpha \beta}$ and the R-symmetry generators ${\cal R}$.  Here, we can think of $\alpha$, $\beta$ as fundamental $\mathfrak{sp}(4)$ indices, and ${\cal M}_{\alpha \beta}$ is a symmetric rank-two tensor.  The supercharges have charge $\pm 1$ under the R-symmetry, so we can represent them as ${\cal Q}_{\alpha \pm}$.  The commutation relations are
 \es{CommutOsp}{
  [{\cal M}_{\alpha \beta}, {\cal M}_{\gamma \delta}]
   &= \omega_{\alpha \delta} {\cal M}_{\beta \gamma}
    +  \omega_{\beta \delta} {\cal M}_{\alpha \gamma} 
    +  \omega_{\alpha \gamma} {\cal M}_{\beta \delta} 
    +  \omega_{\beta \gamma} {\cal M}_{\alpha \delta} \,, \\
  [{\cal Q}_{\alpha \pm}, {\cal M}_{\beta \gamma}] &= 
   \omega_{\alpha \beta} {\cal Q}_{\gamma   \pm} + 
    \omega_{\alpha \gamma} {\cal Q}_{\beta   \pm} \,, \\
   [{\cal R}, {\cal Q}_{\alpha \pm}] &= \pm i {\cal Q}_{\alpha \pm} \,, \\
   \{{\cal Q}_{\alpha +}, {\cal Q}_{\beta -} \} &=  {\cal M}_{\alpha \beta} - i \omega_{\alpha \beta} {\cal R} \,,
 }
where $\omega_{\alpha \beta}$ is the symplectic form, which is antisymmetric in the symplectic indices.

The correspondence between this presentation of the algebra and that in Section~\ref{SUSY} is as follows.  We can identify the spinor indices used in Section~\ref{SUSY} with the symplectic fundamental indices, and we can take the symplectic form to simply be the ${\cal C}$ matrix defined in \cite{Freedman:2012zz}:
 \es{omegaC}{
  \omega_{\alpha \beta} \equiv {\cal C}_{\alpha \beta} \,.
 }
The $\mathfrak{sp}(4)$ generators are the $M_{ab}$ and $P_a$ converted to spinor indices.  If we take
 \es{sp4Gen}{
  {\cal M}_{\alpha \beta} = -\frac{1}{2} M_{ab} \gamma^{ab}_{\alpha \beta} + L P_a \gamma^a_{\alpha \beta} \,,
 }
then the commutation relation in the first line of \eqref{CommutOsp} is obeyed.  For the R-symmetry generator, we simply take ${\cal R} = R$.  And lastly, for the supercharges, if we choose
 \es{calQ}{
  {\cal Q}_{\alpha+} = \sqrt{2L} \left( \Q^1_\alpha - i \Q_{2\alpha} \right) \,, \qquad
   {\cal Q}_{\alpha - } = \sqrt{2L} \left( - i \Q^2_\alpha -  \Q_{1\alpha} \right) \,,
 }
then one can check that all other commutation relations in \eqref{CommutOsp} are satisfied.

\section{When are D-terms exact?}
\label{CLOSEDEXACT}

In Section~\ref{EXACTNESS}, we argued that bulk $\cN=2$ D-terms are exact under some of the $\mathfrak{su}(2|1)_\ell\times\mathfrak{su}(2)_r \subset \mathfrak{osp}(2|4)$ supercharges.  This observation implies that the $S^3$ partition function and all correlation functions of $\mathfrak{su}(2|1)_\ell\times\mathfrak{su}(2)_r $-invariant operators in the dual ${\cal N} = 2$ SCFT are independent of the D-term bulk couplings. In this appendix, we consider the more general question of when D-terms for a supersymmetric algebra are exact under some supercharge of the algebra. As we shall see, for this to be the case it suffices for the algebra to contain a nilpotent operator. We then show that $\cN=1$ D-terms in $\text{AdS}_4$ are not exact, and so the $\cN=1$ $S^3$ partition function will generically depend on them.

Let us consider a supersymmetric theory on some curved manifold $\cM$ which is invariant under some supersymmetry algebra $\mathfrak g$ with $n$ fermionic generators $q_1\,,\dots\,,q_n$. Furthermore, let us deform our theory by some local interaction
 \es{calLGen}{
  \cL = \int_\cM\,d^dx\,\cO(x) \,,
 }
which preserves both the R-symmetries (if any are present) and the isometries of $\cM$. By acting with the supercharges $q_i$ we can construct new deformations of the theory:
\begin{equation}\label{LStates}
\cL\,,\quad q_i\cL\,,\quad q_iq_j\cL \text{ with }i<j\,,\quad\dots\,,\quad q_1\cdots q_n\cL\,.
\end{equation}
Here we have used the $\mathfrak g$ anticommutation relations to order the $q_i$'s so that the supercharges with lower values of $i$ are placed to the left.   In general, it is possible that some of the interactions in \eqref{LStates} simply vanish or, more generally, that not all the interactions in \eqref{LStates} are linearly-independent.     Linear dependencies would necessarily occur when $\cO(x)$ belongs to a short multiplet of the supersymmetry algebra, but they could also be present for certain special $\cO(x)$ that belong to long SUSY multiplets.  Let us consider, however, the ``generic'' case in which no such linear dependencies are present.    In this case, ${\cal O}(x)$ necessarily belongs to a long SUSY multiplet, and the $2^n$ interactions in \eqref{LStates} are a basis for the carrier space $V_{\fcy L}$ of a finite-dimensional representation of $\mathfrak{g}$.\footnote{Note that while local operators transform in infinite-dimensional representations of the supersymmetry algebra, the integrated operators \eqref{LStates} transform in a finite-dimensional representation. For example, the repeated action of bosonic generators of the supersymmetry algebra on ${\cal O}(x)$ yields infinitely-many linearly-independent local operators.  By contrast, acting with the bosonic generators on ${\cal L}$ gives a vanishing result; this is due to the the invariance of ${\cal L}$ under R-symmetry and isometries of ${\cal M}$, as well as the fact that the integrals of total derivatives vanish.} 

We are interested in studying deformations built from long multiplets that preserve the supersymmery algebra $\mathfrak g$. That is, for any $\cL$ as above we want to find some interaction
\begin{equation}
\cL_D = a^{(0)} \cL + \sum_i a^{(1)}_i q_i\cL + \sum_{i<j} a^{(2)}_{ij} q_iq_j\cL  + \dots + a^{(n)}_{1\ldots n} q_1\cdots q_n\cL
\end{equation}
which preserves the supersymmetry algebra $\mathfrak g$ and the isometries of $\cM$, and, in particular, is annihilated by all $q_i$. Such interactions are the most general analogues of D-terms. In the following discussion, we assume that such a SUSY- and isometry-invariant $\cL_D$ can be found, and we study whether or not the partition function depends on it.  

In particular, we will now argue that if $\mathfrak g$ contains some nilpotent supercharge, then $\cL_D$ is exact under this nilpotent supercharge. Without loss of generality we can assume that $q_1$ is such a nilpotent operator. We can use our basis \eqref{LStates} to decompose $V_{\fcy L}$ into two distinct subspaces. Let us define $V_0$ to be the space spanned by deformations 
 \es{qOnL}{
  q_{i_1}\cdots q_{i_m}\cL \text{ for which } i_1>1 \text{ and }i_{k+1}>i_k\,,
 }
and let $q_1V_0$ be the image of $V_0$ under $q_1$. Since both $V_0$ and $q_1V_0$ are $2^{n-1}$ dimensional spaces it immediately follows that $q_1$ is a bijection from $V_0$ to $q_1V_0$. Furthermore, we clearly have $V_{\fcy L} = V_0 \oplus q_1V_0$, and $q_1V_0\subset\text{ker}(q_1)$, and so from this we conclude that  $q_1V_0 = \text{ker}(q_1)$. Thus, any $q_1$-invariant deformation must be $q_1$-exact. In particular, because the D-term $\cL_D$ is annihilated by all supercharges, it must be $q_1$-exact.

In equation \eqref{DtermQExact} of Section~\ref{EXACTNESS}, we saw that the $\cN=2$ D-term in $\text{AdS}_4$ was exact under one of the $\mathfrak{su}(2|1)_\ell$ supercharges.  We showed this by directly constructing the $\text{AdS}_4$ D-term and then breaking the $\mathfrak{osp}(2|4)$ superalgebra down to $\mathfrak{su}(2|1)_\ell\times\mathfrak{su}(2)_r$. Using the results of this appendix, however, gives us a more direct route to this conclusion. The $\mathfrak{su}(2|1)$ superalgebra has nilpotent supercharges,\footnote{The $\mathfrak{su}(2|1)$ algebra has a $\mathfrak{u}(1)_R$ symmetry, under which the fermionic generators $q_i$ and $\bar q_i$ (where $i=1,2$ are $\mathfrak{su}(2)$ indices) are charged. Because no bosonic operator is charged under this $\mathfrak{u}(1)_R$, it follows that both $q_i$ and $\bar q_i$ are nilpotent.} and so it follows that any $\mathfrak{su}(2|1)$-invariant D-term is $\mathfrak{su}(2|1)$-exact. Because long $\mathfrak{osp}(2|4)$ multiplets decompose into long $\mathfrak{su}(2|1)$ multiplets, any the $\mathfrak{osp}(2|4)$ D-term is automatically an $\mathfrak{su}(2|1)$ D-term and therefore is $\mathfrak{su}(2|1)$-exact.

Let us now turn to the case of $\cN=1$ D-terms. The $\text{AdS}_4$ $\cN=1$ supersymmetry algebra is $\mathfrak{osp}(1|4)$.   Since in the ${\cal N} = 2$ case we considered boundary operators that preserve only an $\mathfrak{su}(2|1)_\ell \times \mathfrak{su}(2)_r$ algebra, it is now natural to consider boundary operators preserving an $\mathfrak{osp}(1|2)_\ell \times \mathfrak{su}(2)_r$ subalgebra of $\mathfrak{osp}(1|4)$.  Unlike $\mathfrak{su}(2|1) \cong \mathfrak{osp}(2|2)$, the superalgebra $\mathfrak{osp}(1|2)$ does not contain any nilpotent charges, and so we have to work harder to check whether $\cN=1$ D-terms are exact or not. 

The $\mathfrak{osp}(1|2)$ superalgebra is generated by the two supercharges $q_\pm$, along with the three $\mathfrak{sp}(2)$ generators $X_0$ and $X_\pm$, which satisfy the (anti)-commutator relations 
\begin{equation}\begin{split}
\{q_+,q_-\} &= X_0\,,\qquad \{q_\pm,q_\pm\} = -X_\pm\,,\\
[X_0,q_\pm] &= \pm q_\pm\,,\qquad [X_\pm\,,q_\mp] = -q_\pm\,,\\
[X_0,X_\pm] &= \pm 2X_\pm\,, \qquad [X_+,X_-] = X_0\,.
\end{split}\end{equation}
For a generic local interaction $\cL$ which preserves $\mathfrak{sp}(2)$, the full $\mathfrak{osp}(1|2)$ multiplet of deformations consists of the four linearly-independent states
\begin{equation}
\cL,\qquad q_+\cL\,,\qquad q_-\cL\,,\qquad q_+q_-\cL\,,
\end{equation}
of which only the first and last preserve $\mathfrak{sp}(2)$. The D-term therefore takes the form
\begin{equation}
\cL_D = q_+q_-\cL + \alpha \cL
\end{equation}
for constant $\alpha$ (we can always rescale $\cL_D$ such that the coefficient of $q_+q_-\cL$ is 1). Imposing the condition $q_+\cL_D = 0$, we find that $\alpha = -\frac12$, and so
\begin{equation}
\cL_D = q_+q_-\cL-\frac12\cL\,.
\end{equation}
Although $\cL_D$ is $q_\pm$-invariant, it is manifestly not $q_\pm$-exact. We thus conclude that $\cN=1$ D-terms generically change correlation functions of $\mathfrak{osp}(1|2)_\ell \times \mathfrak{su}(2)_r$-invariant operators of holographic ${\cal N} =1$ SCFTs\@.  

Although we have focused on $\mathfrak{osp}(1|2)$, it is straightforward to extend our arguments to the superalgebra $\mathfrak{osp}(1|2m)$ for any $m\geq1$. In particular, as mentioned above, the ${\cal N} = 1$ $\text{AdS}_4$ algebra is $\mathfrak{osp}(1|4)$, which means that the $S^3$ partition function of an ${\cal N} = 1$ SCFT  (with no insertions) is in general modified by bulk D-terms.  Similarly, the (complexified) $\cN=1$ supersymmetry algebra on $S^4$ is also $\mathfrak{osp}(1|4)$.  Our argument implies that the $S^4$ partition function is modified by D-terms, as previously shown in \cite{Knodel:2014xea}.

\section{Another perspective on the boundary real mass deformation}
\label{REALMASSANOTHER}

In this Appendix, let us present another perspective on the real mass deformation of the boundary SCFTs on $S^3$ that we discussed from a CFT perspective in Section~\ref{REALMASSSECTION}.  As discussed there, the change in the action is given by $\mathfrak{m}$ multiplied by the integral of the linear combination $iJ + K$ of the scalars in a $U(1)$ flavor current multiplet.  (We set the radius of $S^3$ to $1$ here for simplicity.)  In the context of the holographic duality,  a flavor current multiplet on the boundary couples to a dynamical bulk vector multiplet.  Let us denote such a vector multiplet by $(X, \bar X, A_\mu, Y_{ij},  \Omega_i, \Omega^i)$.  The SUSY transformation rules are given in \eqref{SUSYVector} with the upper index $I$ omitted.    Denoting by $(Y^1, Y^2, Y^3) = \vec{Y} = - \frac 12 \vec{\tau}_i{}^j Y_j{}^i $, the variation of $Y^3$ can also be written as
 \es{deltaY3Again}{
  \delta Y^3 &= \frac i4  D_a \left( -\bar \epsilon_1 \gamma^a \Omega_2 -  \bar \epsilon_2  \gamma^a \Omega_1 
  + \bar \epsilon^1 \gamma^a \Omega^2 + \bar \epsilon^2 \gamma^a \Omega^1\right)  \\
   &{}+   \frac{1}{2} \left( \bar \epsilon^2  \Omega_2 +  \bar \epsilon^1  \Omega_1 
  + \bar \epsilon_2  \Omega^2 + \bar \epsilon_1  \Omega^1\right) \,.
 } 

Let us now identify the operator $\int d^3\vec{x} \, \sqrt{g(\vec{x})} \left[ i J(\vec{x}) + K(\vec{x})  \right]$ that couples linearly to the real mass parameter.  One approach is to identify the boundary operators $J(\vec{x})$ and $K(\vec{x})$ as limits of bulk operators from the vector multiplet.    Another approach, which is the one we will follow, is to use supersymmetry:  the real mass deformation should be invariant under $\delta_{A+}$ with $A = 1,2$, and this condition alone will determine the linear term in $\mathfrak{m}$ up to normalization.

Note that the spinors $T_1$ and $T_2$ in \eqref{KillingSpinors} obey the condition $P_L T_A = r \gamma_4 P_R T_A$.  Then, the parameters in \eqref{SolKS1}--\eqref{SolKS2} obey $\epsilon^1 = ir \gamma_4 \epsilon_2$ and $\epsilon^2 = ir \gamma_4 \epsilon_1$.  
Using these relations, we have 
 \es{TranRulesVector2}{
   \delta X &= -\frac{1}{2} ir  \left( \bar \epsilon_2 \gamma_4 \Omega_1 
    + \bar \epsilon_1 \gamma_4 \Omega_2 \right) \,, \qquad 
   \delta \bar X = \frac{1}{2} \left( \bar \epsilon_1 \Omega^1 + \bar \epsilon_2 \Omega^2 \right)  \,, \\
 \delta \left(Y^3  - X - \bar X\right) &=  \frac i4  D_a \left( -\bar \epsilon_1 \gamma^a \Omega_2 -  \bar \epsilon_2  \gamma^a \Omega_1 
  - ir\bar \epsilon_2 \gamma_4 \gamma^a \Omega^2 - ir  \bar \epsilon_1 \gamma_4 \gamma^a \Omega^1\right)  \,.
 }  
Splitting the coordinates into $a = m = 1, 2, 3$ and $a=4$, we have
 \es{deltaAgain}{
\delta \left(Y^3  - X - \bar X \right) &=  \frac i4  D_4 \left( -\bar \epsilon_1 \gamma^4 \Omega_2 -  \bar \epsilon_2  \gamma^4 \Omega_1 
  - ir \bar \epsilon_2  \Omega^2 - ir  \bar \epsilon_1 \Omega^1\right)  
   + D_m \left( \cdots \right) \,.
  }
Thus, 
 \es{deltaAgain2}{
 \delta \left(Y^3  - X - \bar X   -  \frac{1}{1-r^2} \partial_r \left( r  \bar X  + \frac{1}{r} X \right)  \right) = D_m \left( \cdots \right) \,,
 } 
and so the quantity
 \es{MassTerm}{
  {\cal B} \equiv \lim_{r \to 1} \, (1-r)^3 \int_{S_r} d^3 x\, \sqrt{\gamma} 
   \left[   \frac{1}{1-r^2} \partial_r \left( r  \bar X  + \frac{1}{r} X \right) + X + \bar X   - Y^3 \right] 
 }
integrated over a sphere $S_r$ of radius $r$ is supersymmetric with respect to $\delta_{A\pm}$, with $A = 1, 2$.  The overall factor of $(1-r)^3$ was chosen so that the $r\to 1$ limit is finite.  

The operator ${\cal B}$ is, up to normalization, the CFT operator that couples linearly to the real mass parameter:
 \es{RealMassAgain}{
  \int d^3\vec{x} \, \sqrt{g(\vec{x})} \left[ i J(\vec{x}) + K(\vec{x})  \right] \propto {\cal B} \,.
 }
One can determine the overall normalization constant by computing, for instance, two point functions of this operator, but we will not pursue that here.

\section{Flat limit of the hypermultiplet manifold}
\label{FLATLIMIT}

\subsection{General properties}

As mentioned in the main text, in ${\cal N} = 2$ supergravity the hypermultiplet scalar manifold ${\cal M}$ is a quaternionic-K\"ahler manifold of negative Ricci curvature proportional to $=\kappa^2$.  For $n_H$ hypermultiplets, we parameterize ${\cal M}$ by coordinates $q^u$, with $u = 1, \ldots 4 n_H$, and let us denote its metric by $h_{*\, uv}$.  Part of the definition of a quaternionic-K\"ahler manifold is that it is endowed with a triplet of hypercomplex structures $\vec{J}_{*\, u}{}^v$ that are covariantly constant up to an $SU(2)$ rotation:
 \es{HypercompEq}{
  \nabla_w \vec{J}_{*\, u}{}^v + 2 \vec{\omega}_{*\, w} \times \vec{J}_{*\, u}{}^v = 0 \,,
 }
where $\vec{\omega}_{*\, u}$ are functions on the manifold that play the role of the $SU(2)$ connection.  Both the metric $h_{*\, uv}$ and the hypercomplex structures $\vec{J}_{*\, u}{}^v$ can be obtained from a frame field as in \eqref{eq:frame_properties}, but this detail will not be needed in the discussion that follows.

It can be shown that the quaternionic-K\"ahler manifolds are Einstein, and in supergravity the Ricci tensor and Ricci scalar are normalized to be
 \es{RicciSG}{
  R_{*\, uv} = - (n_H + 2) \kappa^2 h_{*\, uv} \,, \qquad 
  R_* = - 4 n_H (n_H + 2) \kappa^2 \,.
 }
Another property of the quaternionic-K\"ahler manifolds is that the $SU(2)$ curvature is proportional to the complex structures, with the coefficient of proportionality related to the Ricci curvature in \eqref{RicciSG}:
 \es{calRSG}{
  \vec{{\cal R}}_{*\, uv} \equiv \partial_u \vec{\omega}_{*\, v} 
   -  \partial_v \vec{\omega}_{*\, u} + 2 \vec{\omega}_{*\, u} \times \vec{\omega}_{*\, v} 
    = - \frac{\kappa^2}{2} \vec{J}_{*\, uv} \,.
 }

\subsection{Flat limit}

Since the curvature of ${\cal M}$ is proportional to $\kappa^2$, when $\kappa \to 0$, the hypermultiplet scalar manifold becomes flat, with the metric and the hypercomplex structures approaching the ones in flat space, as in \eqref{HyperComplexLimit}, which we reproduce here for convenience:
\es{HyperComplexLimitAppendix}{
	h_{*\, uv} = \delta_{uv} + O(\kappa^2) \,, \qquad
	\vec{J}_{*\, uv} = \vec{J}_{uv} + O(\kappa^2) \,,
}
As mentioned in the main text, the $SU(2)$ connection vanishes in this case, but it is important to determine the $O(\kappa^2)$ term in $\vec{\omega}_{*\, u}$.  This can be determined from solving \eqref{calRSG} at leading order in $\kappa^2$.  From this equation we see that $\vec{\omega}_{*\, u} = O(\kappa^2)$, and thus at leading order we can neglect the term quadratic in $\vec{\omega}_*$:
 \es{curvatureLeading}{
  \partial_u \vec{\omega}_{*\, v} 
   -  \partial_v \vec{\omega}_{*\, u}  \approx - \frac{\kappa^2}{2} \vec{J}_{uv} \,.
 }
A solution of this equation is
 \es{omegaLeadingAppendix}{
  \vec{\omega}_{*\, u} = \frac{\kappa^2}{4} \vec{J}_{uv} q^v + O(\kappa^4) \,,
 }
as given in \eqref{omegaLeading}.

\subsection{Projective space and its flat limit}

A more complete derivation of both \eqref{HyperComplexLimitAppendix} and \eqref{omegaLeadingAppendix} can be derived from an explicit example of a quaternionic-K\"ahler manifold ${\cal M}$.  We consider the negatively-curved quaternionic projective space
 \es{Hpn}{
  \mathbb{HP}_{n_H} \equiv \frac{USp(2n_H, 2)}{USp(2n_H) \times SU(2)} 
 }
of quaternionic dimension $n_H$ and, correspondingly, real dimension $4 n_H$.  The constant curvature metric on this space is a quaternionic analog of the Fubini-Study metric, namely: 
 \es{MetricHP}{
  ds^2_* = \frac{dq^u dq^u}{1 - \frac{\kappa^2}{4} q^v q^v}
   + \frac{\kappa^2 \left[ (q^u dq^u)^2 
     + (dq^u\, J_{1 u}{}^w q^w)^2 
      + (dq^u\, J_{2 u}{}^w q^w)^2 + (dq^u\, J_{3 u}{}^w q^w)^2\right] }{4 \left( 1 - \frac{\kappa^2}{4} q^v q^v\right)^2} 
    \,,
 }
where repeated indices are summed over regardless of their up/down placement, and where $J_{1 u}{}^v$, $J_{2 u}{}^v$, and $J_{3 u}{}^v$ are the hypercomplex structures in flat $\R^{4 n_H}$ space obeying $J_{i u}{}^v J_{j v}{}^w = - \delta_{ij} \delta_u^w + \epsilon_{ijk} J_{k u}{}^w$.  For example, $\vec{J}{}_u{}^v$ can be taken to be block diagonal, with each $4 \times 4$ block on the diagonal being equal to the explicit matrices in \eqref{HypercomplexExplicit}. The Ricci curvature derived from the metric \eqref{MetricHP} is as in \eqref{RicciSG}.  The hypercomplex structures $\vec{J}_{*\, u}{}^v$ with one index down and one up are precisely equal to the flat space ones:
 \es{HyperComplexHP}{
  \vec{J}_{*\, u}{}^v = \vec{J}_{u}{}^v \,, 
 }
as can be checked by noticing that they obey the required multiplication relations as well as covariantly-constant property \eqref{HypercompEq} with the connection
 \es{ConnectionHP}{
  \vec{\omega}_{*\, u} = \frac{\kappa^2}{4} \left( 1 - \frac{\kappa^2}{4} q^v q^v \right)  \vec{J}_u{}^w q^w \,.
 }
Here, again, repeated indices are summed over regardless of their placement.

It is then straightforward to expand \eqref{MetricHP}--\eqref{ConnectionHP} in $\kappa^2$ and reproduce \eqref{HyperComplexLimitAppendix} and \eqref{omegaLeadingAppendix}, and even go to higher orders in the $\kappa^2$ expansion if one so desires.

\bibliographystyle{ssg}
\bibliography{S3}

\end{document}